\documentclass[12pt,english,american]{article}
\usepackage[T1]{fontenc}
\usepackage[latin9]{inputenc}
\usepackage{geometry}
\usepackage{color}
\usepackage{babel}
\usepackage{refstyle}
\usepackage{amsthm}
\usepackage{amstext}
\usepackage{amssymb}
\usepackage{mathrsfs}
\usepackage{amsmath}
\usepackage{cancel}  
\usepackage{graphics} 
\usepackage{graphicx}
\usepackage{yfonts}
\usepackage{soul}
\usepackage{tikz}
\usepackage[unicode=true,pdfusetitle,
 bookmarks=true,bookmarksnumbered=false,bookmarksopen=false,
 breaklinks=true,pdfborder={0 0 0},backref=false,colorlinks=true]
 {hyperref}
\hypersetup{
 linkcolor=blue,citecolor=blue, urlcolor=blue}

\makeatletter

\newcommand{\cI}{\mathcal{I}}
\newcommand{\cJ}{\mathcal{J}}
\newcommand{\cK}{\mathcal{K}}

\def\qed{$\Box$\medskip}

\newcommand{\beq}{\begin{equation}}
\newcommand{\eeq}{\end{equation}}
\newcommand{\beqa}{\begin{eqnarray}}
\newcommand{\eeqa}{\end{eqnarray}}
\newcommand{\ben}{\begin{arabicenumerate}}
\newcommand{\een}{\end{arabicenumerate}}

\def\bel{\begin{lem} } 
\def\eel{\end{lem} }
\def\bet{\begin{thm}}
\def\eet{\end{thm}}
\def\bed{\begin{defn}}
\def\eed{\end{defn} }

\def\bec{\begin{cor}}
\def\eec{\end{cor}}
\def\ber{\begin{rem}}
\def\eer{\end{rem}}

\usepackage{enumitem}		
\theoremstyle{plain}
\newtheorem{thm}{\protect\theoremname}[section]
\theoremstyle{definition}
\newtheorem{defn}[thm]{\protect\definitionname}
\theoremstyle{plain}
\newtheorem{prop}[thm]{\protect\propositionname}
\theoremstyle{plain}
\theoremstyle{remark}
\newtheorem{rem}[thm]{\protect\remarkname}
\theoremstyle{plain}
\newtheorem{lem}[thm]{\protect\lemmaname}
\theoremstyle{plain}
\newtheorem{cor}[thm]{\protect\corollaryname}

\usepackage{txfonts,refstyle,xcolor}

\newref{con}{name = Conjecture\ }
\newref{prop}{name = Proposition\ }
\newref{def}{name = Definition\ }
\newref{sec}{name = Section\ }
\newref{sub}{name = Section\ }
\newref{thm}{name = Theorem\ }
\newref{lem}{name = Lemma\ }
\newref{cor}{name = Corollary\ }
\newref{fig}{name = Figure\ }
\newref{rem}{name =  Remark\ }

\usepackage{bbm}
\newcommand{\charf}{\mathbbm{1}}

\usepackage[all]{xy}

\newcommand{\xyR}[1]{%
     \makeatletter
     \xydef@\xymatrixrowsep@{#1}
     \makeatother
}

\newcommand{\xyC}[1]{%
     \makeatletter
     \xydef@\xymatrixcolsep@{#1}
     \makeatother
}

\newcommand{\ncol}[1]{\color{normalcolor}}

\makeatother

\addto\captionsamerican{\renewcommand{\corollaryname}{Corollary}}
\addto\captionsamerican{\renewcommand{\definitionname}{Definition}}
\addto\captionsamerican{\renewcommand{\lemmaname}{Lemma}}
\addto\captionsamerican{\renewcommand{\propositionname}{Proposition}}
\addto\captionsamerican{\renewcommand{\remarkname}{Remark}}
\addto\captionsamerican{\renewcommand{\theoremname}{Theorem}}
\addto\captionsenglish{\renewcommand{\corollaryname}{Corollary}}
\addto\captionsenglish{\renewcommand{\definitionname}{Definition}}
\addto\captionsenglish{\renewcommand{\lemmaname}{Lemma}}
\addto\captionsenglish{\renewcommand{\propositionname}{Proposition}}
\addto\captionsenglish{\renewcommand{\remarkname}{Remark}}
\addto\captionsenglish{\renewcommand{\theoremname}{Theorem}}
\providecommand{\corollaryname}{Corollary}
\providecommand{\definitionname}{Definition}
\providecommand{\lemmaname}{Lemma}
\providecommand{\propositionname}{Proposition}
\providecommand{\remarkname}{Remark}
\providecommand{\theoremname}{Theorem}

\addto\captionsamerican{\renewcommand{\corollaryname}{Corollary}}
\addto\captionsamerican{\renewcommand{\definitionname}{Definition}}
\addto\captionsamerican{\renewcommand{\lemmaname}{Lemma}}
\addto\captionsamerican{\renewcommand{\propositionname}{Proposition}}
\addto\captionsamerican{\renewcommand{\remarkname}{Remark}}
\addto\captionsamerican{\renewcommand{\theoremname}{Theorem}}
\addto\captionsenglish{\renewcommand{\corollaryname}{Corollary}}
\addto\captionsenglish{\renewcommand{\definitionname}{Definition}}
\addto\captionsenglish{\renewcommand{\lemmaname}{Lemma}}
\addto\captionsenglish{\renewcommand{\propositionname}{Proposition}}
\addto\captionsenglish{\renewcommand{\remarkname}{Remark}}
\addto\captionsenglish{\renewcommand{\theoremname}{Theorem}}
\providecommand{\corollaryname}{Corollary}
\providecommand{\definitionname}{Definition}
\providecommand{\lemmaname}{Lemma}
\providecommand{\propositionname}{Proposition}
\providecommand{\remarkname}{Remark}
\providecommand{\theoremname}{Theorem}

\begin{document}

\title{Low energy spectrum of the  XXZ model coupled to a magnetic field}
 \author{Simone Del Vecchio\footnote{Dipartimento di Matematica, Universit\`a degli Studi di Bari, Italy / email: simone.delvecchio@uniba.it}\,,  J\"urg Fr\"ohlich\footnote{Institut f\"ur Theoretiche Physik, ETH-Z\"urich , Switzerland / email: juerg@phys.ethz.ch}\,, Alessandro Pizzo\footnote{Dipartimento di Matematica, Universit\`a di Roma ``Tor Vergata", Italy
/ email: pizzo@mat.uniroma2.it}
, and Alessio Ranallo\footnote{Section de math\'ematiques, Universit\'e de Gen\`eve, Switzerland
/ email: alessio.ranallo@unige.ch}}
  \date{\today}
\maketitle

\abstract{For a class of Hamiltonians of $XXZ$ spin chains in a uniform external  magnetic field that are small 
quantum perturbations of an Ising Hamiltonian, it is shown that the spectral gap above the ground-state energy 
remains strictly positive when the perturbation is turned on, uniformly in the length of the chain. This result is proven 
for perturbations of both the \emph{ferromagnetic} and the \emph{antiferromagnetic} Ising Hamiltonian. In the 
antiferromagnetic case, the external magnetic field is required to be small. For a chain of an even number of sites, 
the two-fold degenerate ground-state energy of the unperturbed antiferromagnetic Hamiltonian may split into two 
energy levels separated by a very small gap.
These results are proven by using a new, quite subtle refinement of a method developed in earlier work and used to 
iteratively block-diagonalize Hamiltonians of systems confined to ever larger subsets of a lattice by using 
\emph{strictly} local unitary conjugations. The new method developed in this paper provides complete control of 
\emph{boundary effects} on the low-energy spectrum of perturbed Ising chains uniformly in their length.
}


\section{Introduction}

In this paper we study short-range perturbations of the Hamiltonian of an Ising chain. An example
covered by our analysis is the celebrated $XXZ$ chain, whose Hamiltonian includes nearest-neighbour 
interactions of quantum spins (with spin $1/2$) with coupling constants of two different strengths, 
a large ``parallel'' one, $J$, in interaction terms among $z$-components of neighboring spins, 
and a small, ``perpendicular'' one in interaction terms among perpendicular ($x$- and $y$-) components. 
In addition, interaction terms of the spins in the chain with an external magnetic field of strength $h$ 
in the $z$-direction may be included in the Hamiltonian.
In this paper we focus our attention on this particular class of models, because they have attracted 
quite a lot of interest. But our methods can be applied to a considerably more general family of models, 
as specified later on; see Remark \ref{rem-bulk}. The results established in this paper cover perturbations of \textit{ferromagnetic} 
($J>0$) and \textit{antiferromagnetic} ($J<0$) Ising Hamiltonians, provided that the perpendicular coupling 
constant is a small parameter as compared to $|J|$. With regard to the strength, $h$, of the magnetic field, 
the regimes we study in this paper depend on the sign of $J$: for $J<0$, $h$ is assumed to be small enough such that 
 an antiferromagnetic ordering of spins is preserved.

Our analysis relies on an iterative block-diagonalization of local Hamiltonians supported in ever longer intervals 
(i.e., subsets of successive sites) of the lattice,  with the help of unitary (Lie Schwinger) conjugations. The
sequence of such conjugations applied to Hamiltonians of subsystems of the chain yields a ``flow'' of transformed 
Hamiltonians that contain effective potentials supported in intervals of arbitrary length. In this respect our method 
of analysis is an elaboration on the one introduced in \cite{FP}, where perturbations of \emph{ultralocal} 
Hamiltonians of quantum chains have been considered; generalizations of the technique to higher dimensional 
lattices, complex coupling constants, and to bosonic Hamiltonians have appeared in \cite{DFPR1}, \cite{DFPR2}, \cite{DFPR3}, and \cite{DFP}. The novelties introduced in the present paper enable us to deal with perturbations 
of some  Hamiltonians, like the one of the antiferromagnetic Ising chain in an external magnetic field,  which do not satisfy all assumptions of most of 
the methods earlier introduced in the literature to study quantum spin chains.  Thus we develop a robust 
method for studying the low-lying energy spectrum of a class of model Hamiltonians with nontrivial ground-state subspaces, describing chains \emph{with general boundary conditions}.

Our main result (see part (b) of {\bf{Theorem}} in Sect. \ref{statement-sect}) concerns small perturbations of the Hamiltonian of \textit{antiferromagnetic} Ising chains, i.e., of the sum of the Ising Hamiltonian and the interaction term with the external magnetic field, 
which,  under natural assumptions on the size of $h$ and for chains with an even number of sites, has a ground-state subspace 
which is two-dimensional. But, in contrast to models such as the celebrated  ``AKLT model" (see \cite{AKLT}), no 
\emph{local topological quantum order} condition (see \cite{BHM}, \cite{MN}) holds. In this respect, a restricted notion 
of the LTQO condition has been devised in \cite{NSY3} for the ground-state subspace of (unperturbed) Hamiltonians 
exhibiting some type of symmetry breaking,  e.g.,  the translation symmetry, but at the price of not controlling the effect 
of the  boundary conditions on the energy spectrum. To our knowledge, the antiferromagnetic $XXZ$ chains in an external 
magnetic field and {\emph{with general boundary conditions}} have not been studied previously using mathematically 
rigorous techniques, with the exception of some special findings within the range of Bethe ansatz techniques; see \cite{F}. 
Recent numerical results based on tensor network renormalization group techniques agree with Bethe ansatz benchmarks 
for some values of the parameters in the Hamiltonian of such chains; see \cite{RW} and references therein.

Our analysis enables us to prove that if $J>0$ (ferromagnetic Hamiltonian), uniformly in the length of the chain, 
a spectral gap (of order $J+h$) above the ground-state energy of the unperturbed Hamiltonian persists when the perpendicular interaction terms are added,  provided the coupling
constant of the latter is sufficiently small. For the antiferromagnetic model, with $J<0$ and sufficiently small $h$, 
we show that, for chains with an odd number of sites there is a gap of order $h$ above the ground-state energy 
provided the  coupling constant $t$ of the perpendicular interaction is sufficiently small compared to $h$, similarly to the ferromagnetic case. For antiferromagnetic chains with an even number of sites, the two-fold degenerate 
ground-state energy of the unperturbed Hamiltonian may split into two energy levels whose difference is, however, 
bounded above by a quantity of order the coupling constant $t$, which is assumed much smaller than $|J|-h$. This splitting is a boundary 
effect. 

\noindent
We stress that our results on small perturbations of the Ising model coupled to a parallel magnetic field hold for short-range but otherwise rather generic perturbations and general boundary conditions. Previous results (see  \cite{NSY3}) concerning the antiferromagnetic chain, based  on the 
Bravyi-Hastings-Michalakis approach, necessitate the use of \textit{periodic} boundary conditions, and,  consequently, the two-fold degeneracy of the ground-state eigenvalue is protected by symmetry.

In order to cope with the fact that the unperturbed Hamiltonian consists of terms that are \emph{local} 
but \emph{not  ultralocal} (i.e., not on-site),   the  conjugations are chosen  strictly local but their supports are enlarged  as compared to  those of the potentials to be block-diagonalized. Consequently,  the problem boils down to estimating the smallness of some commutators associated with the conjugation of unperturbed terms of the Hamiltonian.  In order to provide upper bounds to the operator norms of such commutators,   for the models treated here we can simply take advantage of the triviality of the Ising dynamics.


The analysis presented in this paper only involves spin operators, i.e., no domain-wall representation of
the models is used. Thus, in the interaction terms of the spins with the magnetic field, no \emph{non-local} 
operators appear. The connection of the models studied in this paper to models of interacting fermions 
can be made by using a Klein-Jordan-Wigner transformation to fermionic operators in our analysis of
quantum spin chains. This enables us to draw conclusions on the low-lying energy spectrum of 
one-dimensional systems of (spinless) fermions with Hubbard-type interactions that can be either repulsive 
(corresponding to an antiferromagnetic $XXZ$ spin chain) or attractive (corresponding to a 
ferromagnetic chain). Indeed, our treatment of open boundary conditions imposed on the Hamiltonians of finite chains enables us to study interesting physics 
that is not visible in the thermodynamic limit, anymore.

This paper can be viewed to be a contribution to a research area pertaining to the characterization of 
\textit{``topological phases'';} see, e.g.,  \cite{BN, BH, BHM, DS, K, KT,  LMY, NSY, NSY2, NSY3, H, O, S}, 
which  has been pursued very actively in recent years. In these studies, known techniques and novel 
ones have been tailored to the study of the low-lying energy spectrum of quantum lattice systems.

\noindent
For the ferromagnetic $XXZ$ Hamiltonian of an infinite chain in the absence of an external magnetic field, 
and for an arbitrary ratio greater than $1$ between the ``parallel'' and the ``perpendicular'' coupling constants, 
proofs of a strictly positive spectral energy gap above the ground-state energies can be found in \cite{KN}, 
for spin $\frac{1}{2}$, and in \cite{KNS} for arbitrary spin. Further results on the low lying spectrum  (i.e., on the portion usually referred to as \emph{droplet spectrum}) of this Hamiltonian
have been derived in \cite{NSt}, \cite{NSS}, and \cite{FS}.

\setcounter{equation}{0}

\subsection{Definition of the model}\label{prop-XXZ}

We consider a one-dimensional lattice, $\Lambda,$ consisting of an arbitrary number, $N< \infty$, of sites.
With every site $j \in \Lambda$ we associate a Hilbert space $\mathcal{H}_j \simeq \mathbb{C}^{2}$.
By $\sigma_j=\left(\sigma_j^x, \sigma_j^y, \sigma_j^z\right)$ we denote the Pauli matrices acting on
$\mathcal{H}_j$, for every $j \in \left\{1,\dots, N\right\}$. The Hilbert space of the entire chain is given 
by \begin{equation}\label{tensorprod}
\mathcal{H}^{(N)}:= \bigotimes_{j=1}^{N} \mathcal{H}_{j}\,.
\end{equation}
We consider small, finite-range perturbations of both the ferromagnetic ($J>0$) and the antiferromagnetic  
($J<0$)  Ising Hamiltonian, $H^{0}_\Lambda$, with a magnetic field of strength $h >0$ in the $-z$-direction,
where 
\begin{equation}
H^{0}_\Lambda := -J\sum_{i=1}^{N-1}\sigma_i^z\sigma_{i+1}^z-h\sum_{i=1}^N\sigma_i^z\,.
\end{equation}
In particular, we study the Hamiltonian of the $XXZ$ chain in a magnetic field, which is given by
\begin{equation}\label{XXZ-ham}
K_{\Lambda}\equiv K_{\Lambda}(t):=-J\sum_{i=1}^{N-1}\sigma_i^z\sigma_{i+1}^z+\frac{t}{2}\sum_{i=1}^{N-1}(\sigma_i^x\sigma_{i+1}^x+\sigma_i^y\sigma_{i+1}^y)-h\sum_{i=1}^N\sigma_i^z
\end{equation}
where $t \in \mathbb{R}$ is a coupling constant with $|t|$ small as compared to $|J|$ and, in some cases, to $h$. (We could also
consider tilting the external magnetic field a little.) More precisely, we make the following assumptions:
\begin{itemize}
\item[i)]
For the ferromagnetic coupling, $J>0$,
\begin{equation}\label{ferrocond}
|t|\ll J+h\,.
\end{equation}
\item[ii)]
For the antiferromagnetic  coupling, $J<0$, 
\begin{equation}\label{ass-parameters}
 |t|\ll |J|-h\,,\quad\,h< \frac{|J|}{2};
\end{equation}
 if $\Lambda$ has an odd number of sites, we may also consider $|t|\ll h$; see Remark \ref{h-odd}.
\end{itemize}
\subsubsection{Local Hamiltonians}
To implement the iterative \emph{local} Lie-Schwinger block-diagonalization method, which will be our main
tool to study the low-energy spectrum of the Hamiltonian in (\ref{XXZ-ham}), it is useful to define unperturbed
\textit{local} Hamiltonians associated with intervals $\mathcal{I}\subseteq \Lambda=\left\{1,\dots, N\right\}$ of 
arbitrary length (where an interval is a subset of $\Lambda$ consisting of successive sites).
We say that a self-adjoint operator $A$ is a \textit{local ``observable'' supported} in the interval $\mathcal{I}$ if
\begin{equation} 
A = A' \otimes \charf
\end{equation}
where $A'$ acts on $ \bigotimes_{j\in\mathcal{I} }\,\, \mathcal{H}_{j}\,,$ and $\charf$ is the identity operator on $\bigotimes_{j\notin\mathcal{I} }\,\, \mathcal{H}_{j}\,.$ 
The unperturbed local Hamiltonian associated with $\mathcal{I}$ is
\begin{equation}\label{unper-H0}
H^0_{\mathcal{I}}:=-J\sum_{i\,:\, i,i+1\, \in \mathcal{I}}\sigma_i^z\sigma_{i+1}^z-h\sum_{i\,\in \mathcal{I}}\sigma_i^z\,.
\end{equation}
Since $H^0_{\mathcal{I}}$ is \textit{not} additive under taking the union of adjacent intervals, i.e., 
\begin{equation}
H^0_{\mathcal{I} \cup \mathcal{I}'}\neq H^0_{\mathcal{I}} + H^0_{\mathcal{I}'}
\end{equation}
where $\mathcal{I} \cap \mathcal{I}'$ consists of one single site, 
we will need auxiliary Hamiltonians related to $H^0_{\mathcal{I}}$ ($\mathcal{I}\subset \Lambda$) but enjoying 
the additivity property, in order to control the effective interaction terms created by the 
block-diagonalization algorithm introduced below.  To this end, we 
define
\begin{equation}\label{unper-HC}
H^C_{\mathcal{I}}:=-\sum_{i\,:\, i,i+1\, \in \mathcal{I}}\Big\{J\sigma_i^z\sigma_{i+1}^z+\frac{h}{2}[\sigma^{z}_i+\sigma^{z}_{i+1}]\Big\}\,,
\end{equation}
where the superscript $C$ stands for \emph{``combinatorial''}. 
It is easily verified that
\begin{equation}\label{H0-HC}
H^0_{\mathcal{I}}=H^C_{\mathcal{I}}-\frac{h}{2}\sigma^z_{i_{-}}-\frac{h}{2}\sigma^z_{i_{+}}
\end{equation}
where $i_{\pm}$ are the endpoint sites of the interval $\mathcal{I}$. 


For  an interval $\mathcal{I}$, we denote with $\text{card}(\mathcal{I})$ the number of sites contained in $\mathcal{I}$.
By $|\uparrow\,\rangle$ and $|\downarrow\,\rangle$ we denote the eigenvectors of $\sigma^z$ corresponding
to the eigenvalues $1$ and $-1$, respectively. Similarly, the symbols
\begin{equation}
|\uparrow \cdots \uparrow\,\rangle \quad,\quad |\downarrow \uparrow \cdots \uparrow\,\rangle  \quad,\quad  |\downarrow \uparrow \downarrow \uparrow\cdots \uparrow\,\rangle
\end{equation} 
stand for vectors in $\mathcal{H}_\cI := \bigotimes_{i\in \mathcal{I}} \mathcal{H}_i$ consisting of tensor 
products of $M$ vectors $|\uparrow\,\rangle$ and/or $|\downarrow\,\rangle$.

 In Propositions \ref{propgap} and \ref{propgap-af} below, we identify the ground-states and the spectral gaps 
 above the ground-state energies of the Hamiltonians  $H^0_{\mathcal{I}}$ and $H^C_{\mathcal{I}}$.
\begin{prop}\label{propgap}
Under the assumption that $h$ and $J$ are positive, the Hamiltonians $H^0_{\mathcal{I}}$ and $H^C_{\mathcal{I}}$ have only one ground-state, denoted $\Psi_{\mathcal{I}}$, corresponding to the vector 
\begin{equation}\label{gs-vector}
|\uparrow\uparrow \cdots \uparrow\rangle.
\end{equation}
Moreover, under the condition that $\frac{J}{h}+\frac{3}{2}< \text{card}(\mathcal{I})$,  the spectral gaps above the 
ground-state energies of the Hamiltonians $H^0_{\mathcal{I}}$ and $H^C_{\mathcal{I}}$ are equal to 
$2J+2h$ and $2J+h$, respectively.
\end{prop}
The corresponding proposition for the Hamiltonians with antiferromagnetic exchange coupling constant ($J<0$) 
reads as follows.
\begin{prop}\label{propgap-af}
Let $\text{card}(\mathcal{I})$ be even. Under the assumption  $-J>h>0$, $H^0_{\mathcal{I}}$ and $H^C_{\mathcal{I}}$ have both two ground-states, $\Psi^{A}_{\mathcal{I}}$ and $\Psi^{B}_{\mathcal{I}}$, corresponding to the vectors 
\begin{equation}\label{gs-vector-af-even}
|\uparrow  \downarrow \uparrow \downarrow \uparrow\cdots \uparrow \downarrow\,\rangle\quad,\quad |\downarrow \uparrow  \downarrow \uparrow \downarrow \cdots \downarrow \uparrow\,\rangle\,,
\end{equation}
respectively.
The spectral gap above the ground-state energy is equal to $2|J|-2h $ for $H^0_{\mathcal{I}}$ and to $2|J|-h $ for $H^C_{\mathcal{I}}$. 

Let $\text{card}(\mathcal{I})$ be odd. Under the assumption  $-J>h>0$,  $H^C_{\mathcal{I}}$ has two ground-states, $\Psi^{A}_{\mathcal{I}}$ and $\Psi^{B}_{\mathcal{I}}$, corresponding to the vectors 
\begin{equation}\label{gs-vector-af-odd}
|\uparrow  \downarrow \uparrow \downarrow \uparrow\cdots  \downarrow\uparrow\,\rangle\quad,\quad |\downarrow \uparrow  \downarrow \uparrow \downarrow \cdots  \uparrow\downarrow\,\rangle\,,
\end{equation}
respectively, whereas $\Psi^{A}_{\mathcal{I}}$ is the only ground-state of $H^0_{\mathcal{I}}$. 
For $H^C_{\mathcal{I}}$, the spectral gap above the ground-state energy is equal to  to $2|J|-h $. 
When considering the Hamiltonian $H^0_{\mathcal{I}}$ we call ``spectral gap'' the energy difference between 
the ground-state energy and the spectrum of 
$H^0_{\mathcal{I}}\upharpoonright_{\bigvee^{\perp}\{\Psi^{A}_{\mathcal{I}},\Psi^{B}_{\mathcal{I}}\}}$, where 
$\bigvee^{\perp}\{\Psi^{A}_{\mathcal{I}},\Psi^{B}_{\mathcal{I}}\}$ is the orthogonal complement of the subspace generated by $\Psi^{A}_{\mathcal{I}}$ and $\Psi^{B}_{\mathcal{I}}$. This energy difference is given by $2|J|$. 
Moreover, the distance between the eigenvalue of $H^0_{\mathcal{I}}$ associated with $\Psi^B_\cI$ 
and the spectrum of $H^0_{\mathcal{I}}\upharpoonright_{\bigvee^{\perp}\{\Psi^{A}_{\mathcal{I}},\Psi^{B}_{\mathcal{I}}\}}$ is given by $2|J|-2h$.
\end{prop}

The statements described in Propositions \ref{propgap} and \ref{propgap-af}  are summarized in the table below\footnote{As specified in Proposition \ref{propgap-af},  for an antiferromagnetic chain with an odd number of sites the expression ``spectral gap'' of $H^0_{\mathcal{I}}$ refers to the energy difference between the ground-state energy and the spectrum of $H^0_{\mathcal{I}}\upharpoonright_{\bigvee^{\perp}\{\Psi^{A}_{\mathcal{I}},\Psi^{B}_{\mathcal{I}}\}}$.} 
\begin{table}[h]
\centering
\begin{tabular}{|c|c|c|c|}
\hline
              & $J>0$ &  $J<0$, odd $\#$ of sites & $J<0$,  even $\#$ of sites \\
              \hline
Ground-states of $H^C_{\mathcal{I}}$ &$ |\uparrow\cdots \uparrow \rangle $                   &      $ |\uparrow\downarrow\cdots\downarrow \uparrow \rangle $ and $|\downarrow\uparrow\cdots\uparrow\downarrow\rangle$                             &  $ |\uparrow\downarrow\cdots\downarrow  \rangle $ and $ |\downarrow\uparrow\cdots \uparrow \rangle $                                    \\
\hline
Ground-states of $H^0_{\mathcal{I}}$ & $ |\uparrow\cdots \uparrow \rangle $  &      $ |\uparrow\downarrow\cdots\downarrow \uparrow \rangle $                         &  $ |\uparrow\downarrow\cdots\downarrow  \rangle $ and $ |\downarrow\uparrow\cdots \uparrow \rangle $                                    \\
\hline

Spectral gap of $H^C_{\mathcal{I}}$ & $2J+h$ & $2|J|-h$ & $2|J|-h$\\
\hline
Spectral gap of $H^0_{\mathcal{I}}$ & $2J+2h$  & $2|J|$ & $2|J|-2h $\\
\hline

\end{tabular}
\end{table}

\noindent

\begin{rem}\label{frustration}
We note that, for antiferromagnetic exchange couplings ($J<0$), the Hamiltonian $H^0_{\Lambda}$ does not have the property that $Ker(H^0_{\Lambda})\,\subseteq \,Ker(H^0_{\mathcal{I}})$,  for all $\mathcal{I}\subset \Lambda$. Indeed, the vectors $\Psi^A_{\Lambda}$ and  $\Psi^B_{\Lambda}$ are eigenvectors 
of $H^0_{\mathcal{I}}$ but the corresponding eigenvalues  coincide only if the number of sites of the 
interval $\mathcal{I}$ is even, i.e., they are different whenever the number of sites in $\mathcal{I}$ is odd. In contrast, upon subtracting a $\Lambda-$ dependent constant, the Hamiltonian $H^C_{\Lambda}$   is frustration free according to the usual definition (see \cite{MN}).
\end{rem}

\subsection{Statement of the main result and organization of the paper}\label{statement-sect}
The results proven in this paper are summarized in the theorem below.
\\

\noindent
{\bf{Theorem}}\\
We consider the Hamiltonian $K_{\Lambda}(t)$ of an $XXZ$ model defined in in (\ref{XXZ-ham}) 
on a chain $\Lambda$, with $\text{card}(\Lambda) \equiv N$.




\begin{itemize}
\item[(a)] \emph{If $J>0$ there exists a constant $\bar{t}>0$ depending on $J$ and $h$, but independent of 
$N>\frac{J}{h}+\frac{3}{2}$, such that, for all $|t|<\bar{t}$, the ground-state energy $E_{\Lambda}$ of the Hamiltonian 
$K_{\Lambda}$ in (\ref{XXZ-ham}) is non-degenerate and the spectral gap above the ground-state energy 
is bounded below by $2J+2h-\mathcal{O}(\frac{J}{h}\,|t|)$.}
\item[(b)] \emph{If $J<0$ then, for $|J|>2h$, there exists a $\bar{t}>0$ depending on $J$ and $h$, 
but independent of $N$, such that, for all $|t|<\bar{t}$, the following statements hold. }

\begin{itemize}
\item
\emph{If $\Lambda$ has an odd number of sites, the set $\mathfrak{S}:=\sigma(K_{\Lambda})\cap [E_{\Lambda}\,,\,E_{\Lambda}+2|J|-\mathcal{O}(|t|)]$,  where $\sigma(K_{\Lambda})$ is the spectrum of $K_{\Lambda}$ and $E_{\Lambda}$ its ground-state energy, consists of two points, $E_{\Lambda}$ and $E'_{\Lambda}$,  with $E'_{\Lambda}-E_{\Lambda}=2h-\mathcal{O}(|t|)$, and the spectral projection associated  with $\mathfrak{S}$ is of rank $2$; }
\item
\emph{If $\Lambda$ has an even number of sites, the set $\mathfrak{S}:=\sigma(K_{\Lambda})\cap [E_{\Lambda}\,,\,E_{\Lambda}+2|J|-2h-\mathcal{O}(|t|)]$ consists of at most two points, $E_{\Lambda}$ and $E'_{\Lambda}$,  with $|E'_{\Lambda}-E_{\Lambda}|\leq \mathcal{O}(|t|)$, and the spectral projection associated  with $\mathfrak{S}$ is of rank $2$. }
\end{itemize}
\end{itemize}

\begin{rem}\label{h-odd}
In point (b) of the Theorem above,  the dependence on $h$ of $\bar{t}$ is only required to ensure the gap  $E'_{\Lambda}-E_{\Lambda}$ is given by $2h-\mathcal{O}(|t|)$. More precisely, the allowed range of the coupling constant is $|t|\ll |J|-h$, hence our theorem also applies to a regime where the gap $E'_{\Lambda}-E_{\Lambda}$ may close.
\end{rem}
\begin{rem}\label{rem-bulk}
For $J<0$ (antiferromagnetic chain), similar results hold if the perturbation term $\sum_{i=1}^{N-1}(\sigma_i^x\sigma_{i+1}^x+\sigma_i^y\sigma_{i+1}^y)$ is replaced by an arbitrary \emph{translation-invariant, short-range} perturbation. For $J>0$,  the translation invariance of the perturbation term is not required.
\end{rem}

\begin{rem}
The ferromagnetic $XXZ$ model with $h=0$ is not covered by the theorem formulated above,
because our strategy (for the ferromagnetic chain) uses the non-degeneracy of the ground-state subspace of 
the local Hamiltonians, which holds for $h>0$. However, we can easily deal with the ferromagnetic model
in a vanishing magnetic field ($h=0$) and show that the ground-state energy is doubly degenerate, and 
that the spectral gap above the ground-state energy is bounded below by $2J-\mathcal{O}(|t|)$; see Remark \ref{ferro-0}.
\end{rem}

\begin{rem}
The theorem stated above also holds for the antiferromagnetic $XXZ$ model with spins coupled to a 
staggered magnetic field in the $z$-direction (see \cite{R}), whose Hamiltonian can be obtained from 
the Hamiltonian $K_{\Lambda}$ with ferromagnetic couplings ($J>0$) by a unitary conjugation 
that flips the $\sigma^z$ operators either on all sites with $i$ even, or on all sites with $i$ odd.
\end{rem}

\begin{rem}
We expect that the techniques developed in this paper enable us to extend the theorem stated above to chains of quantum spins
of arbitrary spin $s\geq 1$. Indeed, the spin-$s$ Ising Hamiltonians with spins coupled to a magnetic field in the
$z$-direction have a low-lying energy spectrum very similar to the ones described in Propositions \ref{propgap} 
and \ref{propgap-af} for the ferromagnetic and antiferromagnetic models with $s=\frac{1}{2}$, respectively.  
Since the properties stated in these propositions  
are the only relevant ingredients that will be required, the whole procedure used to prove our main results 
should apply, word-by-word, to the more general class of models alluded to above.
\end{rem}

\noindent
{\bf{Summary of contents.}} 
In Sect.  \ref{Strategy}, we begin with the definition of the local interaction terms (Sect. \ref{local-int}), supported in suitable large
 subsets of lattice sites, on which we will apply our block-diagonalization procedure. 
In Sect. \ref{outline}, we present a brief review of the method originally introduced for perturbations of
ultra-local Hamiltonians. Next, in Sects.~\ref{hooked-sect} and \ref{deg-en-af}, we present the key ideas 
enabling us to cope with complications -- as compared to the Hamiltonians treated in \cite{FP} -- arising
in the implementation of a local block-diagonalization procedure, which are related to: i) the nearest-neighbour 
interaction structure of the Ising Hamiltonian; and ii) the degeneracy (in the antiferromagnetic model) of the 
ground-state energy of the unperturbed Hamiltonians $H^C_{\mathcal{I}}$. 

In Sect. \ref{sec-algo}, we introduce the algorithm that determines the effective potentials of the transformed 
Hamiltonians at each step of the block-diagonalization flow. 

In Sect. \ref{final-section}, we quantitatively control the effective potentials produced along the block-diagonalization
flow (Sect. \ref{normsests}) and analyze the spectrum of the ferromagnetic and antiferromagnetic 
Hamiltonians $K_{\Lambda}(t)$, for arbitrary large $N$ (Sect. \ref{gap}), as described in 
the theorem above.
\\

\noindent
{\bf{Notation}}\\

\noindent
1) We use the same symbol for an operator $O_{\cI}$ acting on $\otimes_{i\in\cI} \mathcal{H}_{i}$ and the corresponding operator acting on the entire Hilbert space $\mathcal{H}^{(N)}$ that is obtained from $O_{\cI}$ 
by tensoring with the identity matrix on the Hilbert spaces of all remaining sites.\\

\noindent
2) With the symbol ``$\subset$'' we denote strict inclusion, otherwise we use the symbol  ``$\subseteq$''. 
\\

\noindent
3) With the symbol $\mathcal{O}(|t|)$ we denote a quantity which, in absolute value, is bounded above 
by $|t|$ multiplied by a constant possibly depending on further parameters entering the definition 
of the models, but independent of the number of sites, $N$, of the chain.
\\

\noindent
4) We denote by $\langle \cdot, \cdot\rangle$ the scalar product on $\mathcal{H}^{(N)}$.
\\

\noindent
5) We denote the identity matrix by $\charf$ or $1$, interchangeably.

\section{Proof strategy}\label{Strategy}
For convenience of notation, and 
without loss of generality, we assume that $t>0$.
\subsection{Local interaction terms and projections}\label{local-int}

\noindent
We consider the term 
\begin{equation}\label{perturbation}
\frac{t}{2}\sum_{i=1}^{N-1}(\sigma_i^x\sigma_{i+1}^x+\sigma_i^y\sigma_{i+1}^y)
\end{equation}
as a small perturbation of the remaining part of the Hamiltonian $K_{\Lambda}$. In order to cope with the fact that the unperturbed Hamilronians $H^0_{\mathcal{I}}$ are not ultralocal,  we  split the operator in (\ref{perturbation}) into terms localized in ($N$-independent) intervals of size set equal to a parameter, $\xi$, and introduce a \emph{macroscopic} lattice with a lattice spacing equal to $\xi$, as explained below (see Definition \ref{rects}). To implementing \emph{strictly} local block-diagonalizations,  in our method a crucial role is played by the intervals associated with the macroscopic lattice  and by a suitable, $\xi$-dependent enlargement of them (see Definition \ref{setJ*} and \ref{overline-setJ*}) needed to take advantage of the finite propagation speed of the dynamics generated by the unperturbed Hamiltonians. Consequently,  $\xi$ has to be chosen,  in general, to depend on the coupling constant $t$ (see, e.g., \cite{AKLT}); but for the models treated in this paper we can actually take $\xi$ to be independent of $t$ since such  propagation speed, i.e., the one associated to $H^0_{\mathcal{I}}$,  is zero. It suffices that the enlargement is larger than or equal to $2$,  thanks to the "zero" propagation speed of the Hamiltonians $H^0_{\mathcal{I}}$; see Lemma \ref{Lemma-LR}. Concretely, it will be set equal to $\frac{\xi}{3}\,(\in \mathbb{N})$. In addition, for the ferromagnetic model, $\xi$ will be chosen sufficiently large according to the constraint in (\ref{constraint-xi}). For the antiferromagnetic model,  $\xi$ is assumed to be an odd integer w.l.o.g., and can be eventually chosen to be equal to $9$; see Lemma \ref{Lemma-LR}.

\noindent
Without loss of generality, we assume that
\begin{equation}
\frac{N-1}{\xi}\,\in \mathbb{N}\,,\, \, \frac{\xi}{3} \, \in \mathbb{N}\,.\label{conditions}
\end{equation}

\noindent
We introduce a \emph{macroscopic} (finite) lattice with left endpoint $X=1$, right endpoint $X=N$, and spacing $\xi$. The $M^{th}$ site of this lattice is the point
\begin{equation}\label{macro-lattice}
1+(M-1)\cdot \xi\,,\quad\text{with}\quad 1\leq  M\leq \frac{(N-1)}{\xi}+1\,.
\end{equation}
The set of successive sites $\mathcal{I}$ at \textit{position} (i.e., starting from) $J=:Q(\mathcal{I})$  
and of \emph{length} $K=:\ell(\mathcal{I})$, in units of $\xi$, is the interval whose endpoints 
coincide with  the sites $M=J$ and $M=J+K$ of the macroscopic lattice.

\begin{defn} \label{rects}
We define $\mathfrak{I}$ to be the set of intervals $\mathcal{I}\subseteq \Lambda$ whose left endpoint 
is the site $1+(J-1)\xi$, for some $J\in\mathbb{N}$, and whose length is given by 
$K \cdot \xi$ (hence $\text{card}(\mathcal{I})=K\cdot \xi +1$), for some $K\in\mathbb{N}$. We set 
$$Q(\mathcal{I}):=J\quad,\quad \ell(\mathcal{I}):=K.$$
Hence the interval $\cI$, with $Q(\mathcal{I})=J$ and $\ell(\mathcal{I})=K$, is 
\begin{equation}
\Big[\,1+(Q(\mathcal{I})-1)\cdot \xi\,,\,1+(Q(\mathcal{I})+\ell(\cI)-1)\cdot \xi\,\Big]\,. 
\end{equation}
\end{defn}

\begin{figure}[h]
\centering
\includegraphics[width=12cm]{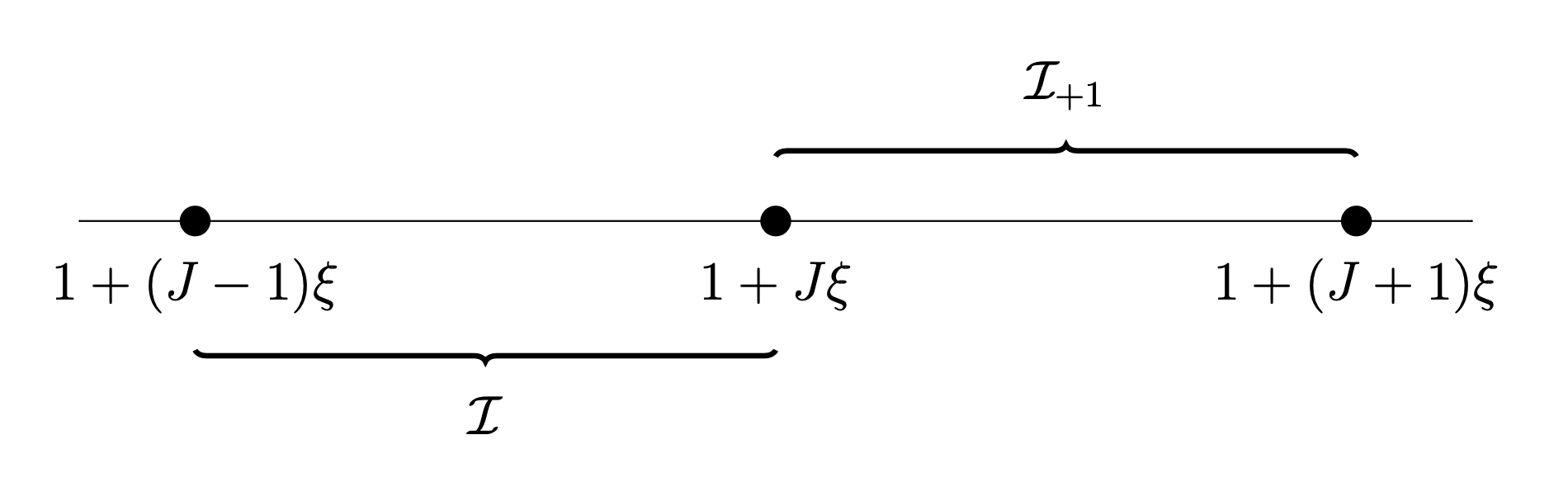}\\
\caption{The picture illustrates the interval $\cI$, with $Q(\cI)=J$ and $\ell(\cI)=1$, and the subsequent 
\mbox{one, $\cI_{+1}$\,.}}
\end{figure}

\noindent
For $\mathcal{I}\in \mathfrak{I}$, with $\ell(\mathcal{I})=1$, we define 
\begin{equation}
V_{\mathcal{I}}:=\frac{1}{2\cdot \xi}\,\sum_{i\,:\,i,i+1\,\in \mathcal{I}}(\sigma_i^x\sigma_{i+1}^x+\sigma_i^y
\sigma_{i+1}^y)\,.
\end{equation}
Our definition of intervals $\mathcal{I}$, with $\ell(\mathcal{I})=1$, imply the bound 
\begin{equation}\label{V-0-bound}
\|\,V_{\mathcal{I}}\,\|\,\leq 1\,.
\end{equation}
In the implementation of the block-diagonalization procedure it is convenient to re-write the Hamiltonian 
$K_{\Lambda}(t)$ by making use of the definitions introduced above, namely
\begin{equation}\label{bare-2}
K_{\Lambda}(t)=H_{\Lambda}^0\,+\,\xi\cdot t\, \sum_{\mathcal{I}\in \mathfrak{I}\,;\, \ell(\mathcal{I})=1}\,V_{\mathcal{I}}\,.\end{equation} 
In addition to (\ref{conditions}), for the ferromagnetic models we require that
\begin{equation}\label{constraint-xi}
( N-\frac{3}{2}> )\, \xi > \frac{J}{h}\,,
\end{equation}
so that Proposition \ref{propgap} holds for all the unperturbed Hamiltonians $H^0_{\mathcal{I}}$.
The block-diagonalization used below is w.r.t. spectral projections supported in intervals 
$\mathcal{I} $, which we define next.
\begin{defn}\label{def-proj}
By $P^{(-)}_{\mathcal{I}}$ we denote the orthogonal projection onto the ground-state subspace of $H^C_{\mathcal{I}}$,  and we set
\begin{equation}\label{vacuum_i}
  P_{\mathcal{I}}^{(+)} := \charf - P^{(-)}_{\mathcal{I}}\,.
\end{equation}
Analogous definitions will be employed for projections associated with other subsets of the lattice $\Lambda$.
\end{defn}
For the following a total ordering relation on the set $\mathfrak{I}$ will turn out to be useful according to which 
shorter intervals precede longer ones. This ordering relation is defined below.
\begin{defn}\label{ordering}
An ordering relation ``$\succ$'' on $\mathfrak{I}$ is specified as follows.
\begin{equation}
\mathcal{I}\succ \mathcal{I}^\prime \quad \text{if}\quad  \ell({\mathcal{I}})>\ell(\mathcal{I}^\prime)\quad \text{or}\quad \text{if}\quad \ell({\mathcal{I}})=\ell(\mathcal{I}^\prime)\quad \text{and}\quad Q(\mathcal{I})>Q(\mathcal{I}^\prime)\,.
\end{equation}
The  symbol $\mathcal{I}_{-1}$ ($\mathcal{I}_{+1}$, resp.) stands for the element of $\mathfrak{I}$ preceding (following,
resp.) $\mathcal{I}$ in the given ordering. For convenience, we define the symbol $\mathcal{I}_0$ to be the element preceding the smallest element of $\mathfrak{I}$ in the given ordering. Note that the biggest element in this ordering is the entire lattice $\Lambda$.
\end{defn}

\subsection{Outline of the block-diagonalization flow}\label{outline}
The study of the low-energy spectrum  of the $XXZ$ Hamiltonians introduced in (\ref{XXZ-ham}) is based on an 
extension and refinement of the local Lie Schwinger block-diagonalization procedure introduced in \cite{FP}. 
Starting from the decomposition of the interaction terms into potentials $V_{\mathcal{I}}$, and taking the 
ordering relation introduced in Definition \ref{ordering} into account, we will construct an iterative block-diagonalization
algorithm based on unitary (Lie-Schwinger) conjugations supported in intervals of the set $\mathfrak{I}$. These conjugations are denoted by $e^{Z_{\mathcal{I}}}$.\\
In the very first step, corresponding to the interval $\mathcal{I}$ with $Q(\mathcal{I})=1$ and $\ell(\mathcal{I})=1$,  the conjugation is such that
\begin{equation}
e^{Z_{\mathcal{I}}}\,(H^0_{\mathcal{I}}+\xi \cdot t\,V_{\mathcal{I}})\,e^{-Z_{\mathcal{I}}}=H^0_{\mathcal{I}}+\sqrt{t}V^{\mathcal{I}}_{\mathcal{I}}\,,
\end{equation}
where the new potential $V^{\mathcal{I}}_{\mathcal{I}}$ is block-diagonal w.r.t. $P^{(-)}_{\mathcal{I}}$ and $P^{(+)}_{\mathcal{I}}:=\charf -P^{(-)}_{\mathcal{I}} $, i.e.,
\begin{equation}
V^{\mathcal{I}}_{\mathcal{I}}\,=\,P^{(-)}_{\mathcal{I}}V^{\mathcal{I}}_{\mathcal{I}}P^{(-)}_{\mathcal{I}}+P^{(+)}_{\mathcal{I}}V^{\mathcal{I}}_{\mathcal{I}}P^{(+)}_{\mathcal{I}}\,.
\end{equation}
It is evident that the action of the conjugation on the remaining terms of the Hamiltonian $K_{\Lambda}$ 
may create new terms. For example, for $\mathcal{I}'$ such that $Q(\mathcal{I}')=2$ and $\ell(\mathcal{I}')=1$,  
we have that
\begin{equation}
e^{Z_{\mathcal{I}}}\,\xi \cdot t\,V_{\mathcal{I}'}\,e^{-Z_{\mathcal{I}}}=\xi \cdot t\,V_{\mathcal{I}'}+\xi \cdot t\,\,\Delta V_{\mathcal{I}\cup \mathcal{I}'}(t)
\end{equation}
where, in general, $\Delta V_{\mathcal{I}\cup \mathcal{I}'}(t)$ is a non-zero operator supported in the longer 
interval $\mathcal{I}\cup\mathcal{I}'$; indeed, $Q(\mathcal{I}\cup\mathcal{I}')=1$ and 
$\ell(\mathcal{I}\cup\mathcal{I}')=2$. 

\noindent
For a Hamiltonian whose unperturbed part is \emph{ultralocal}, i.e., consists of on-site terms only, it is shown 
in \cite{FP} how effective potentials, supported in intervals of arbitrary length belonging to $\mathfrak{I}$, 
are created in subsequent steps of the block-diagonalization, starting from the first sequence of steps in which 
the potentials associated with intervals of length $\ell(\mathcal{I})=1$ are block-diagonalized. The control of their 
norms relies on the fact that the number of growth processes yielding an effective potential supported in an 
interval $\mathcal{I}'$  can be bounded by $const^{\ell(\mathcal{I}')}$. In estimating the norm of a potential
supported in the interval $\mathcal{I}'$, a fractional power of the coupling constant $t$ can be assigned to 
each edge of the interval $\mathcal{I}'$. Indeed, each factor of $Z_{\mathcal{I}}$ appearing in commutators
is proportional to $\xi \cdot t$. Hence, for $t$ sufficiently small, the norm of any effective potential  has power 
law decay in $t$ with an exponent proportional to the length of the interval in which the potential under 
consideration is supported.\\
The block-diagonalization procedure terminates with a final Hamiltonian, unitarily conjugated to the original 
Hamiltonian of the chain, with the property that each effective potential appearing in the final Hamiltonian is  
block-diagonal w.r.t. to the projections in Definition \ref{def-proj} associated with the support of the potential. 
 Hence the final Hamiltonian is block-diagonal w.r.t. to the projections $P^{(-)}_{\Lambda}\,,\, P^{(+)}_{\Lambda}$. Consequently, the low-energy spectrum of the original Hamiltonian can be controlled. For a more detailed 
overview of the block-diagonalization procedure for \textit{ultralocal} unperturbed Hamiltonians see 
Section 2.1 in \cite{DFPRa}.

For the models studied in the present paper, there are several complications arising when one attempts to 
construct the block-diagonalization flow following the strategy in \cite{FP}. Some of these complications 
become already visible in the study of the AKLT model (see \cite{DFPRa}).

\subsection{``Hooked" unperturbed terms.}\label{hooked-sect}
One complication stems from the property of the unperturbed local Hamiltonians $H^0_{\mathcal{I}}$ 
considered in this paper that they are \textit{not ultralocal,} i.e., do not consist only of on-site terms. 
This implies, for example, that in the very first step of the block-diagonalization procedure, i.e., in the 
step corresponding to the interval $\mathcal{I}$ with $Q(\mathcal{I})=\ell(\mathcal{I})=1$, the conjugation 
\begin{equation}
e^{Z_{\mathcal{I}}}\,\Big\{K_{\Lambda}-(H^0_{\mathcal{I}}+\xi \cdot t\,V_{\mathcal{I}})\Big\}\,e^{-Z_{\mathcal{I}}}
\end{equation}
includes terms, such as 
\begin{equation}
e^{Z_{\mathcal{I}}}\,\,\Big[-J\sigma^z_{i_{+}}\sigma^z_{i_{+}+1}\Big]\,e^{-Z_{\mathcal{I}}}\,=\,-J\sigma^z_{i_{+}}\sigma^z_{i_{+}+1}+\xi \cdot t\,\,\sum_{n=1}^{\infty}\frac{1}{n!}ad^n\,Z_{\mathcal{I}}(\frac{-J\sigma^z_{i_{+}}\sigma^z_{i_{+}+1}}{\xi \cdot t\,})\,,
\end{equation}
where $i_{+}$ is the right endpoint of $\mathcal{I}$ in the microscopic lattice, and 
\begin{equation}\label{def-AD}
ad\, A\,(B):=[A\,,\,B]\,,\quad\quad 
ad^n A\,(B):=[A\,,\,ad^{n-1} A\,(B)]\,, \text{  for  }\, n\geq 2\,.
\end{equation}
Hence a new potential is created whose support extends over the enlarged set  $\mathcal{I}\cup \{i_{+}+1\}$. 
In order to gain control over the flow, we have to verify that the contribution to the new potential 
\begin{equation}
\,\sum_{n=1}^{\infty}\frac{1}{n!}ad^n\,Z_{\mathcal{I}}(\frac{-J\sigma^z_{i_{+}}\sigma^z_{i_{+}+1}}{\xi \cdot t\,})
\end{equation}
that needs to be block-diagonalized, i.e., the off-diagonal part w.r.t. the two projections
\begin{equation}
P^{(-)}_{\mathcal{I}\cup \{i_{+}+1\}}\quad\text{and}\quad P^{(+)}_{\mathcal{I}\cup \{i_{+}+1\}},
\end{equation}
has an operator norm bounded by $t^{\,\rho\cdot \ell(\mathcal{I}')}$, where $\rho$ is a universal constant, 
and $\ell(\mathcal{I}')=2$ is the length of the shortest interval in $\mathfrak{I}$ containing the support of 
the effective potential created by the conjugation. In fact, this decay holds trivially for all the terms in the 
series with the exception of the leading one, i.e., except for the off-diagonal part of
\begin{equation}\label{hook-1}
ad\,Z_{\mathcal{I}}(\frac{-J\sigma^z_{i_{+}}\sigma^z_{i_{+}+1}}{\xi \cdot t\,})\,,
\end{equation}
which we will refer to as a \emph{``hooked''} unperturbed term created in the conjugation generated by the
operator $Z_{\mathcal{I}}$. To control the size of the term in (\ref{hook-1}) we are forced to change 
the strategy in \cite{FP} by introducing intervals $\mathcal{I}^*$ and $\overline{\mathcal{I}^*}$, which are \emph{enlargements}  of  $\mathcal{I}$, and defining suitable corresponding operators $Z_{\mathcal{I}^*}$.
\begin{defn}\label{setJ*}
On $\mathfrak{I}$ we define the operation $*$ assigning to each interval $\cI \in \mathfrak{I} $, $\cI \neq \Lambda$, a larger interval $\cI^*$ contained in the lattice. 
$\cI^*$ is defined in the following way 
\begin{equation}
\mathcal{I}^*=\Big\{\,i \in \Lambda \cap [\,1+(Q(\cI)-\frac{4}{3})\cdot \xi,\,1+(Q(\cI)-\frac{2}{3}+\ell(\cI))\cdot \xi\,]\,\Big\}\, .
\end{equation}
Moreover, we denote by $\mathfrak{I}^*$ the image of $\mathfrak{I}$ under the map $*$.

\end{defn} 
\begin{figure}[h]
\centering
\includegraphics[width=\textwidth]{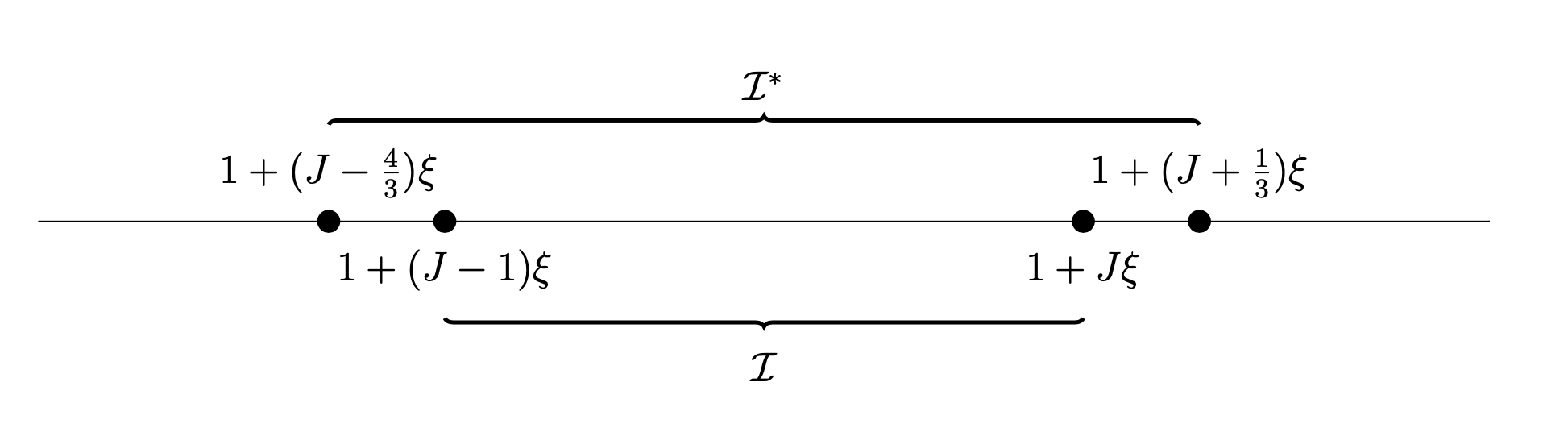}
\caption{The picture displays the relation between $\cI$, with $Q(\cI)=J$ and $\ell(\cI)=1$,  and $\cI^\ast$.}
\end{figure}

\begin{defn}\label{overline-setJ*}
For each $\mathcal{I}^*\in \mathfrak{I}^*$ we define $\overline{\mathcal{I}^*}$ the interval obtained from $\mathcal{I}^*\in \mathfrak{I}^*$ by joining the nearest two sites (if present) belonging to the lattice, both on the right and  on the left.  We call these sites $i^*_{+}+ 1$, $i^*_{+}+ 2$, $i^*_{-}-1$,  and $i^*_{-}- 2$, where $i^*_{\pm}$ are the sites corresponding to the right and the left endpoint of $\mathcal{I}^*$, respectively.
\end{defn}

\begin{figure}[h]
\centering
\includegraphics[width=12cm]{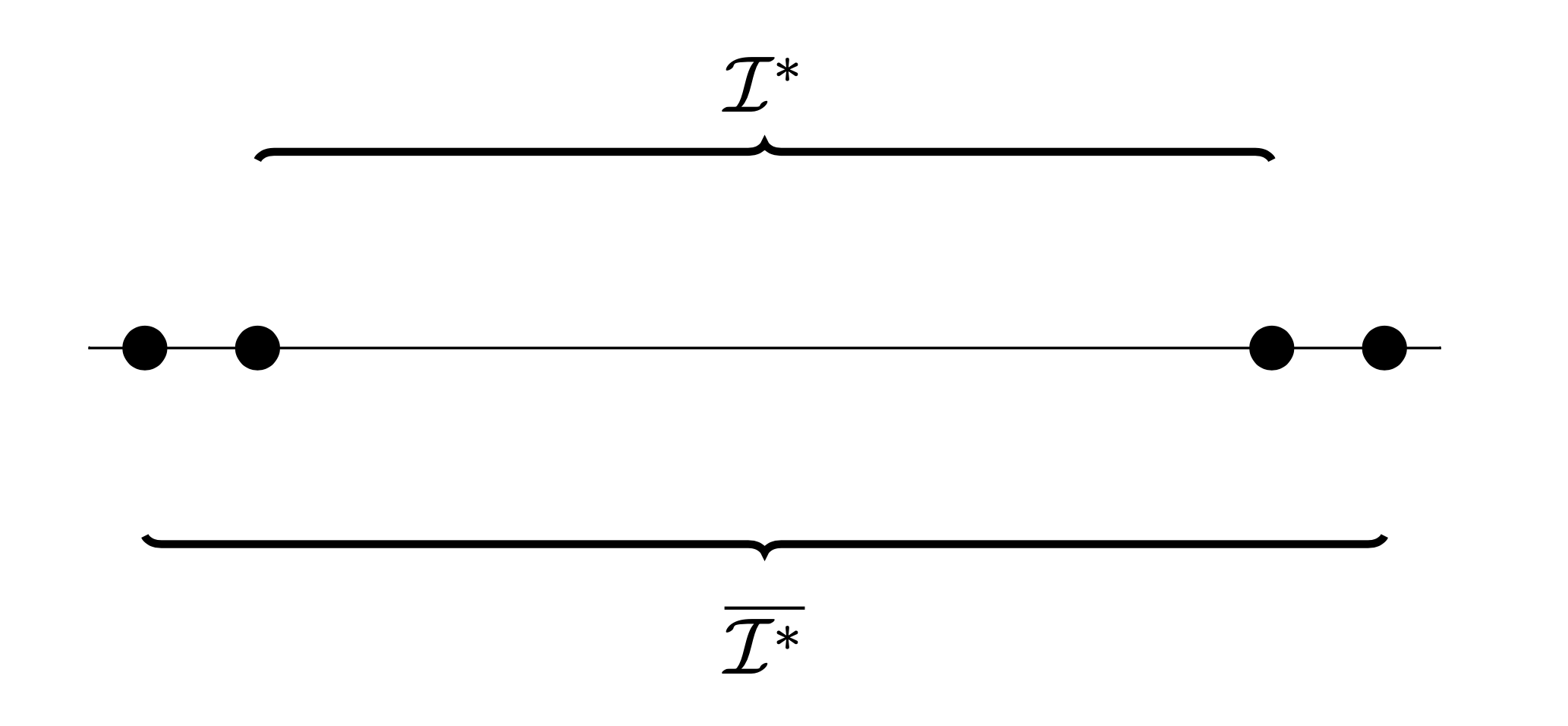}
\caption{The picture displays the relation between $\cI^\ast$ and $\overline{\cI^*}$. }
\end{figure}

The rationale behind the use of enlarged intervals is as follows. Consider the very first step as described above. In order to block-diagonalize a potential supported in $\mathcal{I}$, we consider the unperturbed operator $H^0_{\mathcal{I}^*}$ and the unitary operator $Z_{\mathcal{I}^*}$ (defined in Section \ref{conj-formulae}) such that
\begin{equation}
e^{Z_{\mathcal{I}^*}}\,(H^0_{\mathcal{I}^*}+\xi \cdot t\,V_{\mathcal{I}})\,e^{-Z_{\mathcal{I}^*}}=:H^0_{\mathcal{I}^*}+\xi \cdot t\,V'_{\mathcal{I}^*}
\end{equation}
where by construction $V'_{\mathcal{I}^*}$ is block-diagonal w.r.t. $P^{(-)}_{\mathcal{I}^*}\,,\, P^{(+)}_{\mathcal{I}^*}$.
The counterpart of the term in (\ref{hook-1}) is 
\begin{equation}\label{hook-1*}
ad\,Z_{\mathcal{I}^*}(\frac{-J\sigma^z_{i^\ast_{+}}\sigma^z_{i^\ast_{+}+1}}{\xi \cdot t})\,.
\end{equation}
Since the leading order term in the operator $Z_{\mathcal{I}^*}$ corresponds (in the ferromagnetic case) to
\begin{equation}
\frac{1}{H^0_{\mathcal{I}^*}-E_{\mathcal{I}^*}}P^{(+)}_{\mathcal{I}^*}\,\xi \cdot t\,\,V_{\mathcal{I}}\,P^{(-)}_{\mathcal{I}^*}-h.c.
\end{equation}
where $E_{\mathcal{I}^*}$ is defined in (\ref{def-E}), the core of the argument is  showing that the operator norm of the commutator
\begin{equation}\label{commutator-robinson}
P^{(+)}_{\overline{\mathcal{I}^*}}\,\Big[\frac{1}{H^0_{\mathcal{I}^*}-E_{\mathcal{I}^*}}P^{(+)}_{\mathcal{I}^*}\,V_{\mathcal{I}}\,P^{(-)}_{\mathcal{I}^*}\,,\,-J\sigma^z_{i^\ast_{+}}\sigma^z_{i^\ast_{+}+1}\Big]P^{(-)}_{\overline{\mathcal{I}^*}}
\end{equation}
decays as  $\mathcal{O}((\xi \cdot t)^{\frac{1}{2}}\cdot \|V_{\mathcal{I}}\|)$, as $t\to 0$,  where we exploit that $i_{+}^*$ is at a distance larger than $2$  from the set $\mathcal{I}$ in microscopic units, and that the propagation speed of the unperturbed dynamics (i.e., the one generated by $H^0_{\mathcal{I}^*}$) is zero.

\noindent
An analogous procedure holds for the subsequent block-diagonalization steps.

\begin{rem}
In a general context the treatment of the  commutator in (\ref{commutator-robinson}) requires a Lieb- Robinson bound (see \cite{LR}) on the propagation speed of the unperturbed dynamics and a $t$-dependent coarse graining. The latter amounts to choose $\xi$ to be $t$-dependent; see, e,g., \cite{AKLT}.
\end{rem}

\begin{defn}\label{enl-tilde}
For each $\mathcal{I}^* \in \mathfrak{I}^*$ we define $\widetilde{\mathcal{I}^*}\in \mathfrak{I}$ the smallest interval in $\mathfrak{I}$  containing the interval $\mathcal{I}^*$.
\end{defn}

\begin{rem}\label{bound.-bulk}
Concerning the enlarged intervals introduced in Definitions \ref{setJ*},  \ref{overline-setJ*}, and \ref{enl-tilde},  in the next sections it is helpful to distinguish \emph{boundary-} from \emph{bulk-intervals}. The first set consists of those intervals $\mathcal{I}$ for which one of the endpoints coincides with an endpoint of $\Lambda$. For this reason the corresponding enlargements $\mathcal{I}^*$, $\overline{\mathcal{I}^*}$ and $\widetilde{\mathcal{I}^*}$ involve only one side of the interval.
\end{rem}

\begin{figure}[h]
\centering
\includegraphics[width=13cm]{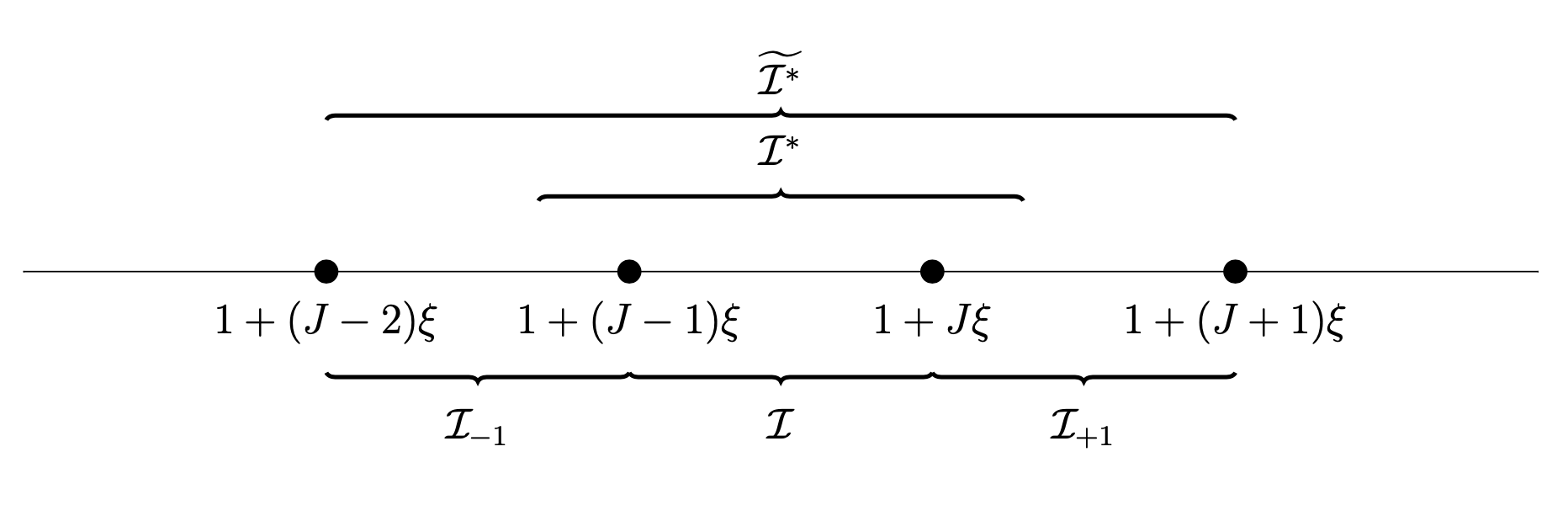}
\caption{The picture shows how $\cI$, with $Q(\cI)=J$ and $\ell(\cI)=1$, relates to $\widetilde{\cI^\ast}$.}
\end{figure}

\subsection{Degeneracy of the ``bulk'' ground-state energy level in the antiferromagnetic case}\label{deg-en-af}
The second complication appearing in the study of the antiferromagnetic case is related to the fact that the ground-state subspace of $H^0_{\mathcal{I}}$ is degenerate if the number of sites is even,  but the ground-state  energy splits into two levels as soon as the interval $\mathcal{I}$ is enlarged by one more site. This yields some technical difficulties to show that, by adding the perturbation (which is by assumption translation invariant; see Remark \ref{rem-bulk}),   if the two-degenerate energy level splits the resulting ones remain in fact very close one to each other, where the small gap is due to boundary effects. 

The control of this possible energy splitting is carried out in detail in Section \ref{gap-af},  since some definitions are needed. Here we just explain the underlying property we shall use.  We call  it degeneracy of the \emph{bulk} ground-state energy.

\noindent
Consider three intervals $\mathcal{I}\,,\,\mathcal{J}$ and $\mathcal{J}'$, with $\mathcal{J}\,,\,\mathcal{J}' \subset \mathcal{I}$,  and where $\mathcal{J}\,,\,\mathcal{J}'$ are two successive intervals  of same length, i.e., 
\begin{equation}
Q(\mathcal{J})=J\,,\,Q(\mathcal{J}')=J+1 \quad,\quad \ell(\mathcal{J})=\ell(\mathcal{J}')\,.
\end{equation} 
Next we consider two operators $W_{\mathcal{J}}$ and $W_{\mathcal{J}'}$ with the property $W_{\mathcal{J}'}=\tau_{1}(W_{\mathcal{J}})$ where $\tau_n$ is the natural shift by $n$ edges in the macroscopic lattice (for details, see Definition \ref{def-transl}). Then, for a macroscopic unit corresponding to $\xi$ which is by assumption an odd natural number,  the following property holds
\begin{equation}\label{prop-pairs}
\langle \Psi^A_{\mathcal{I}}\,,\,W_{\mathcal{J}} \Psi^A_{\mathcal{I}}\rangle =\langle \Psi^B_{\mathcal{I}}\,,\,W_{\mathcal{J}'} \Psi^B_{\mathcal{I}}\rangle \quad,\quad \langle \Psi^B_{\mathcal{I}}\,,\,W_{\mathcal{J}} \Psi^B_{\mathcal{I}}\rangle =\langle \Psi^A_{\mathcal{I}}\,,\,W_{\mathcal{J}'} \Psi^A_{\mathcal{I}}\rangle\,.
\end{equation}
We assume that each potential $W_{\mathcal{J}}$ is block-diagonal w.r.t. $$P^{(-)\,A}_{\mathcal{J}}\,,\,P^{(-)\,B}_{\mathcal{J}}\,,\,P^{(+)}_{\mathcal{J}}\,,$$
i.e.,
\begin{equation}\label{W}
W_{\mathcal{J}}=P^{(-)\,A}_{\mathcal{J}}W_{\mathcal{J}}P^{(-)\,A}_{\mathcal{J}}+P^{(-)\,B}_{\mathcal{J}}W_{\mathcal{J}}P^{(-)\,B}_{\mathcal{J}}+P^{(+)}_{\mathcal{J}}W_{\mathcal{J}}P^{(+)}_{\mathcal{J}}\,,
\end{equation}
where $P^{(-)\,A}_{\mathcal{J}}$ and $P^{(-)\,B}_{\mathcal{J}}$ are the spectral projections onto the subspaces generated by $\Psi^{A}_{\mathcal{J}}$ and $\Psi^{B}_{\mathcal{J}}$, respectively. The importance of having   operators $W_\cJ$ block-diagonalized as displayed in (\ref{W}) is discussed in Remark \ref{block-discuss}. Next we consider
 the Hamiltonian
\begin{equation}
\mathfrak{G}_{\mathcal{I}}:=H^0_{\mathcal{I}}+\sum'_{\mathcal{J}\subset \mathcal{I}}W_{\mathcal{J}}\,,
\end{equation}
where the symbol $'$ means that we sum over an even number of intervals which are paired, in the sense that they come into pairs consisting of an  interval $\mathcal{J}$  and its translated $\mathcal{J}'$. Then, by using the property in (\ref{prop-pairs}),  we have
\begin{equation}
\langle \Psi^A_{\mathcal{I}}\,,\, \mathfrak{G}_{\mathcal{I}}\Psi^A_{\mathcal{I}}\rangle =\langle \Psi^B_{\mathcal{I}}\,,\,\mathfrak{G}_{\mathcal{I}} \Psi^B_{\mathcal{I}}\,\rangle\,.
\end{equation}
\setcounter{equation}{0}
\section{The block-diagonalization algorithm}\label{sec-algo}
We implement the procedure outlined in Sections \ref{outline} and \ref{hooked-sect} by an algorithm described in Section \ref{algo}. In order to do this we need some definitions collected in the next section.
\subsection{Conjugation formulae}\label{conj-formulae}
The expressions that we are going to define  enter the definition of the operators appearing in the transformed Hamiltonian in step $\cI$, which will turn out to be of the form
\begin{eqnarray}\label{ham-conj}
K_{\Lambda}^{\,\cI}(t)
&:= &H^0_{\Lambda}+{\xi \cdot t}\sum_{\mathcal{J}\preceq \mathcal{I}}V^{\,\mathcal{I}}_{\overline{\mathcal{J}^*}}+{\xi \cdot t}\sum_{\mathcal{J}\succ\mathcal{I}}V^{\,\mathcal{I}}_{\mathcal{J}}
\end{eqnarray}
where the reader can notice that
the intervals labelling the potentials $V$ are of two types:
\begin{itemize}
\item[i)]
if $\cJ \preceq \cI$ then $V$ is labeled by intervals $\overline{\mathcal{J}^*}$; 
\item[ii)]if $\cJ \succ \cI$ the corresponding $V$ is labeled by $\mathcal{J}$. 
\end{itemize}
This distinction is due to the fact that the first type of potentials, i.e., those corresponding to $\cJ \preceq \cI$, are block-diagonalized, and the block-diagonalization is w.r.t. the two projections $P^{(-)}_{\overline{\mathcal{J}^*}}$, $P^{(+)}_{\overline{\mathcal{J}^*}}$ (see Definition \ref{def-proj}), consequently they can be written as
\begin{equation}
V^{\,\cI}_{\overline{\mathcal{J}^*}}=P^{(+)}_{\overline{\mathcal{J}^*}}\,V^{\,\cI}_{\overline{\mathcal{J}^*}}\,P^{(+)}_{\overline{\mathcal{J}^*}}+P^{(-)}_{\overline{\mathcal{J}^*}}V^{\,\cI}_{\overline{\mathcal{J}^*}}P^{(-)}_{\overline{\mathcal{J}^*}}\,.
\end{equation}
\noindent
In Section \ref{algo} we explain how these potentials emerge, step by step, as byproducts of the block-diagonalization flow.
In this respect, the algorithm described in Definition \ref{def-interactions-multi} prescribes that, in step $ \cI$, the potential $V^{\,\cI_{-1}}_{\mathcal{I}}$ 
gets transformed to a block-diagonalized potential $V^{\,\cI}_{\overline{\mathcal{I}^*}}$ which  does not coincide but includes the leading order term  of the Lie-Schwinger series (for details see point b) in Definition \ref{def-interactions-multi}),
\begin{equation}\label{L-S}
\sum_{j=1}^{\infty}(\xi \cdot t)^{j-1}\,(V^{\cI_{-1}}_{\mathcal{I}^*})^{\text{diag}}_j\,,
\end{equation}
where the operators $(V^{\cI_{-1}}_{\mathcal{I}^*})_j$ are defined below; here ``$\text{diag}$" means diagonal part w.r.t. to the two projections $P^{(+)}_{\mathcal{I}^*}\,,\,P^{(-)}_{\mathcal{I}^*}$. The key identity which we exploit is 
\begin{equation}\label{conj-G}
e^{Z_{\mathcal{I}^*}}\,(G_{\mathcal{I}^*}+{\xi \cdot t} V_{\mathcal{I}}^{\,\cI_{-1}})\,e^{-Z_{\mathcal{I}^*}}=G_{\mathcal{I}^*}+{\xi \cdot t }\sum_{j=1}^{\infty}(\xi\cdot t)^{j-1}\,(V^{\cI_{-1}}_{\mathcal{I}^*})^{\text{diag}}_j
\end{equation}
where $G_{\mathcal{I}^*}$, $(V^{\cI_{-1}}_{\mathcal{I}^*})^{\text{diag}}_j$, and $Z_{\mathcal{I}^*}$ are defined below:
\begin{enumerate}
\item[1)]
 \begin{eqnarray}\label{expression-G}
G_{\mathcal{I}^*}&:=& H_{\mathcal{I}^*}^{0}+{\xi \cdot t}\,\,\sum_{\overline{\mathcal{J}^*}\subset \mathcal{I}^*} V^{\cI_{-1}}_{\overline{\mathcal{J}^*}}\,,
\end{eqnarray} 
\noindent
for which we remark  that if $\mathcal{I}$ is a bulk-interval (see Remark \ref{bound.-bulk}) then the sum in (\ref{expression-G}) does \emph{not} include those  $\mathcal{J}^\ast$  sharing one of the endpoints with $\mathcal{I}^\ast$,  consequently $G_{\mathcal{I}^*}$ is localized\footnote{If $\mathcal{I}$ is a boundary-interval the intervals $\mathcal{J}\subset \mathcal{I}$  sharing  one of the endpoints of $\Lambda$ with $\mathcal{I}$ are included in the sum in (\ref{expression-G}), but $G_{\mathcal{I}^*}$ is still localized in  $\mathcal{I}^*$.} in $\mathcal{I}^*$;

%
%
\item[2)]
\begin{equation}\label{formula-V_1}
(V^{\cI_{-1}}_{\mathcal{I}^*})_1:=V^{\cI_{-1}}_{\mathcal{I}}\,,
\end{equation}
and, for $j\geq 2$, 
\begin{eqnarray}
&&(V^{\cI_{-1}}_{\mathcal{I}^*})_j\,:=\label{formula-V_j}\\
\quad& &\sum_{p\geq 2, r_1\geq 1 \dots, r_p\geq 1\, ; \, r_1+\dots+r_p=j}\frac{1}{p!}\text{ad}\,(Z_{\mathcal{I}^*})_{r_1}\Big(\text{ad}\,(Z_{\mathcal{I}^*})_{r_2}\dots (\text{ad}\,(Z_{\mathcal{I}^*})_{r_p}(G_{\mathcal{I}^*}))\dots \Big)+\nonumber \\
&&\sum_{p\geq 1, r_1\geq 1 \dots, r_p\geq 1\, ; \, r_1+\dots+r_p=j-1}\frac{1}{p!}\text{ad}\,(Z_{\mathcal{I}^*})_{r_1}\Big(\text{ad}\,(Z_{\mathcal{I}^*})_{r_2}\dots (\text{ad}\,(Z_{\mathcal{I}^*})_{r_p}((V^{\cI_{-1}}_{\mathcal{I}^*})_1)\dots \Big)\, . \nonumber
\end{eqnarray}
where the $ad$ (adjoint action) has been defined in (\ref{def-AD}); 
\item[3)]
\begin{equation}\label{S-definition-1}
Z_{\mathcal{I}^*}:=\sum_{j=1}^{\infty}(\xi\cdot t)^{j}(Z_{\mathcal{I}^*})_j\,,
\end{equation}
where the terms $(Z_{\mathcal{I}^*})_j$ are defined accordingly to the rank of $P^{(-)}_{\mathcal{I}^*}$ , i.e., we distinguish two cases, $J>0$ (ferromagnetic behavior) and $J<0$ (antiferromagnetic behavior).
\begin{itemize}
\item[\underline{J>0}]
In this case, the ground-state subspace of $H^C_{\mathcal{I}^*}$ is one-dimensional, thus $P^{(-)}_{\mathcal{I}^*}$ is a rank one orthogonal projection onto the subspace generated by $\Psi_{I^*}$. The $(Z_{\mathcal{I}^*})_j$ are defined recursively as follows
\begin{equation}
\label{S-definition-2-0}
(Z_{\mathcal{I}^*})_j
:=\frac{1}{G_{\mathcal{I}^*}-E_{\mathcal{I}^*}}P^{(+)}_{\mathcal{I}^*}\,(V^{\cI_{-1}}_{\mathcal{I}^*})_j\,P^{(-)}_{\mathcal{I}^*} -h.c.
\end{equation}
where 
\begin{equation}\label{def-E}
E_{\mathcal{I}^*}:=\, \langle \Psi_{\mathcal{I}^*}\, , \,\Big\{H^0_{\cI^*}+ {\xi \cdot t }\,\sum_{\overline{\mathcal{J}^*}\subset \mathcal{I}^*} \, V^{\cI_{-1}}_{\overline{\mathcal{J}^*}}\Big\}\, \Psi_{\mathcal{I}^*} \rangle\,.
\end{equation}
\item[\underline{J<0}]
In this case, the ground-state subspace of $H^C_{\mathcal{I}^*}$ is two-dimensional, thus $P^{(-)}_{\mathcal{I^*}}$ is a rank two orthogonal projection, i.e., $P^{(-)}_{\mathcal{I}^*}=P^{(-),A}_{\mathcal{I}^*}+P^{(-),B}_{\mathcal{I}^*}$ (see their definition below (\ref{W})),  thus the operators $(Z_{\mathcal{I}^*})_j$ are defined recursively as follows
\begin{equation}
\label{S-definition-2-0}
(Z_{\mathcal{I}^*})_j
:=\Big\{\frac{1}{G_{\mathcal{I}^*}-E^A_{\mathcal{I}^*}}P^{(+)}_{\mathcal{I}^*}\,(V^{\cI_{-1}}_{\mathcal{I}^*})_j\,P^{(-),A}_{\mathcal{I}^*} + \frac{1}{G_{\mathcal{I}^*}-E^B_{\mathcal{I}^*}}P^{(+)}_{\mathcal{I}^*}\,(V^{\cI_{-1}}_{\mathcal{I}^*})_j\,P^{(-),B}_{\mathcal{I}^*}\Big\} -h.c.
\end{equation}
where 
\begin{equation}\label{def-E-A}
E^A_{\mathcal{I}^*}:=\, \langle \Psi^A_{\mathcal{I}^*}\, , \,\Big\{H^0_{\cI^*}+ {\xi \cdot t}\,\sum_{\overline{\mathcal{J}^*}\subset \mathcal{I}^*} \, V^{\cI_{-1}}_{\overline{\mathcal{J}^*}}\Big\}\, \Psi^A_{\mathcal{I}^*} \rangle,
\end{equation}
\begin{equation}\label{def-E-B}
E^B_{\mathcal{I}^*}:=\, \langle \Psi^B_{\mathcal{I}^*} \, , \,\Big\{H^0_{\cI^*}+ {\xi \cdot t}\,\sum_{\overline{\mathcal{J}^*}\subset \mathcal{I}^*} \, V^{\cI_{-1}}_{\overline{\mathcal{J}^*}}\Big\}\, \Psi^B_{\mathcal{I}^*} \rangle.
\end{equation}
We shall prove inductively that $\Psi^A_{\mathcal{I}^*}$ and $\Psi^B_{\mathcal{I}^*}$ are eigenvectors of $G_{\mathcal{I}^*}$; this implies the identity  in (\ref{conj-G}) with the given definition of $Z_{\mathcal{I}^*}$; see \cite{DFFR}.
\end{itemize}
We point out that the construction of $Z_{\mathcal{I}^*}$ requires the control of some (depending on the sign of $J$) of the following gaps:
\begin{equation}\label{resolvents-F}
\text{for}\quad J>0\,,\quad \inf \text{spec} \,[(G_{\mathcal{I}^*}-E_{\mathcal{I}^*})P^{(+)}_{\mathcal{I}^*}]\,;
\end{equation}
\begin{equation}\label{resolvents-AF}
\text{for}\quad J<0\,,\quad\,\inf \text{spec} \,[(G_{\mathcal{I}^*}-E^{A}_{\mathcal{I}^*})P^{(+)}_{\mathcal{I}^*}]\,,\, \inf \text{spec} \,[(G_{\mathcal{I}^*}-E^{B}_{\mathcal{I}^*})P^{(+)}_{\mathcal{I}^*}]\,.
\end{equation} 
We discuss these quantities in Section \ref{gap}.
\end{enumerate}
\begin{rem}\label{block-discuss}
In the antiferromagnetic case, the definition of $Z_{\cI^*}$ requires  the decomposition of $P^{(-)}\mathcal{H}^N$ into the (one-dimensional) subspaces corresponding to the ranges of  $P_{\cI^*}^{(-),A}$ and $P_{\cI^*}^{(-),B}$. Differently from the case where a \emph{local quantum topological order} condition holds (see \cite{DFPRa}), here  we cannot design an algorithm  such that $G_{\cI^*}\,P^{(-)}_{\mathcal{I}^*}=\hat{E}_{\cI^*}\,P^{(-)}_{\mathcal{I}^*}$,  for some $\hat{E}_{\cI^*}$, so as to essentially reduce the block-diagonalization to the usual one where $P^{(-)}_{\mathcal{I}^*}$ is of rank $1$.

This feature makes the control of the two gaps in (\ref{resolvents-AF}) challenging (see Lemma \ref{gap-bound}) and not feasible without a structure where the potentials $V^{\cI_{-1}}_{\overline{\mathcal{J}^*}}$ in (\ref{expression-G}) are block-diagonalized also w.r.t. $P_{\overline{\mathcal{J}^*}}^{(-),A}$ and $P_{\overline{\mathcal{J}^*}}^{(-),B}$. The control of the two gaps  in (\ref{resolvents-AF}) is also possible thanks to the estimate of the energy difference $E^A_{\mathcal{I}^*}-E^B_{\mathcal{I}^*}$  which turns out to be $\mathcal{O}(t)$; see Lemma \ref{diffenergy}. In the proof of Lemma \ref{diffenergy},   the block-diagonalization of the potentials $V^{\cI_{-1}}_{\overline{\mathcal{J}^*}}$ w.r.t. $P_{\overline{\mathcal{J}^*}}^{(-),A}$ and $P_{\overline{\mathcal{J}^*}}^{(-),B}$ is crucial to exploit the argument explained in Section \ref{deg-en-af}.  Thanks to the structure of the \emph{enlargements} described above,  the block-diagonalization property (also of the Hamiltonian $G_{\cI^*}$) with respect to $P_{\cI^*}^{(-),A}$ and $P_{\cI^*}^{(-),B}$ is easily granted by prescription b) of the algorithm in Definition \ref{def-interactions-multi}.
\end{rem}




\subsection{The algorithm: definition and consistency}\label{algo}

Next Definition \ref{pot(0,N)} provides the basis of the iteration yielding the effective potentials $V^{\mathcal{I}}_{\mathcal{J}}$, step by step,  by applying the algorithm in Definition \ref{def-interactions-multi}, where the "steps" are associated with the intervals $\mathcal{I}\in \mathfrak{I}$, and the ordering of the steps follows the ordering relation of the intervals set in Definition \ref{ordering}.

\begin{defn}\label{pot(0,N)} 
We set:
\begin{itemize}
\item for $\mathcal{J}\in \mathfrak{I}$ such that $\ell(\mathcal{J})=1$,
\begin{equation}\label{0-pot-bis}
V_{\mathcal{J}}^{\mathcal{I}_0}:=V_{\mathcal{J}}\Big(=\frac{1}{2\cdot \xi}\,\sum_{i\,:\,i,i+1\,\in \mathcal{J}}(\sigma_i^x\sigma_{i+1}^x+\sigma_i^y\sigma_{i+1}^y)\Big)\,;
\end{equation}
\item for $\mathcal{J}\in \mathfrak{I}$ with $\ell(\mathcal{J})\geq 2$,
\begin{equation}\label{0-pot}
V_{\mathcal{J}}^{\mathcal{I}_0}:=0.
\end{equation}
\end{itemize}
\end{defn} 

In the sequel,  $i^\ast_{-}, i^\ast_{+}$ are the sites in the microscopic lattice corresponding to the two endpoints  of the interval $\mathcal{I}^\ast$. In (\ref{diaghop2}), (\ref{b-21}), and (\ref{b-212}) one of the two hooked terms is absent whenever $i_{-}$ or $i_{+}$ (endpoints of $\mathcal{I}$) coincides with the left or with the right endpoint of $\Lambda$, respectively.

\begin{defn}\label{def-interactions-multi}
For the subsequent steps we set:
\begin{itemize}
\item[a-1)]
if  $\mathcal{I} \prec \mathcal{J}$, and $\mathcal{I}^*\nsubseteq \mathcal{J}$, 
\begin{equation}\label{a-0}
V^{\mathcal{I}}_{\mathcal{J}}:=V^{\mathcal{I}_{-1}}_{\mathcal{J}}\,;
\end{equation}

\item[a-2)]
if  $\mathcal{I} \succ \mathcal{J}$,
\begin{equation}\label{a-bis}
V^{\mathcal{I}}_{\overline{\mathcal{J}^*}}:=V^{\mathcal{I}_{-1}}_{\overline{\mathcal{J}^*}}\,;
\end{equation}
\item[b)]
if $\mathcal{I}=\mathcal{J}$,
\begin{eqnarray}\label{L-S-series}
V^{\mathcal{I}}_{\overline{\mathcal{J}^*}}&:=& P^{(+)}_{\mathcal{I}^*} V^{\mathcal{I}_{-1}}_{\mathcal{I}}P^{(+)}_{\mathcal{I}^*} +  P^{(-)}_{\mathcal{I}^*} V^{\mathcal{I}_{-1}}_{\mathcal{I}}P^{(-)}_{\mathcal{I}^*}\label{firstorder}\\
&&+P^{(+)}_{\overline{\mathcal{I}^*}}\Big(ad\,Z_{\mathcal{I}^*}(\,\frac{\sigma^z_{i^\ast_- -1}\sigma^z_{i^\ast_-}}{\xi \cdot t}+\frac{\sigma^z_{i^\ast_{+}}\sigma^z_{i^\ast_{+}+1}}{\xi \cdot t})\Big)P^{(+)}_{\overline{\mathcal{I}^*}}\,; \label{diaghop2}
\end{eqnarray}
\item[c)]
if $\mathcal{I}^* \subset \mathcal{J}$, first we introduce some symbols referring to three sets of intervals
entering the  formula for $V^{\mathcal{I}}_{\mathcal{J}}$ and used in  (\ref{A-map-1-bis}), (\ref{Valgo3}), and (\ref{Valgo67}), respectively,
\begin{eqnarray}
[\mathcal{G}^{\mathcal{I}}_{\mathcal{J}}]_1 &:=&\Big\{ \, \mathcal{K} \in\mathfrak{I} \,\,\vert \,\,\mathcal{K} \succ \mathcal{I} \,,\,\mathcal{K} \cap \mathcal{I}^*\neq \emptyset , \,\\
&&\,\mathcal{K} \neq \mathcal{J} \,,\,\text{and} \,\, \widetilde{\mathcal{I}^* }\cup \mathcal{K} =\mathcal{J} \,\,\Big\}\,\nonumber 
\end{eqnarray}
\begin{eqnarray}
[ \mathcal{G}^{\mathcal{I}}_{\mathcal{J}}]_2  &:=& 
\Big\{ \, \mathcal{K}^* \in\mathfrak{I}^* \,\,\vert \,\,\mathcal{I} \succ \mathcal{K}\,,\,\mathcal{K}^* \cap \mathcal{I}^*\neq \emptyset \,, \mathcal{K}^* \nsubset\mathcal{I}^*\nonumber\\ 
&& \text{and} \,\,
\widetilde{\mathcal{I}^*}\cup \widetilde{\mathcal{K}^*}=\mathcal{J}\,\,\Big\} \,\nonumber
\end{eqnarray}
\begin{eqnarray}
[ \mathcal{G}^{\mathcal{I}}_{\mathcal{J}}]_3 &:=& 
\Big\{ \, \mathcal{K}^*\in\mathfrak{I}^*\,\,\vert \,\, \mathcal{K}^*\subset\mathcal{I}^*,\,  \overline{\mathcal{K}^*}\nsubset \mathcal{I}^*\,\Big\} \,.\nonumber
\end{eqnarray}
Next, we write the definition of the potential $V^{\mathcal{I}}_{\mathcal{J}}$, which, apart from the term in (\ref{identity-c}), results from growth processes where some potentials are hooked in the conjugation implemented by $e^{Z_{\mathcal{I}^*}} $ (see  (\ref{A-map-1-bis}), (\ref{Valgo3}), and (\ref{Valgo67})), and from collecting higher order terms both of the Lie-Schwinger series (see (\ref{lshighorder})) and of the hooked Ising terms (see (\ref{b-212}) and  (\ref{b-21})) which were not included in b):
\begin{eqnarray}\label{construction-conn}
V^{\mathcal{I}}_{\mathcal{J}} &:= & e^{Z_{\mathcal{I}^*}}\,V^{\mathcal{I}_{-1}}_{\mathcal{J}}\,e^{-Z_{\mathcal{I}^*}}\label{identity-c}\\
& &+\sum_{\mathcal{K} \in [\mathcal{G}^{\mathcal{I}}_{\mathcal{J}}]_1}\,\sum_{n=1}^{\infty}\frac{1}{n!}\,ad^{n}Z_{\mathcal{I}^*}(V^{\mathcal{I}_{-1}}_{\mathcal{K}})\label{A-map-1-bis}\\
& &+\sum_{\mathcal{K}^*\in [\mathcal{G}^{\mathcal{I}}_{\mathcal{J}}]_2}\,\sum_{n=1}^{\infty}\frac{1}{n!}\,ad^{n}Z_{\mathcal{I}^*}(V^{\mathcal{I}_{-1}}_{\overline{\mathcal{K}^*}})\label{Valgo3}\\
& &+\delta_{\widetilde{\mathcal{I}^*}=\mathcal{J}} \sum_{\mathcal{K}^*\in [\mathcal{G}^{\mathcal{I}}_{\mathcal{J}}]_3}\,\sum_{n=1}^{\infty}\frac{1}{n!}\,ad^{n}Z_{\mathcal{I}^*}(V^{\mathcal{I}_{-1}}_{\overline{\mathcal{K}^*}})\label{Valgo67}\\
& &+\delta_{\widetilde{\mathcal{I}^*}=\mathcal{J}} [\sum_{m=2}^{\infty}\,(\xi \cdot t)^{m-1}(V^{\mathcal{I}_{-1}}_{\mathcal{I}^*})^{\text{diag}_{\,\mathcal{I}^*}}_m]\label{lshighorder}\\
& &+\delta_{\widetilde{\mathcal{I}^*}=\mathcal{J}}\Big(\sum_{n=2}^{\infty}\frac{1}{n!}\,ad^{n}Z_{\mathcal{I}^*}(\,{\frac{\sigma^z_{i^\ast_{-}-1}\sigma^z_{i^\ast_{-}}}{\xi \cdot t}}+{\frac{\sigma^z_{i^\ast_{+}}\sigma^z_{i^\ast_{+}+1}}{\xi \cdot t}})\Big) \label{b-21}\\
&&+\delta_{\widetilde{\mathcal{I}^*}=\mathcal{J}}[P^{(-)}_{\overline{\mathcal{I}^*}}\Big(ad\,Z_{\mathcal{I}^*}(\,\frac{\sigma^z_{i^\ast_{-}-1}\sigma^z_{i^\ast_{-}}}{\xi \cdot t}+{\frac{\sigma^z_{i^\ast_{+}}\sigma^z_{i^\ast_{+}+1}}{\xi \cdot t}})\Big)P^{(+)}_{\overline{\mathcal{I}^*}}+h.c.]\,. \label{b-212}
\end{eqnarray}
\end{itemize}
\end{defn}

\noindent
 \begin{rem}
The fact that the operators defined in the algorithm, namely $V^{\cI}_{\cJ},\, V^{\cI}_{\overline{\cJ^*}}$, and $ Z_{\cI^*}$, are all well-defined bounded operators requires the main technical results contained in this paper. More precisely, we iteratively use, at each step, Theorem \ref{th-norms}, Lemma \ref{control-LS}, and, in the antiferromagnetic case, Lemma \ref{Gstructure}.
\end{rem} 
\noindent

The next theorem states that the Hamiltonian in (\ref{ham-conj}) is  obtained by successive conjugations, of $K_{\Lambda}$,  generated by the operators $Z_{\mathcal{J}^*}$,  with $\mathcal{J}\preceq \mathcal{I}$,  up to $\mathcal{J}=\mathcal{I}$. For this purpose, it is enough that, for each step of the block-diagonalization, the algorithm is consistent with the conjugation of the Hamiltonian $K_{\Lambda}^{\mathcal{I}_{-1}}(t)$ implemented by the operator $Z_{\mathcal{I}^*}$.
\begin{thm}\label{consistency}
 Given the algorithm in Definition \ref{def-interactions-multi}, the Hamiltonian $K_{\Lambda}^{\,\mathcal{I}}(t)$ defined in  (\ref{ham-conj})  satisfies
\begin{equation}\label{equality-hams}
K_{\Lambda}^{\,\mathcal{I}}(t)=e^{Z_{\mathcal{I}^*}}\,K_{\Lambda}^{\,\mathcal{I}_{-1}}(t)\,e^{-Z_{\mathcal{I}^*}}.
\end{equation}
\end{thm}

\noindent
\emph{Proof}

We consider the conjugation of each term in the expression below 
\begin{eqnarray}
& &e^{Z_{\mathcal{I}^*}}\,K_{\Lambda}^{\,\mathcal{I}_{-1}}(t)\,e^{-Z_{\mathcal{I}^*}}\\
&= &e^{Z_{\mathcal{I}^*}}\,\Big[\,H^0_{\Lambda}+{\xi \cdot t}\sum_{\mathcal{J}\prec \mathcal{I}}V^{\,\mathcal{I}_{-1}}_{\overline{\mathcal{J}^*}}+\xi \cdot t\sum_{\mathcal{J}\succeq\mathcal{I}}V^{\,\mathcal{I}_{-1}}_{\mathcal{J}}\Big]\, e^{-Z_{\mathcal{I}^*}}\label{Kconj}
\end{eqnarray}
and we re-assemble the obtained operators  according to the rules of Definition \ref{def-interactions-multi},  so as to get the Hamiltonian $K_{\Lambda}^{\,\mathcal{I}}(t)$.  The final result follows from combining the observations below.

\begin{itemize}
\item[i)]
For  $\mathcal{J} \cap \mathcal{I}^*=\emptyset$ and for $\mathcal{J}^* \cap \mathcal{I}^*=\emptyset$, 
\begin{eqnarray}
e^{Z_{\mathcal{I}^*}}\,V^{\mathcal{I}_{-1}}_{\mathcal{J}}\, e^{-Z_{\mathcal{I}^*}}&=&V^{\mathcal{I}_{-1}}_{\mathcal{J}}=:V^{\mathcal{I}}_{\mathcal{J}}\\
e^{Z_{\mathcal{I}^*}}\,V^{\mathcal{I}_{-1}}_{\overline{\mathcal{J}^*}}\,e^{-Z_{\mathcal{I}^*}}&=&V^{\mathcal{I}_{-1}}_{\overline{\mathcal{J}^*}}=:V^{\mathcal{I}}_{\overline{\mathcal{J}^*}}
\end{eqnarray}
hold respectively; the identities above follow from a-1) and a-2) in  Definition \ref{def-interactions-multi}.
\item[ii)]
From the definition of the unperturbed Hamiltonian $G_{\mathcal{I}^*}$ (see (\ref{expression-G})),  we observe that
\begin{eqnarray}
& &e^{Z_{\mathcal{I}^*}}\,\,(H_{\mathcal{I}^*}^{0}+{\xi \cdot t}\,\sum_{\mathcal{J}\subset\mathcal{I}^*}\,\, V^{\mathcal{I}_{-1}}_{\overline{\mathcal{J}^*}}+\xi \cdot t\,V^{\mathcal{I}_{-1}}_{\mathcal{I}})\,e^{-Z_{\mathcal{I}^*}}\,\label{2.75}\\
&=&e^{Z_{\mathcal{I}^*}}\Big\{H_{\mathcal{I}^*}^{0}+{\xi \cdot t}\,\,\sum_{\overline{\mathcal{J}^*}\subset \mathcal{I}^*} V^{\mathcal{I}_{-1}}_{\overline{\mathcal{J}^*}}+\xi \cdot t\,V^{\mathcal{I}_{-1}}_{\mathcal{I}} \\
&&\quad +\xi \cdot t\sum_{\mathcal{J}^*\subset \mathcal{I}^*\, ; \,\overline{\mathcal{J}^*}\nsubset \mathcal{I}^*}V^{\mathcal{I}_{-1}}_{\overline{\mathcal{J}^*}} \Big\}e^{-Z_{\mathcal{I}^*}}\nonumber\\
&=&e^{Z_{\mathcal{I}^*}}\Big\{G_{\mathcal{I}^*}+\xi \cdot t\,V^{\mathcal{I}_{-1}}_{\mathcal{I}}+\xi \cdot t\,\sum_{\mathcal{J}^*\subset \mathcal{I}^*\, ; \,\overline{\mathcal{J}^*}\nsubset \mathcal{I}^*}V^{\mathcal{I}_{-1}}_{\overline{\mathcal{J}^*}} \Big\}e^{-Z_{\mathcal{I}^*}}\nonumber\,;
\end{eqnarray}
next we use (\ref{conj-G}) and get (recall that $G_{\mathcal{I}^*}=H_{\mathcal{I}^*}^{0}+\xi \cdot t\,\,\sum_{\overline{\mathcal{J}^*}\subset \mathcal{I}^*} V^{\mathcal{I}_{-1}}_{\overline{\mathcal{J}^*}}$)
\begin{eqnarray}
(\ref{2.75})&= & H_{\mathcal{I}^*}^{0}+{\xi \cdot t}\,\,\sum_{\overline{\mathcal{J}^*}\subset \mathcal{I}^*} V^{\mathcal{I}_{-1}}_{\overline{\mathcal{J}^*}}+\xi \cdot t\sum_{m=1}^{\infty}{(\xi\cdot t)^{m-1}}(V^{\mathcal{I}_{-1}}_{\mathcal{I}^*})^{\text{diag}_{\,\mathcal{I}^*}}_m\, \label{G-ident}\\
&&+e^{Z_{\mathcal{I}^*}}\xi \cdot t\sum_{\mathcal{J}^*\subset \mathcal{I}^*\, ; \,\overline{\mathcal{J}^*}\nsubset \mathcal{I}^*}V^{\mathcal{I}_{-1}}_{\overline{\mathcal{J}^*}} e^{-Z_{\mathcal{I}^*}}\nonumber\\
&= & H_{\mathcal{I}^*}^{0}+\xi \cdot t\,\,\sum_{\mathcal{J}\subset \mathcal{I}^*} V^{\mathcal{I}}_{\overline{\mathcal{J}^*}}\\
& &+\xi \cdot t\,((\ref{firstorder}))+\xi \cdot t\,((\ref{lshighorder}))+
\xi \cdot t\,((\ref{Valgo67}))\,,\label{sum-line}
\end{eqnarray}
where for the second  identity we have used that
\begin{equation}
\xi \cdot t\,\,\sum_{\mathcal{J}\subset \mathcal{I}^*} V^{\mathcal{I}}_{\overline{\mathcal{J}^*}}={\xi \cdot t}\,\,\sum_{\overline{\mathcal{J}^*}\subset \mathcal{I}^*} V^{\mathcal{I}_{-1}}_{\overline{\mathcal{J}^*}}+\xi \cdot t\sum_{\mathcal{J}^*\subset \mathcal{I}^*\, ; \,\overline{\mathcal{J}^*}\nsubset \mathcal{I}^*}V^{\mathcal{I}_{-1}}_{\overline{\mathcal{J}^*}}
\end{equation}
and
\begin{eqnarray}
& &\xi \cdot t\sum_{m=1}^{\infty}{(\xi\cdot t)^{m-1}}(V^{\mathcal{I}_{-1}}_{\mathcal{I}^*})^{\text{diag}_{\,\mathcal{I}^*}}_m+\Big\{e^{Z_{\mathcal{I}^*}}\xi \cdot t\sum_{\mathcal{J}^*\subset \mathcal{I}^*\, ; \,\overline{\mathcal{J}^*}\nsubset \mathcal{I}^*}V^{\mathcal{I}_{-1}}_{\overline{\mathcal{J}^*}} e^{-Z_{\mathcal{I}^*}}-\xi \cdot t\sum_{\mathcal{J}^*\subset \mathcal{I}^*\, ; \,\overline{\mathcal{J}^*}\nsubset \mathcal{I}^*}V^{\mathcal{I}_{-1}}_{\overline{\mathcal{J}^*}}\Big\}\nonumber\\
&=&\xi \cdot t\,(\ref{firstorder})+\xi \cdot t\,(\ref{lshighorder})+\xi \cdot t\,(\ref{Valgo67})\nonumber
\end{eqnarray}
where (\ref{lshighorder}) and (\ref{Valgo67}) are referred to case c) of Definition \ref{def-interactions-multi} for a potential $V^{\mathcal{I}}_{\mathcal{J}'}$ with  $\mathcal{J}' \equiv \widetilde{\mathcal{I}^*}$.
\item[iii)]
With regard to the terms $V^{\mathcal{I}_{-1}}_{\mathcal{J}}$, with $\mathcal{I}^* \subset \mathcal{J}$, we have 
\begin{equation}
e^{Z_{\mathcal{I}^*}}\,V^{\mathcal{I}_{-1}}_{\mathcal{J}}\,e^{-Z_{\mathcal{I}^*}}=(\ref{identity-c}).
\end{equation}
\item[iv)]
With regard to the terms $V^{\mathcal{I}_{-1}}_{\mathcal{J}}$, with $ \mathcal{I}^* \cap \mathcal{J}\neq \emptyset $ and $\mathcal{I}^* \nsubset \mathcal{J}$, $\mathcal{J}\nsubset \mathcal{I}^* $ , it follows that
\begin{equation}
e^{Z_{\mathcal{I}^*}}\,V^{\mathcal{I}_{-1}}_{\mathcal{J}}\,e^{-Z_{\mathcal{I}^*}}=V^{\mathcal{I}_{-1}}_{\mathcal{J}}+\sum_{n=1}^{\infty}\frac{1}{n!}\,ad^{n}Z_{\mathcal{I}^*}(V^{\mathcal{I}_{-1}}_{\mathcal{J}})\,,
\end{equation}
and we observe that the first term on the right hand side is $V^{\mathcal{I}}_{\mathcal{J}}$ (see case a-1) Definition \ref{def-interactions-multi}), while the second term is a contribution to $V^{\mathcal{I}}_{\mathcal{J}^\prime}$, according to the rule in (\ref{A-map-1-bis}), for $\mathcal{J}^\prime\equiv \mathcal{J}\cup \widetilde{\mathcal{I}^*}$.
\item[v)] With regard to the terms $V^{\mathcal{I}_{-1}}_{\overline{\mathcal{J}^*}}$, we notice that they are present in (\ref{Kconj}) only for $\mathcal{I}_{-1}\succeq \mathcal{J}$. Hence, we have to consider the occurrences $\mathcal{I}\succ \mathcal{J}$ which we discuss below:
\begin{itemize}
\item  the case $\mathcal{J}^* \cap \mathcal{I}^*=\emptyset$ has been already discussed in i); 
\item the case $\mathcal{J}^* \subset \mathcal{I}^*$ has been already discussed in ii); 
\item  if $\mathcal{J}^* \cap \mathcal{I}^*\neq \emptyset$ and $\mathcal{J}^* \nsubset \mathcal{I}^*$ we write
\begin{equation}
e^{Z_{\mathcal{I}^*}} V^{\mathcal{I}_{-1}}_{\overline{\mathcal{J}^*}}\,e^{-Z_{\mathcal{I}^*}}=V^{\mathcal{I}_{-1}}_{\overline{\mathcal{J}^*}}+\sum_{n=1}^{\infty}\frac{1}{n!}\,ad^{n}Z_{\mathcal{I}^*}(V^{\mathcal{I}_{-1}}_{\overline{\mathcal{J}^*}})
\end{equation}
where the first term corresponds to $V^{\mathcal{I}}_{\overline{\mathcal{J}^*}}$ by a-2) of Definition \ref{def-interactions-multi}, and the other terms, i.e., 
$$\sum_{n=1}^{\infty}\frac{1}{n!}\,ad^{n}Z_{\mathcal{I}^*}(V^{\mathcal{I}_{-1}}_{\overline{\mathcal{J}^*}})\,,$$
contribute to $V^{\mathcal{I}}_{\mathcal{J}^\prime}$, with $\widetilde{\mathcal{I}^*}\cup \widetilde{\mathcal{J}^*}\equiv \mathcal{J}'$, according to (\ref{Valgo3}).
\end{itemize}
\item[vi)] With regard to the terms in the unperturbed Hamiltonian $H^0_{\Lambda}$ which are supported in a microscopic interval $(i,i+1)$ overlapping with $\mathcal{I}^*$, but not contained in it, we have
\begin{eqnarray}
&&e^{Z_{\mathcal{I}^*}}\,\,(\frac{\sigma^z_{i^\ast_{+}}\sigma^z_{i^\ast_{+}+1}}{\xi \cdot t}+\frac{\sigma^z_{i^\ast_{-}-1}\sigma^z_{i^\ast_{-}}}{\xi \cdot t})\,e^{-Z_{\mathcal{I}^*}}\nonumber\\
&=&\frac{\sigma^z_{i^\ast_{+}}\sigma^z_{i^\ast_{+}+1}}{\xi \cdot t}+\sum_{n=1}^{\infty}\frac{1}{n!}\,ad^{n}Z_{\mathcal{I}^*\equiv \mathcal{J}^*}(\frac{\sigma^z_{i^\ast_{+}} \sigma^z_{i^\ast_{+}+1}}{\xi \cdot t})\nonumber\\
&=&\frac{\sigma^z_{i^\ast_{+}}\sigma^z_{i^\ast_{+}+1}}{\xi \cdot t}+(\ref{diaghop2})+(\ref{b-21})+(\ref{b-212})\,,\nonumber
\end{eqnarray}
where the first term in the last line contributes (once multiplied by $-\xi \cdot t\, J$) to $H^0_\Lambda$.
\item[vii)] All the terms in $H^0_{\Lambda}$ which are supported in intervals with empty intersection with $\mathcal{I}^*$ are left invariant by the conjugation.
\end{itemize} 
 \hfill \qed
 
In the final part of this section, we show that the potentials produced by the algorithm of Definition \ref{def-interactions-multi} are covariant under translations in the bulk.
\begin{defn}\label{def-transl}
Let $\mathbb{Z}\ni r \mapsto U_{[r]}$, where $U_{[r]}$ is a unitary operator on $\mathcal{H}_\Lambda$ and  $[r]$ is the equivalence class of $r$ mod $N$ (i.e., $r\mapsto [r]\in\mathbb{Z}/{N\mathbb{Z}}$), be the action of the additive group of integers factorizing through the natural unitary action of the finite, cyclic translation group $\mathbb{Z}/N\mathbb{Z}$ on the Hilbert space of the chain,   $\Lambda$, with $N$ sites. We define 
$$\tau_k(\cdot):=U_{[k\,\xi]}\, (\cdot) \, U^*_{[k\,\xi]}\, ,$$ 
with $k \in \mathbb{Z}$ and likewise $k \xi$ (recall that $\xi \in \mathbb{N}$). We also use the same symbol $\tau_k$ for the translation by $[k \xi ]\in \mathbb{Z}/N\mathbb{Z}$ of the sites of the chain.
 \end{defn}
 
In the following, given  an interval $\cJ$, we denote with $\mathfrak{j}_{\ell}(\mathcal{J})$ the leftmost site of the interval, analogously $\mathfrak{j}_{r}(\mathcal{J})$ stands for the rightmost site of the interval. This notation will be needed again later in Section \ref{gap-af}.

\begin{prop}\label{trans-cov}
For any intervals $\mathcal{I}$, $\mathcal{J}$, and for any $k\in\mathbb{Z}\cap [0,\, N-1]$ such that 
$$\mathfrak{j}_{\ell}(\cJ)>1\,,\, \text{and}\quad \mathfrak{j}_{r}(\mathcal{J})+k \xi\,<N,$$
we have
 \begin{enumerate}
\centering
\item[i)] $\tau_k(V^{\cI}_{\overline{\mathcal{J}^*}})= V^{(\cI)_{+k}}_{\tau_k(\overline{\mathcal{J}^*})}\quad\quad \text{ if }\cI\succeq \cJ\,;$

\item[ii)] $\tau_k(V^{\cI}_{\mathcal{J}})= V^{(\cI)_{+k}}_{{\tau_k(\mathcal{J}})}  \quad\quad \text{ if }\cI\prec \cJ\,.$
\end{enumerate}

\end{prop}

\noindent
\emph{Proof}

We make use of an induction on the index $\mathcal{I}$ labelling the steps of the iteration. For the basis of the induction, namely the case $\cI=(\cI_0)_{+1}$, since for every $\cJ$ satisfying the hypotheses of the statement it holds $(\cI_0)_{+1} \prec \cJ$, only $ii)$ above is to be proven. In order to prove it, we invoke a-1) in Definition \ref{def-interactions-multi} and we deduce that  $V^{(\cI_0)_{+1}}_{\mathcal{J}}=V^{\cI_0}_{\mathcal{J}}$,  $V^{(\cI_0)_{+k+1}}_{\tau_k(\mathcal{J})}=V^{\cI_0}_{\tau_k(\mathcal{J})}$ for $\ell(\mathcal{J})=1$, and $V^{(\cI_0)_{+1}}_{\mathcal{J}}=V^{(\cI_0)_{+1}}_{\tau_k(\mathcal{J})}=0$ for $\ell(\mathcal{J})>1$; thus the statement follows by the translation covariance of the potentials in the Hamiltonian $K_{\Lambda}$ which yields $\tau_{k}(V^{\cI_0}_{\mathcal{J}})=V^{\cI_0}_{\tau_k(\mathcal{J})}$.



Now suppose that the statement holds in step $\cI_{-1}$. We observe that the various cases a), b), and c) of the algorithm depend on the relative position between $\mathcal{I}$ and $\mathcal{J}$.
\noindent

If $\mathcal{I}$ and $\mathcal{J}$ are in a relative position such that case a) of the algorithm has to be used to express the potential $V^{\cI}_{\overline{\mathcal{J}^*}}$ in terms of $V^{\cI_{-1}}_{\overline{\mathcal{J}^*}}$ (resp.  $V^{\cI}_{\mathcal{J}}$ in terms of $V^{\cI_{-1}}_{\mathcal{J}}$), we only have to show that  case a) also applies to $V^{(\cI)_{+k}}_{\tau_k(\overline{\mathcal{J}^*})}$ (resp. $V^{(\cI)_{+k}}_{\tau_k(\mathcal{J})}$), since then the statement will follow by the inductive hypothesis. Indeed, if $\cI \succ \cJ$ (i.e., case a-2) applies to $V^{\cI}_{\overline{\mathcal{J}^*}})$, the relation $(\cI)_{+k} \succ \tau_k(\cJ)$ follows, thus case a-2) applies to $V^{(\cI)_{+k}}_{\tau_k(\overline{\mathcal{J}^*})}$. If  $\mathcal{I} \prec \mathcal{J}$ and $\mathcal{I}^*\nsubseteq \mathcal{J}$ (i.e., case a-1) applies to $V^{\cI}_{\mathcal{J}})$, we want to show that  $(\mathcal{I})_{+k} \prec \tau_k(\mathcal{J})$ and $(\mathcal{I})_{+k}^*\nsubseteq \tau_k(\mathcal{J})$ (recall the constraint on $k$ in the statement). We check the latter for $k=1$ so that the statement for general $k$ follows by iteration. The relation $\mathcal{I}_{+1} \prec \tau_1(\mathcal{J})$ is obvious, while for $\mathcal{I}_{+1}^*\nsubseteq \tau_1(\mathcal{J})$ the only slightly nontrivial case is when $\cI$ contains the right endpoint of the lattice. In this case $\mathcal{I}_{+1}^*\nsubseteq \tau_1(\mathcal{J})$ holds since $\mathcal{I}_{+1}^*$ contains the left endpoint of the boundary of the lattice while $\tau_1(\mathcal{J})$ does not. 

\noindent
If $\mathcal{I}$ and $\mathcal{J}$ are in a relative position such that case b) of the algorithm has to be applied to $V^{\cI}_{\overline{\mathcal{J}^*}}$ then case b) applies also to $V^{(\cI)_{+k}}_{\tau_k(\overline{\mathcal{J}^*})}$ since in this case $\tau_k(\cI)=\cI_{+k}$. Consequently,  the statement follows by inspecting formulas (\ref{firstorder})-(\ref{diaghop2}) together with (\ref{S-definition-1})-(\ref{formula-V_j}) and by the inductive hypothesis; in this respect we remark that the endpoints of the lattice $\Lambda$ cannot belong to the sets $\mathcal{J}$ and $\tau_k(\mathcal{J})$ (as assumed in the statement)  in order to exploit the translation covariance of formula (\ref{firstorder})-(\ref{diaghop2}). 

\noindent
The proof is analogous if  $\mathcal{I}$ and $\mathcal{J}$ are in a relative position such that case c) of the algorithm applies to $V^{\cI}_{\mathcal{J}}$.
\qed

\begin{rem}
A result analogous to Proposition \ref{trans-cov} holds for negative $k$.
\end{rem}

\setcounter{equation}{0}

\section{Operator norms and control of the flow}\label{final-section}
The control of the algorithm designed in the previous section requires an elaborate proof by induction which concerns the operator norm of the effective potentials, the convergence of the series yielding the operators $Z_{\mathcal{I}^*}$, and a bound from below on the spectral gaps in (\ref{resolvents-F}) and (\ref{resolvents-AF}).
For this purpose, we split the proof into different parts which will be merged as ingredients in the proof of  Theorem \ref{th-norms}. In Section \ref{gap}  we provide the argument  to control the gaps  in (\ref{resolvents-F}) and (\ref{resolvents-AF}), for which we have to distinguish the cases $J>0$ and $J<0$. 


\subsection{Gap estimates}\label{gap}
In  estimating the gap above the ground-state energy for the local Hamiltonians $G_{\mathcal{I}^\ast_{+1}}$,
%
%
for both the ferromagnetic and the antiferromagnetic case our standing assumption is 
\begin{equation}\label{ind-ass}
\|V^{\cI}_{\overline{\mathcal{J}^*}}\|\,\leq C_{J,h}\cdot\frac{(\xi\cdot t)^{\frac{\ell(\cJ)-1}{8}}}{(\ell(\cJ))^2}
\end{equation}
for all $\cJ \preceq \cI$. Here $C_{J,h}$ is a quantity dependent on $J$ and $h$; see Lemma \ref{control-LS}. (The proof of this operator norm estimate will be the content of Theorem \ref{th-norms} and Lemma \ref{control-LS} in Section \ref{normsests}).
Our results and the related arguments heavily depend on the features of the model; as said above  we present them treating the ferromagnetic and antiferromagnetic cases separately. 

\subsubsection{Spectral Gap of $G_{\mathcal{I}^*_{+1}}$ in the ferromagnetic case}\label{ferrogap}

In this case, the strategy for estimating the spectral gap of $G_{\mathcal{I}^*_{+1}}$ is similar to the one in \cite{FP} due to the nondegeneracy  of the ground-state energy of the local unperturbed Hamiltonians.  Nonetheless we spell it out in detail for the convenience of the reader. Our proof relies on the following considerations.
\begin{itemize}
\item[(i)] For  $\overline{\cJ^*} \subset \cI^*_{+1}$, $V^{\cI}_{\overline{{\mathcal{J}^*}}}$ is a block-diagonalized potential with respect to the projections $P^{(-)}_{\mathcal{I}^*_{+1}}, P^{(+)}_{\mathcal{I}^*_{+1}}$, i.e.,
\begin{eqnarray}
V^{\cI}_{\overline{\mathcal{J}^*}}&:= & P^{(+)}_{\mathcal{I}^*_{+1}}  V^{\cI}_{\overline{\mathcal{J}^*}} P^{(+)}_{\mathcal{I}^*_{+1}} + P^{(-)}_{\mathcal{I}^*_{+1}}  V^{\cI}_{\overline{\mathcal{J}^*}} P^{(-)}_{\mathcal{I}^*_{+1}}\,.  \nonumber
\end{eqnarray}
Indeed, by inspecting the definition of $V^{\cI}_{\overline{\mathcal{J}^*}}$, see formulae (\ref{firstorder})-(\ref{diaghop2}) where $\cI$ coincides with $ \cJ$, this easily follows from $P^{(+)}_{\overline{\cJ^*}}P^{(-)}_{\cI^*_{+1}}=0$ and
$$P^{(+)}_{\cI^*_{+1}}P^{(-)}_{\cJ^*}V^{\cJ_{-1}}_{\mathcal{J}}P^{(-)}_{\cJ^*}P^{(-)}_{\cI^*_{+1}}=P^{(+)}_{\cI^*_{+1}}\langle V^{\cJ_{-1}}_{\mathcal{J}} \rangle_{\Psi_{\cJ^*}}P^{(-)}_{\cJ^*}P^{(-)}_{\cI^*_{+1}}=\langle V^{\cJ_{-1}}_{\mathcal{J}} \rangle_{\Psi_{\cJ^*}}\, P^{(+)}_{\cI^*_{+1}}P^{(-)}_{\cI^*_{+1}}=0\,,$$
where we recall that for $\langle V \rangle_{\Psi}$ stands for $\langle \Psi\,,\,V\,\Psi\rangle$,  for any vector $\Psi$ and operator $V$.
In particular this implies that $G_{\mathcal{I}^*_{+1}}$ is block-diagonal with respect to the projections $P^{(-)}_{\mathcal{I}^*_{+1}}, P^{(+)}_{\mathcal{I}^*_{+1}}$.
\item[(ii)]

 Denoting $(\mathcal{I}^*_{+1})_K:=\{\cJ\in\mathfrak{I}: \ell(\mathcal{J})=K,\,\, \overline{\cJ^*}\subset \mathcal{I}^*_{+1}\}$ and $\mathscr{I}_{K}:=\{ \,i\in \bigcup\limits_{\cJ\in(\mathcal{I}^*_{+1})_K} \overline{\cJ^*}\}$, we have
\begin{eqnarray}
\sum_{\cJ\in (\mathcal{I}^*_{+1})_K}(H^C_{\overline{\mathcal{J}^*}}-\langle \, H^C_{\overline{\mathcal{J}^*}}\,\rangle_{\Psi_{\overline{\cJ^*}}})&=&\sum_{\cJ\in (\mathcal{I}^*_{+1})_K}\,\sum_{i,i+1\,\in \overline{\mathcal{J}^*}}\,(H^C_{i,i+1}-\langle\,H^C_{i,i+1}\,\rangle_{\Psi_{i,i+1}})
\label{sum-H-in}\\
&\leq &(K+1) \sum_{i,i+1\,\in \mathscr{I}_{K}} (H^C_{i,i+1}-\langle\, H^C_{i,i+1}\,\rangle_{\Psi_{i,i+1}}) \,\\
\nonumber
&=&(K+1)\, (H^C_{\mathscr{I}_{K}}-\langle\,H^C_{\mathscr{I}_{K}}\,\rangle_{\Psi_{\mathscr{I}_{K}}})
\end{eqnarray}
where $H^C_{i,i+1}$ and $\Psi_{i,i+1}$ are $H^C_{\mathcal{I}}$ and $\Psi_{\mathcal{I}}$ for $\mathcal{I}\equiv \{i,i+1\}$, consequently $H^C_{\mathscr{I}_{K}}=\sum_{i,i+1\,\in \mathscr{I}_{K}}H^C_{i,i+1}$ by definition, with $\Psi_{\mathscr{I}_{K}}$ its  ground-state.
\end{itemize}
\noindent
Next, assuming  the bound in (\ref{ind-ass})  and making use of the inequality
\begin{equation}\label{P+/HC}
P^{(+)}_{\overline{\mathcal{J}^*}}\leq \frac{1}{2J+h}\,(H^C_{\overline{\mathcal{J}^*}}-\langle\, H^C_{\overline{\mathcal{J}^*}}\,\rangle_{\Psi_{\overline{\cJ^*}}})\,,
\end{equation}
where we have used that $2J+h$ is the spectral gap above the ground-state energy of $H^C_{\overline{\mathcal{J}^*}}$, 
we can estimate
\begin{equation}
\pm P^{(+)}_{\overline{\mathcal{J}^*}}\, V^{\cI}_{\overline{\mathcal{J}^*}} \,P^{(+)}_{\overline{\mathcal{J}^*}} \leq  \frac{{C_{J,h}}}{2J+h} \cdot (\xi\cdot t)^{\frac{\ell(\cJ)-1}{8}}\,(H^C_{\overline{\mathcal{J}^*}}-\langle\,H^C_{\overline{\mathcal{J}^*}}\,\rangle_{\Psi_{\overline{\cJ^*}}})\,.\label{fin-est-V-F}
\end{equation}
The inequality in (\ref{fin-est-V-F}) combined with point (ii) above yields
\begin{equation}\label{ferrogapest}
\pm \sum_{\cJ\in(\mathcal{I}^*_{+1})_K} P^{(+)}_{\overline{\mathcal{J}^*}}\, V^{\cI}_{\overline{\mathcal{J}^*}} \,P^{(+)}_{\overline{\mathcal{J}^*}} \leq  \frac{C_{J,h}}{2J+h} \cdot (\xi\cdot t)^{\frac{\ell(\cJ)-1}{8}}\,(K+1)\, (H^C_{\mathcal{I}^*_{+1}}-\langle\,H^C_{\mathcal{I}^*_{+1}}\,\rangle_{\Psi_{\mathcal{I}^*_{+1}}})\,,
\end{equation}
since $\mathscr{I}_{K}\subset \mathcal{I}^*_{+1}$ for all $K$ and $H^C_{\overline{\mathcal{J}^*}}-\langle\, H^C_{\overline{\mathcal{J}^*}}\,\rangle_{\Psi_{\overline{\cJ^*}}}\leq H^C_{\mathcal{I}^*_{+1}}-\langle\,H^C_{\mathcal{I}^*_{+1}}\,\rangle_{\Psi_{\mathcal{I}^*_{+1}}}$.

\noindent
With these ingredients, the proof of the next lemma can then be easily derived.
\begin{lem}\label{gap-bound-ferr}
Assuming that the bound in (\ref{ind-ass}) holds in step $\cI$ of the block-diagonalization, and choosing $t>0$ small enough such that
\begin{equation}
\Big\{{ 1}-\frac{4 C_{J,h}}{2J+h}\cdot \xi\cdot t-\frac{2 C_{J,h}}{2J+h}\cdot  \xi\cdot t \sum_{l=3}^{\infty}l\cdot (\xi\cdot t)^{\frac{l-2}{8}}\Big\}>0\,,
\end{equation}
the inequality
\begin{equation}
P^{(+)}_{\mathcal{I}^*_{+1}}\,(G_{{\mathcal{I}^*_{{+1}}}}-E_{{\mathcal{I}^*_{{+1}}}})\,P^{(+)}_{{\mathcal{I}^*_{{+1}}}}
\geq\,(2J+2h)\Big\{1-\frac{4C_{J,h}}{2J+h}\cdot  \xi\cdot t-\frac{2C_{J,h}}{2J+h}\cdot  \xi\cdot t \sum_{l=3}^{\infty}l\cdot (\xi\cdot t)^{\frac{l-2}{8}}\Big\}\,P^{(+)}_{\mathcal{I}^*_{{+1}}} \label{final-eq-1}
\end{equation}
holds,  where 
\begin{equation}\label{def-E-bis}
E_{\mathcal{I}^*_{{+1}}}:=\,\langle\, G_{{\mathcal{I}^*_{{+1}}}}\,\rangle_{\Psi_{\cI^*_{+1}}}\,.
\end{equation}


\end{lem}

\noindent
\emph{Proof}

By definition of $G_{{\mathcal{I}^*_{{+1}}}}$ and using the identity
$$ V^{\cI}_{\overline{\mathcal{J}^*}}= P^{(+)}_{\overline{\mathcal{J}^*}} V^{\cI}_{\overline{\mathcal{J}^*}}P^{(+)}_{\overline{\mathcal{J}^*}} +P^{(-)}_{\overline{\mathcal{J}^*}} V^{\cI}_{\overline{\mathcal{J}^*}}P^{(-)}_{\overline{\mathcal{J}^*}} $$ for all $\overline{\mathcal{J}^*}\subset \mathcal{I}^*_{{+1}}$, we can write
\begin{eqnarray}
P^{(+)}_{\mathcal{I}^*_{{+1}}}\,G_{{\mathcal{I}^*_{{+1}}}}\,P^{(+)}_{{\mathcal{I}^*_{{+1}}}}&=& P^{(+)}_{\mathcal{I}^*_{{+1}}}\,(H_{\mathcal{I}^*_{{+1}}}^{0}+\xi\cdot t \,\,\sum_{\overline{\mathcal{J}^*}\subset \mathcal{I}^*_{{+1}}} V^{\cI}_{\overline{\mathcal{J}^*}})\, P^{(+)}_{\mathcal{I}^*_{{+1}}}\nonumber\\
&=&P^{(+)}_{\mathcal{I}^*_{{+1}}}\,(H_{\mathcal{I}^*_{{+1}}}^{0}+\xi\cdot t \,\,\sum_{\overline{\mathcal{J}^*}\subset \mathcal{I}^*_{{+1}}} P^{(+)}_{\overline{\mathcal{J}^*}}V^{\cI}_{\overline{\mathcal{J}^*}}\, P^{(+)}_{\overline{\mathcal{J}^*}})P^{(+)}_{\mathcal{I}^*_{{+1}}}\\
&&+P^{(+)}_{\mathcal{I}^*_{{+1}}}\,\xi\cdot t\,\,\,\sum_{\overline{\mathcal{J}^*}\subset \mathcal{I}^*_{{+1}}}  \langle\,V^{\cI}_{\overline{\mathcal{J}^*}}\,\rangle_{\Psi_{\overline{\cJ^*}}}\, (\charf-P^{(+)}_{\overline{\mathcal{J}^*}})\,P^{(+)}_{\mathcal{I}^*_{{+1}}}.
\end{eqnarray}
Now, using (\ref{ferrogapest})  above together with 
\begin{equation}\label{Hc<H0}
H^C_{\mathcal{I}^*_{{+1}}}-\langle\,H^C_{\mathcal{I}^*_{{+1}}}\,\rangle_{\Psi_{\cI^*_{+1}}} \leq H^0_{\mathcal{I}^*_{{+1}}}-\langle\,H^0_{\mathcal{I}^*_{{+1}}}\,\rangle_{\Psi_{\cI^*_{+1}}}\,,
\end{equation}
(which trivially follows from (\ref{H0-HC}) and (\ref{gs-vector}))
we get 
\begin{eqnarray}
\nonumber
&& \pm \sum_{\overline{\mathcal{J}^* }\subset \mathcal{I}^*_{{+1}}}\,P^{(+)}_{\overline{\mathcal{J}^*}}\,   V^{\cI}_{\overline{\mathcal{J}^*}}   \,P^{(+)}_{\overline{\mathcal{J}^*}} \\
&\leq & \frac{C_{J,h}}{2J+h}\cdot \sum_{\ell(\cJ)=1}^{\ell(\cI)-1} (\xi\cdot t)^{\frac{\ell(\cJ)-1}{8}}\,(\ell(\cJ)+1)\,(H^0_{\mathcal{I}^*_{{+1}}}-\langle\,H^0_{\mathcal{I}^*_{{+1}}}\,\rangle_{\Psi_{\cI^*_{+1}}}) \,.\label{bound-G}
\end{eqnarray}
Thus, with the help of (\ref{P+/HC}), (\ref{Hc<H0}), and using the identity
$$ H^0_{\mathcal{I}^*_{{+1}}} -E_{{\mathcal{I}^*_{+1}}}=H^0_{\mathcal{I}^*_{{+1}}}-\langle\,H^0_{\mathcal{I}^*_{{+1}}}\,\rangle_{\Psi_{\cI^*_{+1}}}-\xi\cdot t\,\,\,\sum_{\overline{\mathcal{J}^*}\subset \mathcal{I}^*_{{+1}}}  \langle\,V^{\cI}_{\overline{\mathcal{J}^*}}\,\rangle_{\Psi_{\overline{\cJ^*}}}\,,$$  we can conclude that
\begin{eqnarray}
\nonumber
&&P^{(+)}_{\mathcal{I}^*_{+1}}\,(G_{{\mathcal{I}^*_{+1}}}-E_{{\mathcal{I}^*_{+1}}})\,P^{(+)}_{{\mathcal{I}^*_{+1}}}\\
\nonumber
&&\geq P^{(+)}_{\mathcal{I}^*_{+1}}\Big(1- \frac{2C_{J,h}}{2J+h}\cdot \xi\cdot t\sum_{K'=1}^\infty (\xi\cdot t)^{\frac{K'-1}{8}}\,(K'+1)\,(H^0_{\mathcal{I}^*_{+1}}-\langle\,H^0_{\mathcal{I}^*_{{+1}}}\,\rangle_{\Psi_{\cI^*_{+1}}})\Big)P^{(+)}_{\mathcal{I}^*_{+1}} \nonumber\\
\nonumber
&&\geq (2J+2h) \Big(1- \frac{2C_{J,h}}{2J+h}\cdot \xi\cdot t\sum_{K'=1}^\infty (\xi\cdot t)^{\frac{K'-1}{8}}\,(K'+1)\Big)\,P^{(+)}_{\mathcal{I}^*_{+1}} 
\end{eqnarray}
where for the last inequality we have used 
$$P^{(+)}_{\mathcal{I}^*_{+1}}(H^0_{\mathcal{I}^*_{{+1}}}-\langle\,H^0_{\mathcal{I}^*_{{+1}}}\,\rangle_{\Psi_{\cI^*_{+1}}})P^{(+)}_{\mathcal{I}^*_{+1}}\geq (2J+2h)P^{(+)}_{\mathcal{I}^*_{+1}}
$$
which follows from Proposition \ref{propgap}.
\qed
\subsubsection{Low-lying spectrum of $G_{\mathcal{I}^*_{+1}}$ in the antiferromagnetic case}\label{gap-af}

The discussion is carried out in three steps. 
\begin{itemize}
\item[I)] In Lemma \ref{Gstructure} we show that, due to the definition of $V^{\mathcal{J}}_{\overline{\mathcal{J}^*}}$ in point b) of the algorithm (see (\ref{firstorder})-(\ref{diaghop2})), the operator 
$G_{\mathcal{I}^*_{+1}}$ is not only block-diagonal with respect to $P^{(-)}_{\mathcal{I}^*_{+1}}, P^{(+)}_{\mathcal{I}^*_{+1}}$ but also w.r.t. $P^{(-),A}_{\mathcal{I}^*_{+1}}, P^{(-),B}_{\mathcal{I}^*_{+1}}$;
\item[II)]  {The result is then used in Lemma \ref{diffenergy} where we estimate the difference $ |E^B_{{\mathcal{I}^*_{{+1}}}}-E^A_{{\mathcal{I}^*_{{+1}}}}| $} of the two lowest eigenvalues of $G_{\mathcal{I}^*_{+1}}$;

\item[III)] In Lemma \ref{gap-bound} we finally estimate the distance between the spectrum of $P^{(+)}_{\mathcal{I}^*_{+1}}G_{\mathcal{I}^*_{+1}}P^{(+)}_{\mathcal{I}^*_{+1}}$ and the two lowest eigenvalues of $G_{\mathcal{I}^*_{+1}}$, with the help of Lemma \ref{gapestantif} which provides an intermediate result.
\end{itemize}

In the following, due to the structure of the two lowest-energy eigenvectors of the local unperturbed Hamiltonians, it will be useful to split the set consisting of intervals $\mathcal{J}^*$ such that $\overline{\mathcal{J}^*}\subset \mathcal{I}^*_{+1}$, i.e., 
\begin{equation}\label{set-int}
(\mathcal{I}^*_{+1})_{\text{int}} := \left\{ \mathcal{J}^*\in\mathfrak{I}^* \, :  \overline{\mathcal{J}^*}\subset \mathcal{I}^*_{+1}\right\}\,,
\end{equation}
into two sets which are defined  below, where $\mathfrak{j}_{\ell}(\mathcal{J})$ stands for the microscopic coordinate of the leftmost site of the interval $\mathcal{J}$:
\begin{eqnarray}\label{def-ev}
&& (\mathcal{I}^*_{+1})_{\text{ev}} := \left\{ \mathcal{J}^*\in\mathfrak{I}^* \, :  \overline{\mathcal{J}^*}\subset \mathcal{I}^*_{+1}, |\mathfrak{j}_{\ell}(\mathcal{J}^*)-\mathfrak{j}_{\ell}(\mathcal{I}_{+1}^*)| \text{ is even }\, \right\},\\
&& (\mathcal{I}^*_{+1})_{\text{odd}}:=\left\{\mathcal{J}^*\in\mathfrak{I}^* \, :  \overline{\mathcal{J}^*}\subset \mathcal{I}^*_{+1}, |\mathfrak{j}_{\ell}(\mathcal{J}^*)-\mathfrak{j}_{\ell}(\mathcal{I}_{+1}^*)| \text{ is odd }\,\right\}\,.\label{def-odd}
\end{eqnarray}
\begin{rem}\label{switch}
We observe that, by definition of the vectors $\Psi^A_{\mathcal{I}}$ and $\Psi^B_{\mathcal{I}}$, and of the sets $(\mathcal{I}^*_{+1})_{\text{ev}}\,,\,(\mathcal{I}^*_{+1})_{\text{odd}}$, the following identities hold true:

\noindent
if $\mathcal{J}^*\in(\mathcal{I}^*_{+1})_{\text{ev}}$
\begin{equation}
\langle V_{\mathcal{J}^*} \rangle_{\Psi^{A/B}_{\mathcal{I}^*_{+1}}}=\langle V_{\mathcal{J}^*} \rangle_{\Psi^{A/B}_{\mathcal{J}^*}}
\end{equation}
if $\mathcal{J}^*\in(\mathcal{I}^*_{+1})_{\text{odd}}$
\begin{equation}
\langle V_{\mathcal{J}^*} \rangle_{\Psi^{A/B}_{\mathcal{I}^*_{+1}}}=\langle V_{\mathcal{J}^*} \rangle_{\Psi^{B/A}_{\mathcal{J}^*}}\,.
\end{equation}
\end{rem}
In the lemma below we prove a property that can be seen as a weak form of  the LTQO condition for the ground-state subspace of $H^C_{\mathcal{I}^*_{+1}}$. It consists in the absence of off-diagonal terms of the effective potentials $V^\mathcal{I}_{\overline{\mathcal{J}^*}}$ w.r.t. $P^{(-),A}_{\mathcal{I}^*_{+1}}$ and $ P^{(-),B}_{\mathcal{I}^*_{+1}}$. This property results from using enlarged intervals in the block-diagonalization procedure. 
\begin{lem}\label{Gstructure}
For any $\mathcal{I},\cJ\in\mathfrak{I}$ with $\overline{\mathcal{J}^*}\subset\cI^*_{+1}$, we have the following:
\begin{itemize}
\item[(a)]$V^\mathcal{I}_{\overline{\mathcal{J}^*}}$ is block-diagonal with respect to $P^{(-),A}_{\overline{\mathcal{J}^*}}, P^{(-),B}_{\overline{\mathcal{J}^*}}, P^{(+)}_{\overline{\mathcal{J}^*}}$ (see their definitions below (\ref{W}));
\item[(b)] $V^\mathcal{I}_{\overline{\mathcal{J}^*}}$ is block-diagonal with respect to $P^{(-),A}_{\mathcal{I}^*_{+1}}, P^{(-),B}_{\mathcal{I}^*_{+1}}, P^{(+)}_{\mathcal{I}^*_{+1}}$. 
\end{itemize}
Consequently, $G_{\mathcal{I}^*_{+1}}$ is block-diagonal with respect to $P^{(-),A}_{\mathcal{I}^*_{+1}}, P^{(-),B}_{\mathcal{I}^*_{+1}}, P^{(+)}_{\mathcal{I}^*_{+1}}$.
\end{lem}

\noindent
\emph{Proof}\\

The proofs of points (a) and (b) are identical; thus we show only (b), that is we prove that
\begin{eqnarray}
&&P^{(+)}_{\cI^*_{+1}} V^{\cI}_{\overline{\cJ^*}} P^{(-)}_{\cI^*_{+1}}=0,\label{first}\\
&& P^{(-),A}_{\cI^*_{+1}} V^{\cI}_{\overline{\cJ^*}} P^{(-), B}_{\cI^*_{+1}}=0. \label{second}
\end{eqnarray}
We recall that (see point b) in Definition \ref{def-interactions-multi}) for $\mathcal{I}\succeq \mathcal{J}$
\begin{eqnarray}
V^{\mathcal{I}}_{\overline{\mathcal{J}^*}}=V^{\mathcal{J}}_{\overline{\mathcal{J}^*}}&:=& P^{(+)}_{\mathcal{J}^*} V^{\mathcal{J}_{-1}}_{\mathcal{J}}P^{(+)}_{\mathcal{J}^*} +  P^{(-)}_{\mathcal{J}^*} V^{\mathcal{J}_{-1}}_{\mathcal{J}}P^{(-)}_{\mathcal{J}^*}\label{firstorder1}\\
&&+P^{(+)}_{\overline{\mathcal{J}^*}}\Big(adZ_{\mathcal{J}^*}(\,\frac{\sigma^z_{i^\ast_- -1}\sigma^z_{i^\ast_-}}{\xi\cdot t}+\frac{\sigma^z_{i^\ast_{+}}\sigma^z_{i^\ast_{+}+1}}{\xi\cdot t})\Big)P^{(+)}_{\overline{\mathcal{J}^*}},
\end{eqnarray}
and we observe the following relations:
\begin{enumerate}
\item[(i)] $P^{(-)}_{\mathcal{I}^*_{+1}}P^{(+)}_{\mathcal{J}^*}=P^{(-)}_{\mathcal{I}^*_{+1}}P^{(+)}_{\overline{\mathcal{J}^*}}=0$ since $\overline{\mathcal{J}^*}\subset\cI^*_{+1}$ by assumption;
\item[(ii)] $P^{(-),A}_{{\mathcal{J}^*}}V^{\cJ_{-1}}_{\mathcal{J}}P^{(-),B}_{{\mathcal{J}^*}}=0$, due to the fact that $V^{\cJ_{-1}}_{\cJ}$ is localized in $\mathcal{J}$, which is strictly contained in ${\mathcal{J}^*}$; hence $$P^{(-)}_{{\mathcal{J}^*}}V^{\cJ_{-1}}_{\mathcal{J}}P^{(-)}_{{\mathcal{J}^*}}=\langle V^{\cJ_{-1}}_{\mathcal{J}}\rangle_{\Psi^A_{\cJ^*}}P^{(-),A}_{{\mathcal{J}^*}}+ \langle V^{\cJ_{-1}}_{\mathcal{J}}\rangle_{\Psi^B_{\cJ^*}} P^{(-),B}_{{\mathcal{J}^*}};$$

\item[(iii)] If $\cJ^*\in(\mathcal{I}^*_{+1})_{\text{ev}}$,\quad $P^{(-)}_{\mathcal{I}^*_{+1}} P^{(-),A}_{{\mathcal{J}^*}}=P^{(-),A}_{\mathcal{I}^*_{+1}}$ and $P^{(-)}_{\mathcal{I}^*_{+1}} P^{(-),B}_{{\mathcal{J}^*}}=P^{(-),B}_{\mathcal{I}^*_{+1}}$.\\
If $\cJ^*\in(\mathcal{I}^*_{+1})_{\text{odd}}$,\quad $P^{(-)}_{\mathcal{I}^*_{+1}} P^{(-),A}_{{\mathcal{J}^*}}=P^{(-),B}_{\mathcal{I}_{+1}^*}$ and $P^{(-)}_{\mathcal{I}^*_{+1}} P^{(-),B}_{{\mathcal{J}^*}}=P^{(-),A}_{\mathcal{I}^*_{+1}}.$
\item[(iv)]  If $\cJ^*\in(\mathcal{I}^*_{+1})_{\text{ev}}$,\quad $P^{(-),A}_{\mathcal{I}^*_{+1}} P^{(-),A}_{{\mathcal{J}^*}}=P^{(-),A}_{\mathcal{I}^*_{+1}}$\,,\,\, $P^{(-),B}_{\mathcal{I}^*_{+1}} P^{(-),B}_{{\mathcal{J}^*}}=P^{(-),B}_{\mathcal{I}^*_{+1}}$\,, and $P^{(-),A}_{\mathcal{I}^*_{+1}} P^{(-),B}_{{\mathcal{J}^*}}=0$.\\
If $\cJ^*\in(\mathcal{I}^*_{+1})_{\text{odd}}$,\quad $P^{(-),A}_{\mathcal{I}^*_{+1}} P^{(-),A}_{{\mathcal{J}^*}}=0$\,,\,\, $P^{(-),B}_{\mathcal{I}^*_{+1}} P^{(-),B}_{{\mathcal{J}^*}}=0$\,, and $P^{(-),A}_{\mathcal{I}^*_{+1}} P^{(-),B}_{{\mathcal{J}^*}}=P^{(-),A}_{\mathcal{I}^*_{+1}}$.
\end{enumerate}
Thus, using (\ref{firstorder1}),
\begin{eqnarray}
&& P^{(+)}_{\cI^*_{+1}} V^{\cI}_{\overline{\cJ^*}} P^{(-)}_{\cI^*_{+1}}\nonumber\\
&=&P^{(+)}_{\cI^*_{+1}}  P^{(-)}_{\mathcal{J}^*} V^{\mathcal{J}_{-1}}_{\mathcal{J}}P^{(-)}_{\mathcal{J}^*} P^{(-)}_{\cI^*_{+1}}\nonumber\\
&=& P^{(+)}_{\cI^*_{+1}} \Big(\langle V^{\cJ_{-1}}_{\mathcal{J}}\rangle_{\Psi^A_{\cJ^*}}P^{(-),A}_{{\mathcal{J}^*}}+ \langle V^{\cJ_{-1}}_{\mathcal{J}}\rangle_{\Psi^B_{\cJ^*}} P^{(-),B}_{{\mathcal{J}^*}}\Big)P^{(-)}_{\cI^*_{+1}}=0\nonumber
\end{eqnarray}
which proves (\ref{first}), where the first equality is due to item (i), the second to item (ii), and the last one to item (iii). 

Concerning (\ref{second}), 
\begin{eqnarray}
&&P^{(-),A}_{\cI^*_{+1}} V^{\cI}_{\overline{\cJ^*}} P^{(-), B}_{\cI^*_{+1}}\nonumber \\
&=&  P^{(-),A}_{\cI^*_{+1}}  P^{(-)}_{\mathcal{J}^*} V^{\mathcal{J}_{-1}}_{\mathcal{J}}P^{(-)}_{\mathcal{J}^*}P^{(-), B}_{\cI^*_{+1}}\nonumber\\
&=&  P^{(-),A}_{\cI^*_{+1}}  (\langle V^{\cJ_{-1}}_{\mathcal{J}}\rangle_{\Psi^A_{\cJ^*}}P^{(-),A}_{{\mathcal{J}^*}}+ \langle V^{\cJ_{-1}}_{\mathcal{J}}\rangle_{\Psi^B_{\cJ^*}} P^{(-),B}_{{\mathcal{J}^*}})P^{(-),B}_{\cI^*_{+1}}=0\,, \nonumber
\end{eqnarray}
where the first equality follows from item (i), the second from item (ii), and the last one from item (iv).
\qed

In the next lemma we make use of the argument (see Section \ref{deg-en-af})  regarding what we refer to as \emph{degeneracy of the bulk ground-state energy}.
\begin{lem}\label{diffenergy}
Let \begin{equation}\label{def-EA-bis}
E^A_{\mathcal{I}^*_{{+1}}}:=\langle\, G_{{\mathcal{I}^*_{{+1}}}}\,\rangle_{\Psi^A_{\cI^*_{+1}}},
\end{equation}
and  $$E^B_{\mathcal{I}^*_{{+1}}}:=\langle\, G_{{\mathcal{I}^*_{{+1}}}}\,\rangle_{\Psi^B_{\cI^*_{+1}}}.$$
If  $\mathcal{I}^*$ has an odd number of sites, then 
$$ E^B_{{\mathcal{I}^*_{{+1}}}}-E^A_{{\mathcal{I}^*_{{+1}}}} = 2h + O(\xi \cdot t).$$
If $\cI^*$ has an even number of sites, then 
$$|E^B_{\mathcal{I}^*_{+1}}-E^A_{\mathcal{I}^*_{+1}}| \leq O(\xi \cdot t).$$
\end{lem}

\noindent
\emph{Proof}

If the set $\cI^*$ has odd cardinality, then $$\langle\,H^0_{\mathcal{I}^*_{{+1}}}\,\rangle_{\Psi^B_{\cI^*_{+1}}} - \langle\,H^0_{\mathcal{I}^*_{{+1}}}\,\rangle_{\Psi^A_{\cI^*_{+1}}}=2h,$$ whereas  if $\cI^*$ has even cardinality $$\langle\,H^0_{\mathcal{I}^*_{{+1}}}\,\rangle_{\Psi^B_{\cI^*_{+1}}} - \langle\,H^0_{\mathcal{I}^*_{{+1}}}\,\rangle_{\Psi^A_{\cI^*_{+1}}}=0.$$
Next we consider two subsets of  $(\cI^*_{+1})_{\text{ev}}$ and $(\cI^*_{+1})_{\text{odd}}$, respectively:

\begin{equation}\label{ev-prime}
(\cI^*_{+1})'_{\text{ev}}:=\{\cJ^*\in (\cI^*_{+1})_{\text{ev}}: \tau_{-1}(\overline{\cJ^*})\subset \cI^*_{+1}\}
\end{equation}
 and 
 \begin{equation}\label{odd-prime}
 (\cI^*_{+1})'_{\text{odd}}:=\{\cJ^*\in (\cI^*_{+1})_{\text{odd}}: \tau_{1}(\overline{\cJ^*})\subset \cI^*_{+1}\}\,.
 \end{equation}
The set $(\cI^*_{+1})'_{\text{ev}}$, respectively $(\cI^*_{+1})'_{\text{odd}}$, consists of intervals that are still contained in $\cI^*_{+1}$ when shifted by $\tau_{k}$ with $k=-1$, respectively $k=1$ (see Definition \ref{def-transl}). We also observe that
\begin{equation}\label{translints}
\tau_{-1}((\cI^*_{+1})'_{\text{ev}})=(\cI^*_{+1})'_{\text{odd}}\,.
\end{equation}
If  $\mathcal{I}^*$ has an odd number of sites, by splitting $(\cI^*_{+1})_{\text{int}}$ (see (\ref{set-int})) into $(\cI^*_{+1})_{\text{ev}}\cup (\cI^*_{+1})_{\text{odd}}$, and by using the properties in Remark \ref{switch},  we can write

\begin{eqnarray}
&&E^B_{{\mathcal{I}^*_{{+1}}}}-E^A_{{\mathcal{I}^*_{{+1}}}}\\
&=&\langle\,H^0_{\mathcal{I}^*_{{+1}}}\,\rangle_{\Psi^B_{\cI^*_{+1}}} - \langle\,H^0_{\mathcal{I}^*_{{+1}}}\,\rangle_{\Psi^A_{\cI^*_{+1}}} +\xi\cdot t\cdot \sum_{\overline{\cJ^*}\subset\cI^*_{+1}}(\langle\,V^{\cI}_{\overline{\mathcal{J}^*}}\,\rangle_{\Psi^B_{\mathcal{I}^*_{+1}}}-\langle V^{\cI}_{\overline{\mathcal{J}^*}}\,\rangle_{\Psi^A_{\mathcal{I}^*_{+1}}})\\
&= & 2h +\xi\cdot t\cdot\left( \sum_{\cJ^*\in(\cI^*_{+1})_{\text{ev}}}(\langle\,V^{\cI}_{\overline{\mathcal{J}^*}}\,\rangle_{\Psi^B_{\overline{\mathcal{J}^*}}}-\langle V^{\cI}_{\overline{\mathcal{J}^*}}\,\rangle_{\Psi^A_{\overline{\mathcal{J}^*}}})+\sum_{\cJ^*\in(\cI^*_{+1})_{\text{odd}}}(\langle\,V^{\cI}_{\overline{\mathcal{J}^*}}\,\rangle_{\Psi^A_{\overline{\mathcal{J}^*}}}-\langle V^{\cI}_{\overline{\mathcal{J}^*}}\,\rangle_{\Psi^B_{\overline{\mathcal{J}^*}}})\right)\,.\nonumber 
\end{eqnarray}
Next, by splitting  $(\cI^*_{+1})_{\text{ev/odd}}$ into $(\cI^*_{+1})'_{\text{ev/odd}}\cup [(\cI^*_{+1})_{\text{ev/odd}}\setminus (\cI^*_{+1})'_{\text{ev/odd}}]$, we can estimate

\begin{eqnarray}
& &|E^B_{{\mathcal{I}^*_{{+1}}}}-E^A_{{\mathcal{I}^*_{{+1}}}}-2h|\nonumber \\
&\leq&\xi\cdot t\cdot\Big|\left(\sum_{\cJ^*\in(\cI^*_{+1})_{\text{ev}}'}(\langle\,V^{\cI}_{\overline{\mathcal{J}^*}}\,\rangle_{\Psi^B_{\overline{\mathcal{J}^*}}}-\langle V^{\cI}_{\overline{\mathcal{J}^*}}\,\rangle_{\Psi^A_{\overline{\mathcal{J}^*}}})+\sum_{\cJ^*\in(\cI^*_{+1})_{\text{odd}}'}(\langle\,V^{\cI}_{\overline{\mathcal{J}^*}}\,\rangle_{\Psi^A_{\overline{\mathcal{J}^*}}}-\langle V^{\cI}_{\overline{\mathcal{J}^*}}\,\rangle_{\Psi^B_{\overline{\mathcal{J}^*}}})\right)\Big|\label{cancel}\\
& &+\xi\cdot t\cdot\sum_{\cJ^*\in (\cI^*_{+1})_{\text{ev}}\setminus (\cI^*_{+1})_{\text{ev}}'}|(\langle\,V^{\cI}_{\overline{\mathcal{J}^*}}\,\rangle_{\Psi^B_{\overline{\mathcal{J}^*}}}-\langle V^{\cI}_{\overline{\mathcal{J}^*}}\,\rangle_{\Psi^A_{\overline{\mathcal{J}^*}}})|+\xi\cdot t\cdot\sum_{\cJ^*\in (\cI^*_{+1})_{\text{odd}}\setminus(\cI^*_{+1})_{\text{odd}}'}|(\langle\,V^{\cI}_{\overline{\mathcal{J}^*}}\,\rangle_{\Psi^A_{\overline{\mathcal{J}^*}}}-\langle V^{\cI}_{\overline{\mathcal{J}^*}}\,\rangle_{\Psi^B_{\overline{\mathcal{J}^*}}})|\,.\nonumber
\end{eqnarray}
Now we claim that the terms in (\ref{cancel}) cancel out. In order to show this we recall (\ref{translints}) and  observe that, for each $\cJ^*\in(\cI^*_{+1})_{\text{ev}}'$, we have: 
\begin{itemize}
\item by invoking Proposition \ref{trans-cov} 
\begin{equation}\label{id-cov-1}
\langle\,V^{\cI}_{\overline{\mathcal{J}^*}}\,\rangle_{\Psi^{A/B}_{\overline{\mathcal{J}^*}}}=\langle\,V^{\cI_{-1}}_{\tau_{-1}(\overline{\mathcal{J}^*})}\,\rangle_{\Psi^{A/B}_{\tau_{-1}(\overline{\mathcal{J}^*})}}\,;
\end{equation}
\item due to the rules in Definition \ref{def-interactions-multi},  a potential that has been block-diagonalized does not change along the flow; since $\mathcal{I}\succ \mathcal{J}$, the potential $V^{\tau_{-1}(\cI)}_{\tau_{-1}(\overline{\mathcal{J}^*})}$ is already block-diagonalized and, consequently,
\begin{equation}\label{id-cov-2}
\langle\,V^{\cI_{-1}}_{\tau_{-1}(\overline{\mathcal{J}^*})}\,\rangle_{\Psi^{A/B}_{\tau_{-1}(\overline{\mathcal{J}^*})}}=\langle\,V^{\cI}_{\tau_{-1}(\overline{\mathcal{J}^*})}\,\rangle_{\Psi^{A/B}_{\tau_{-1}(\overline{\mathcal{J}^*})}}\,,
\end{equation}
where $\tau_{-1}(\overline{\mathcal{J}^*})$ is indeed an interval belonging to $(\cI^*_{+1})^{'}_{\text{odd}}$.
\end{itemize}
The identities in (\ref{id-cov-1}) and (\ref{id-cov-2}) imply that with each term in the first sum of (\ref{cancel}) we can associate another one in the second sum that is exactly its opposite.

Then we can write
\begin{eqnarray}
& &|E^B_{{\mathcal{I}^*_{{+1}}}}-E^A_{{\mathcal{I}^*_{{+1}}}}-2h|\\
&\leq &\xi\cdot t\cdot\sum_{\cJ^*\in (\cI^*_{+1})_{\text{ev}}\setminus (\cI^*_{+1})_{\text{ev}}'}|(\langle\,V^{\cI}_{\overline{\mathcal{J}^*}}\,\rangle_{\Psi^B_{\overline{\mathcal{J}^*}}}-\langle V^{\cI}_{\overline{\mathcal{J}^*}}\,\rangle_{\Psi^A_{\overline{\mathcal{J}^*}}})|+\xi\cdot t\cdot\sum_{\cJ^*\in (\cI^*_{+1})_{\text{odd}}\setminus(\cI^*_{+1})_{\text{odd}}'}|(\langle\,V^{\cI}_{\overline{\mathcal{J}^*}}\,\rangle_{\Psi^A_{\overline{\mathcal{J}^*}}}-\langle V^{\cI}_{\overline{\mathcal{J}^*}}\,\rangle_{\Psi^B_{\overline{\mathcal{J}^*}}})|\nonumber\\
&\leq  &4 \cdot \xi\cdot t\cdot\sum_{K=1}^\infty C_{J,h}\cdot \frac{(\xi\cdot t)^{\frac{K-1}{8}}}{K^2}\,,
\end{eqnarray}
where for the last inequality we use that there is at most one interval of length $K$ for each sum and its norm is bounded by $C_{J,h}\cdot \frac{(\xi\cdot t)^{\frac{K-1}{8}}}{K^2}$ by (\ref{ind-ass}).
This proves the inequality in the statement of the lemma. An analogous argument applies if $\mathcal{I}^*_{+1}$ has an even number of sites.
\qed

Similarly to the ferromagnetic case, we derive

\begin{lem}\label{gapestantif}
Assuming that the bound in (\ref{ind-ass}) holds in step $\cI$ of the block-diagonalization, the following holds true
\begin{enumerate}
\item If the number of sites of $\cI^*_{+1}$ is odd
\begin{eqnarray}
\label{first-lem-3.5}
\pm \sum_{\overline{\mathcal{J}^*}\subset \mathcal{I}^*_{{+1}}} P^{(+)}_{\overline{\mathcal{J}^*}}V^{\cI}_{\overline{\mathcal{J}^*}}\, P^{(+)}_{\overline{\mathcal{J}^*}}
\leq  \sum_{\ell(\cJ)=1}^{\ell(\cI)-1}\, \frac{C_{J,h}}{2|J|-h}\cdot (\xi\cdot t)^{\frac{\ell(\cJ)-1}{8}}\,(\ell(\cJ)+1)\,\Big(H^0_{\mathcal{I}^*_{{+1}}}-\langle\,H^0_{\mathcal{I}^*_{{+1}}}\,\rangle_{\Psi^{A}_{\cI^*_{+1}}}\Big)\,,\quad
\end{eqnarray}
and
\begin{equation}
\label{first-lem-3.5b}
P^{(+)}_{\overline{\mathcal{I}^*_{+1}}}\leq \frac{1}{2|J|}\,\Big(H^0_{\overline{\mathcal{I}^*_{+1}}}-\langle\,H^0_{\overline{\mathcal{I}^*_{+1}}}\,\rangle_{\Psi^A_{\overline{\cI^*_{+1}}}}\Big)\,,
\end{equation}
while, concerning  $\Psi^B_{\cI^*_{+1}}$, 
\begin{eqnarray}
\label{second-lem-3.5}
\pm \sum_{\overline{\mathcal{J}^*}\subset \mathcal{I}^*_{{+1}}} P^{(+)}_{\overline{\mathcal{J}^*}}V^{\cI}_{\overline{\mathcal{J}^*}}\, P^{(+)}_{\overline{\mathcal{J}^*}}
\leq   \sum_{\ell(\cJ)=1}^{\ell(\cI)-1}\, \frac{C_{J,h}}{2|J|-h}\cdot (\xi\cdot t)^{\frac{\ell(\cJ)-1}{8}}\,(\ell(\cJ)+1)\,\Big(H^0_{\mathcal{I}^*_{{+1}}}-\langle\,H^0_{\mathcal{I}^*_{{+1}}}\,\rangle_{\Psi^{B}_{\cI^*_{+1}}} + 2h\Big)\,;\quad
\end{eqnarray}

\item if the number of sites of $\cI^*_{+1}$ is even
\begin{eqnarray}
\label{third-lem-3.5}
\pm \sum_{\overline{\mathcal{J}^*}\subset \mathcal{I}^*_{{+1}}} P^{(+)}_{\overline{\mathcal{J}^*}}V^{\cI}_{\overline{\mathcal{J}^*}}\, P^{(+)}_{\overline{\mathcal{J}^*}}
\leq  \sum_{\ell(\cJ)=1}^{\ell(\cI)-1}\, \frac{C_{J,h}}{2|J|-h}\cdot (\xi\cdot t)^{\frac{\ell(\cJ)-1}{8}}\,(\ell(\cJ)+1)\,\Big(H^0_{\mathcal{I}^*_{{+1}}}-\langle\,H^0_{\mathcal{I}^*_{{+1}}}\,\rangle_{\Psi^{A/B}_{\cI^*_{+1}}}+h\Big)\,,\quad 
\end{eqnarray}
and
\begin{equation}\label{third-lem-3.5b}
P^{(+)}_{\overline{\mathcal{I}^*_{+1}}}\leq \frac{1}{2|J|-2h}\,\Big(H^0_{\overline{\mathcal{I}^*_{+1}}}-\langle\,H^0_{\overline{\mathcal{I}^*_{+1}}}\,\rangle_{\Psi^A_{\overline{\cI^*_{+1}}}}\Big)\,.
\end{equation}

\end{enumerate}

\end{lem}

\noindent
\emph{Proof}

From the definitions of $H^0_{\mathcal{I}}$ and $H^C_{\mathcal{I}}$ in (\ref{unper-H0}) and (\ref{unper-HC}), respectively, we easily deduce:

\noindent
if $\mathcal{I}^*_{+1}$ contains an even number of sites
\begin{equation}\label{geqA}
H^C_{\mathcal{I}^*_{+1}} - \langle H^C_{\mathcal{I}^*_{+1}}\rangle_{\Psi^A_{\cI^*_{+1}}} \leq H^0_{\mathcal{I}^*_{+1}} - \langle H^0_{\mathcal{I}^*_{+1}}\rangle_{\Psi^A_{\cI^*_{+1}}}+h\,
\end{equation}
and
\begin{equation}\label{geqB-0}
H^C_{\mathcal{I}^*_{+1}} - \langle H^C_{\mathcal{I}^*_{+1}}\rangle_{\Psi^B_{\cI^*_{+1}}} \leq H^0_{\mathcal{I}^*_{+1}} - \langle H^0_{\mathcal{I}^*_{+1}}\rangle_{\Psi^B_{\cI^*_{+1}}}+h\,;
\end{equation}
if $\mathcal{I}^*_{+1}$ contains an odd number of sites
\begin{equation}\label{geqA-0}
H^C_{\mathcal{I}^*_{+1}} - \langle H^C_{\mathcal{I}^*_{+1}}\rangle_{\Psi^A_{\cI^*_{+1}}} \leq H^0_{\mathcal{I}^*_{+1}} - \langle H^0_{\mathcal{I}^*_{+1}}\rangle_{\Psi^A_{\cI^*_{+1}}}\,
\end{equation}
\begin{equation}\label{geqB}
H^C_{\mathcal{I}^*_{+1}} - \langle H^C_{\mathcal{I}^*_{+1}}\rangle_{\Psi^B_{\cI^*_{+1}}} \leq H^0_{\mathcal{I}^*_{+1}} - \langle H^0_{\mathcal{I}^*_{+1}}\rangle_{\Psi^B_{\cI^*_{+1}}}+2h\,.
\end{equation}
Analogously to (\ref{ferrogapest}), we can show that
\begin{equation}\label{AFHC}
\pm \sum_{\cJ\in(\mathcal{I}^*_{+1})_K} P^{(+)}_{\overline{\mathcal{J}^*}}\, V^{\cI}_{\overline{\mathcal{J}^*}} \,P^{(+)}_{\overline{\mathcal{J}^*}} \leq  \frac{C_{J,h}}{2|J|-h} \cdot (\xi\cdot t)^{\frac{\ell(\cJ)-1}{8}}\,(K+1)\, \Big(H^C_{\mathcal{I}^*_{+1}}-\langle\,H^C_{\mathcal{I}^*_{+1}}\,\rangle_{\Psi^{A/B}_{\mathcal{I}^*_{+1}}}\Big),
\end{equation}
where $(\mathcal{I}^*_{+1})_K$ is defined in item $(ii)$ of Subsection $\ref{ferrogap}$. 

\noindent
Finally, by combining (\ref{geqA})-(\ref{geqB-0}) and (\ref{geqA-0})-(\ref{geqB}) with (\ref{AFHC}), we get the inequalities in (\ref{first-lem-3.5}), (\ref{second-lem-3.5}), and (\ref{third-lem-3.5}).

Depending on the odd/even case the spectral gap changes according to Proposition \ref{propgap}, and consequently the inequalities (\ref{first-lem-3.5b}) and (\ref{third-lem-3.5b}) follow.

\qed

In the lemma below we use all the crucial ingredients: 1) the translation covariance of the model combined with the antiferromagnetic structure of the two groundstates  of the unperturbed local Hamiltonians $H^C_\mathcal{I}$; 2) the block-diagonalization of the effective potentials $V^\mathcal{I}_{\overline{\mathcal{J}^*}}$ with respect to the triples of projections displayed in (a) and (b) of Lemma \ref{Gstructure}; 3) the spectral gap of the unperturbed local Hamiltonians.

\begin{lem}\label{gap-bound}
 Assuming that the bound in (\ref{ind-ass}) holds in step $\cI$ of the block-diagonalization, there exists $\bar{t}>0$ small enough such that $\forall t \leq \bar{t}$
\begin{equation}
P^{(+)}_{\mathcal{I}^*_{{+1}}}\,(G_{{\mathcal{I}^*_{{+1}}}}-E^{A/B}_{{\mathcal{I}^*_{{+1}}}})\,P^{(+)}_{{\mathcal{I}^*_{{+1}}}}
\geq\,(2|J|-2h)\cdot \Big[1- \mathcal{O}(\xi \cdot t ) \Big]\,P^{(+)}_{\mathcal{I}^*_{{+1}}}\,.\label{final-eq-1}
\end{equation}
\end{lem}

\noindent
\emph{Proof}

We recall that thanks to point (a) of Lemma \ref{Gstructure}
\begin{eqnarray}
P^{(+)}_{\mathcal{I}^*_{{+1}}}\,G_{{\mathcal{I}^*_{{+1}}}}\,P^{(+)}_{{\mathcal{I}^*_{{+1}}}} &=& P^{(+)}_{\mathcal{I}^*_{{+1}}}\,(H_{\mathcal{I}^*_{{+1}}}^{0}+{\xi\cdot t }\,\,\sum_{\overline{\mathcal{J}^*}\subset \mathcal{I}^*_{{+1}}} V^{\cI}_{\overline{\mathcal{J}^*}})\, P^{(+)}_{\mathcal{I}^*_{{+1}}}\label{abc}\\
&=&P^{(+)}_{\mathcal{I}^*_{{+1}}}\,(H_{\mathcal{I}^*_{{+1}}}^{0}+{\xi\cdot t }\,\,\sum_{\overline{\mathcal{J}^*}\subset \mathcal{I}^*_{{+1}}} P^{(+)}_{\overline{\mathcal{J}^*}}V^{\cI}_{\overline{\mathcal{J}^*}}\, P^{(+)}_{\overline{\mathcal{J}^*}})\,P^{(+)}_{\mathcal{I}^*_{{+1}}}\label{a}\\
&& + P^{(+)}_{\mathcal{I}^*_{{+1}}} {\xi\cdot t}\,(\,\,\sum_{\overline{\mathcal{J}^*}\subset \mathcal{I}^*_{+1}} \langle\,V^{\cI}_{\overline{\mathcal{J}^*}}\,\rangle_{\Psi^B_{\overline{\mathcal{J}^*}}}\,P^{(-),B}_{\overline{\mathcal{J}^*}})\, P^{(+)}_{\mathcal{I}^*_{+1}} \label{b}\\
&&+ P^{(+)}_{\mathcal{I}^*_{+1}}\xi \cdot t\,(\,\,\sum_{\overline{\mathcal{J}^*}\subset \mathcal{I}^*_{+1}} \langle\,V^{\cI}_{\overline{\mathcal{J}^*}}\,\rangle_{\Psi^A_{\overline{\mathcal{J}^*}}}P^{(-),A}_{\overline{\mathcal{J}^*}})\,P^{(+)}_{\mathcal{I}^*_{+1}}\,.\label{c}
\end{eqnarray}
Using the definitions in (\ref{def-ev}) and (\ref{def-odd}), we can write:
\begin{eqnarray}
& &P^{(+)}_{\mathcal{I}^*_{{+1}}}\,G_{{\mathcal{I}^*_{{+1}}}}\,P^{(+)}_{{\mathcal{I}^*_{{+1}}}}\\
&=&P^{(+)}_{\mathcal{I}^*_{{+1}}}\,(H_{\mathcal{I}^*_{{+1}}}^{0}+{\xi\cdot t}\,\,\sum_{\overline{\mathcal{J}^*}\subset \mathcal{I}^*_{{+1}}} P^{(+)}_{\overline{\mathcal{J}^*}}V^{\cI}_{\overline{\mathcal{J}^*}}\, P^{(+)}_{\overline{\mathcal{J}^*}})\,P^{(+)}_{\mathcal{I}^*_{{+1}}} \\
&& + P^{(+)}_{\mathcal{I}^*_{{+1}}} {\xi\cdot t }\,(\,\,\sum_{\mathcal{J}^*\in(\cI^*_{+1})_{\text{ev}}}  \langle\, V^{\cI}_{\overline{\mathcal{J}^*}}\,\rangle_{\Psi^A_{\overline{\cJ^*}}} \,P^{(-),A}_{\overline{\mathcal{J}^*}})\, P^{(+)}_{\mathcal{I}^*_{{+1}}} + P^{(+)}_{\mathcal{I}^*_{{+1}}}\xi\cdot t\,(\,\,\sum_{\mathcal{J}^*\in(\cI^*_{+1})_{\text{odd}}} \langle\, V^{\cI}_{\overline{\mathcal{J}^*}}\,\rangle_{\Psi^A_{\overline{\cJ^*}}} P^{(-),A}_{\overline{\mathcal{J}^*}})\,P^{(+)}_{\mathcal{I}^*_{{+1}}} \quad\quad\quad \\
& &+ P^{(+)}_{\mathcal{I}^*_{{+1}}} {\xi\cdot t}\,(\,\,\sum_{\mathcal{J}^*\in(\cI^*_{+1})_{\text{odd}}}  \langle\, V^{\cI}_{\overline{\mathcal{J}^*}}\,\rangle_{\Psi^B_{\overline{\cJ^*}}} \,P^{(-),B}_{\overline{\mathcal{J}^*}})\, P^{(+)}_{\mathcal{I}^*_{{+1}}} + P^{(+)}_{\mathcal{I}^*_{{+1}}}\xi\cdot t\,(\,\sum_{\mathcal{J}^*\in(\cI^*_{+1})_{\text{ev}}} \langle\, V^{\cI}_{\overline{\mathcal{J}^*}}\,\rangle_{\Psi^B_{\overline{\cJ^*}}}  P^{(-),B}_{\overline{\mathcal{J}^*}})\,P^{(+)}_{\mathcal{I}^*_{{+1}}}\, .
\end{eqnarray}
Next, for each $\mathcal{J}^*\in(\cI^*_{+1})_{\text{ev}}$ and $\mathcal{J}^*\in(\cI^*_{+1})_{\text{odd}}$, we add and subtract
\begin{equation*}
\langle\, V^{\cI}_{\overline{\mathcal{J}^*}}\,\rangle_{\Psi^A_{\overline{\cJ^*}}}P^{(-),B}_{\overline{\mathcal{J}^*}}\quad,\quad
\langle\, V^{\cI}_{\overline{\mathcal{J}^*}}\,\rangle_{\Psi^B_{\overline{\cJ^*}}}P^{(-),A}_{\overline{\mathcal{J}^*}}\,,
\end{equation*}
respectively.
Thus we obtain
\begin{eqnarray}
& &P^{(+)}_{\mathcal{I}^*_{+1}}\,G_{\mathcal{I}^*_{+1}}\,P^{(+)}_{\mathcal{I}^*_{+1}} \\
&=& P^{(+)}_{\mathcal{I}^*_{{+1}}}\,(H_{\mathcal{I}^*_{{+1}}}^{0}+{\xi\cdot t }\,\,\sum_{\overline{\mathcal{J}^*}\subset \mathcal{I}^*_{{+1}}} P^{(+)}_{\overline{\mathcal{J}^*}}V^{{\cI}}_{\overline{\mathcal{J}^*}}\, P^{(+)}_{\overline{\mathcal{J}^*}})P^{(+)}_{\mathcal{I}^*_{{+1}}}\label{gapfirstline}\\
&& + P^{(+)}_{\mathcal{I}^*_{{+1}}} {\xi\cdot t }\,\Big[\,\,\sum_{\mathcal{J}^*\in(\cI^*_{+1})_{\text{ev}}}  (\langle\, V^{\cI}_{\overline{\mathcal{J}^*}}\,\rangle_{\Psi^B_{\overline{\cJ^*}}} -\langle\, V^{\cI}_{\overline{\mathcal{J}^*}}\,\rangle_{\Psi^A_{\overline{\cJ^*}}}) \,P^{(-),B}_{\overline{\mathcal{J}^*}}\Big]\,P^{(+)}_{\mathcal{I}^*_{{+1}}} \label{diff1}\\
&&+ P^{(+)}_{\mathcal{I}^*_{{+1}}}\xi\cdot t\,(\,\,\sum_{\mathcal{J}^*\in(\cI^*_{+1})_{\text{odd}}} \langle\, V^{\cI}_{\overline{\mathcal{J}^*}}\,\rangle_{\Psi^B_{\overline{\cJ^*}}}P^{(-)}_{\overline{\mathcal{J}^*}})\,P^{(+)}_{\mathcal{I}^*_{{+1}}}\label{energy1}\\
&& + P^{(+)}_{\mathcal{I}^*_{{+1}}} {\xi\cdot t }\,\Big[\,\,\sum_{\mathcal{J}^*\in(\cI^*_{+1})_{\text{odd}}}  (\langle\, V^{\cI}_{\overline{\mathcal{J}^*}}\,\rangle_{\Psi^A_{\overline{\cJ^*}}}-\langle\, V^{\cI}_{\overline{\mathcal{J}^*}}\,\rangle_{\Psi^B_{\overline{\cJ^*}}}) \,P^{(-),A}_{\overline{\mathcal{J}^*}}\Big]\, P^{(+)}_{\mathcal{I}^*_{{+1}}}\label{diff2}\\
&&+ P^{(+)}_{\mathcal{I}^*_{{+1}}}\xi\cdot t\,(\,\,\sum_{\mathcal{J}^*\in(\cI^*_{+1})_{\text{ev}}} \langle\, V^{\cI}_{\overline{\mathcal{J}^*}}\,\rangle_{\Psi^A_{\overline{\cJ^*}}}  P^{(-)}_{\overline{\mathcal{J}^*}})\,P^{(+)}_{\mathcal{I}^*_{{+1}}}\,,\label{energy2}
\end{eqnarray}
where we have used $$\langle\, V^{\cI}_{\overline{\mathcal{J}^*}}\,\rangle_{\Psi^{A/B}_{\overline{\cJ^*}}}\Big\{P^{(-),B}_{\overline{\mathcal{J}^*}}+P^{(-),A}_{\overline{\mathcal{J}^*}}\Big\}=\langle\, V^{\cI}_{\overline{\mathcal{J}^*}}\,\rangle_{\Psi^{A/B}_{\overline{\cJ^*}}}\,P^{(-)}_{\overline{\mathcal{J}^*}}\,.$$
The rest of the proof is separated into two parts, namely the study of $(\ref{diff1})+(\ref{diff2})$ and of $(\ref{energy1})+(\ref{energy2})$, respectively, and in a conclusion where we collect our partial estimates and finally prove the result stated in (\ref{final-eq-1}).
\\

\noindent
\underline{\emph{Study of the terms $(\ref{diff1})+(\ref{diff2})$}}
\\

We intend to show some cancellations in the expression above. For this purpose we observe that if $\mathcal{J}^*\in(\cI^*_{+1})_{\text{odd}}'\subset (\cI^*_{+1})_{\text{odd}}$ (see the definition of $(\cI^*_{+1})_{\text{odd}}'$ in (\ref{odd-prime})) then
\begin{itemize}
\item[(i).a]  $$P^{(-),A}_{\overline{\mathcal{J}^*}}P^{(-),A}_{\overline{\mathcal{J}^*}\cup \tau_1(\overline{\mathcal{J}^*})}=P^{(-),A}_{\overline{\mathcal{J}^*}\cup \tau_1(\overline{\mathcal{J}^*})}\,,$$
\item[(ii).a]   $$P^{(-),A}_{\overline{\mathcal{J}^*}}P^{(-),B}_{\overline{\mathcal{J}^*}\cup \tau_1(\overline{\mathcal{J}^*})}=0\,.$$
\end{itemize}
Analogously, for $\mathcal{J}^*\in(\cI^*_{+1})_{\text{ev}}'\subset (\cI^*_{+1})_{\text{ev}}$, the following hold
\begin{itemize}
\item[(i).b]  $$P^{(-),B}_{\overline{\mathcal{J}^*}}P^{(-),B}_{\overline{\mathcal{J}^*}\cup \tau_{-1}(\overline{\mathcal{J}^*})}=0\,,$$
\item[(ii).b]   $$P^{(-),B}_{\overline{\mathcal{J}^*}}P^{(-),A}_{\overline{\mathcal{J}^*}\cup \tau_{-1}(\overline{\mathcal{J}^*})}=P^{(-),A}_{\overline{\mathcal{J}^*}\cup \tau_{-1}(\overline{\mathcal{J}^*})}\,.$$
\end{itemize}
%
Therefore, regarding each summand in (\ref{diff2}) associated with an interval  $\mathcal{J}^*\in(\cI^*_{+1})_{\text{odd}}'$, we decompose the identity into
\begin{equation}\label{ident-1}
\charf = P^{(-),A}_{\overline{\mathcal{J}^*}\cup \tau_1(\overline{\mathcal{J}^*})} + P^{(-),B}_{\overline{\mathcal{J}^*}\cup \tau_1(\overline{\mathcal{J}^*})}+ P^{(+)}_{\overline{\mathcal{J}^*}\cup \tau_1(\overline{\mathcal{J}^*})}\,;
\end{equation}
analogously, for an interval  $\mathcal{J}^*\in(\cI^*_{+1})_{\text{ev}}'$ in (\ref{diff1}) we use
\begin{equation}\label{ident-2}
\charf = P^{(-),A}_{\overline{\mathcal{J}^*}\cup \tau_{-1}(\overline{\mathcal{J}^*})} + P^{(-),B}_{\overline{\mathcal{J}^*}\cup \tau_{-1}(\overline{\mathcal{J}^*})}+ P^{(+)}_{\overline{\mathcal{J}^*}\cup \tau_{-1}(\overline{\mathcal{J}^*})}\,.
\end{equation}
Then we split $(\cI^*_{+1})_{\text{ev/odd}}$ into $(\cI^*_{+1})'_{\text{ev/odd}}\cup [(\cI^*_{+1})_{\text{ev/odd}}\setminus (\cI^*_{+1})'_{\text{ev/odd}}]$, and we substitute (\ref{ident-2}) into (\ref{diff1}) and (\ref{ident-1}) into (\ref{diff2}). 
For each $\cJ \in (\cI^*_{+1})'_{\text{ev}}$, the corresponding term in (\ref{diff1}) is
\begin{eqnarray}
&&P^{(+)}_{\mathcal{I}^*_{{+1}}}\Big(\langle\, V^{\cI}_{\overline{\mathcal{J}^*}}\,\rangle_{\Psi^B_{\overline{\cJ^*}}} -\langle\, V^{\cI}_{\overline{\mathcal{J}^*}}\,\rangle_{\Psi^A_{\overline{\cJ^*}}}\Big) \,P^{(-),B}_{\overline{\mathcal{J}^*}}\,P^{(+)}_{\mathcal{I}^*_{{+1}}}\\
&=&P^{(+)}_{\mathcal{I}^*_{{+1}}}\Big(\langle\, V^{\cI}_{\overline{\mathcal{J}^*}}\,\rangle_{\Psi^B_{\overline{\cJ^*}}} -\langle\, V^{\cI}_{\overline{\mathcal{J}^*}}\,\rangle_{\Psi^A_{\overline{\cJ^*}}}\Big) \,P^{(-),B}_{\overline{\mathcal{J}^*}}\charf P^{(+)}_{\mathcal{I}^*_{{+1}}}\\
&=& P^{(+)}_{\mathcal{I}^*_{{+1}}}\Big(\langle\, V^{\cI}_{\overline{\mathcal{J}^*}}\,\rangle_{\Psi^B_{\overline{\cJ^*}}} -\langle\, V^{\cI}_{\overline{\mathcal{J}^*}}\,\rangle_{\Psi^A_{\overline{\cJ^*}}}\Big) \,P^{(-),B}_{\overline{\mathcal{J}^*}} P^{(-),A}_{\overline{\mathcal{J}^*}\cup \tau_{-1}(\overline{\mathcal{J}^*})}P^{(+)}_{\mathcal{I}^*_{{+1}}}\label{sempl1} \\
&&+P^{(+)}_{\mathcal{I}^*_{{+1}}}\Big(\langle\, V^{\cI}_{\overline{\mathcal{J}^*}}\,\rangle_{\Psi^B_{\overline{\cJ^*}}} -\langle\, V^{\cI}_{\overline{\mathcal{J}^*}}\,\rangle_{\Psi^A_{\overline{\cJ^*}}}\Big) \,P^{(-),B}_{\overline{\mathcal{J}^*}}P^{(-),B}_{\overline{\mathcal{J}^*}\cup \tau_{-1}(\overline{\mathcal{J}^*})}P^{(+)}_{\mathcal{I}^*_{{+1}}}\label{sempl2}\\
&&+P^{(+)}_{\mathcal{I}^*_{{+1}}}\Big(\langle\, V^{\cI}_{\overline{\mathcal{J}^*}}\,\rangle_{\Psi^B_{\overline{\cJ^*}}} -\langle\, V^{\cI}_{\overline{\mathcal{J}^*}}\,\rangle_{\Psi^A_{\overline{\cJ^*}}}\Big) \,P^{(-),B}_{\overline{\mathcal{J}^*}} P^{(+)}_{\overline{\mathcal{J}^*}\cup \tau_{-1}(\overline{\mathcal{J}^*})}P^{(+)}_{\mathcal{I}^*_{{+1}}}\,.
\end{eqnarray}
Using property (ii).b above, (\ref{sempl1}) is equal to $$P^{(+)}_{\mathcal{I}^*_{{+1}}} \Big(\langle\, V^{\cI}_{\overline{\mathcal{J}^*}}\,\rangle_{\Psi^B_{\overline{\cJ^*}}} -\langle\, V^{\cI}_{\overline{\mathcal{J}^*}}\,\rangle_{\Psi^A_{\overline{\cJ^*}}}\Big) \,P^{(-),A}_{\overline{\mathcal{J}^*}\cup \tau_{-1}(\overline{\cJ^*})}P^{(+)}_{\mathcal{I}^*_{{+1}}}\,,$$
while, using property (i).b above, (\ref{sempl2}) is equal to $0$. Thus, we get
\begin{eqnarray}
& &P^{(+)}_{\mathcal{I}^*_{{+1}}}\Big(\langle\, V^{\cI}_{\overline{\mathcal{J}^*}}\,\rangle_{\Psi^B_{\overline{\cJ^*}}} -\langle\, V^{\cI}_{\overline{\mathcal{J}^*}}\,\rangle_{\Psi^A_{\overline{\cJ^*}}}\Big) \,P^{(-),B}_{\overline{\mathcal{J}^*}}P^{(+)}_{\mathcal{I}^*_{{+1}}}\\
&=& P^{(+)}_{\mathcal{I}^*_{{+1}}} \Big(\langle\, V^{\cI}_{\overline{\mathcal{J}^*}}\,\rangle_{\Psi^B_{\overline{\cJ^*}}} -\langle\, V^{\cI}_{\overline{\mathcal{J}^*}}\,\rangle_{\Psi^A_{\overline{\cJ^*}}}\Big) \,P^{(-),A}_{\overline{\mathcal{J}^*}\cup \tau_{-1}(\overline{\cJ^*})}P^{(+)}_{\mathcal{I}^*_{{+1}}}\\
\nonumber
&&+P^{(+)}_{\mathcal{I}^*_{{+1}}}\Big( \langle\, V^{\cI}_{\overline{\mathcal{J}^*}}\,\rangle_{\Psi^B_{\overline{\cJ^*}}} -\langle\, V^{\cI}_{\overline{\mathcal{J}^*}}\,\rangle_{\Psi^A_{\overline{\cJ^*}}}\Big)P^{(-),B}_{\overline{\mathcal{J}^*}} P^{(+)}_{\overline{\mathcal{J}^*}\cup \tau_{-1}(\overline{\mathcal{J}^*})}P^{(+)}_{\mathcal{I}^*_{{+1}}}.
\end{eqnarray}

Analogously, from the properties in (i).a and (ii).a above, we deduce that, for each $\cJ \in (\cI^*_{+1})'_{\text{odd}}$, the corresponding term in (\ref{diff2}) is
\begin{eqnarray}
\nonumber
& &P^{(+)}_{\mathcal{I}^*_{{+1}}}\Big(\langle\, V^{\cI}_{\overline{\mathcal{J}^*}}\,\rangle_{\Psi^A_{\overline{\cJ^*}}} -\langle\, V^{\cI}_{\overline{\mathcal{J}^*}}\,\rangle_{\Psi^B_{\overline{\cJ^*}}}\Big) \,P^{(-),A}_{\overline{\mathcal{J}^*}}P^{(+)}_{\mathcal{I}^*_{{+1}}}\\
&=&P^{(+)}_{\mathcal{I}^*_{{+1}}}  \Big(\langle\, V^{\cI}_{\overline{\mathcal{J}^*}}\,\rangle_{\Psi^A_{\overline{\cJ^*}}} -\langle\, V^{\cI}_{\overline{\mathcal{J}^*}}\,\rangle_{\Psi^B_{\overline{\cJ^*}}}\Big) \,P^{(-),A}_{\overline{\mathcal{J}^*}\cup \tau_{1}(\overline{\cJ^*})}P^{(+)}_{\mathcal{I}^*_{{+1}}} \\
\nonumber
&&+P^{(+)}_{\mathcal{I}^*_{{+1}}}\Big( \langle\, V^{\cI}_{\overline{\mathcal{J}^*}}\,\rangle_{\Psi^A_{\overline{\cJ^*}}} -\langle\, V^{\cI}_{\overline{\mathcal{J}^*}}\,\rangle_{\Psi^B_{\overline{\cJ^*}}}\Big)P^{(-),A}_{\overline{\mathcal{J}^*}} P^{(+)}_{\overline{\mathcal{J}^*}\cup \tau_{1}(\overline{\mathcal{J}^*})}P^{(+)}_{\mathcal{I}^*_{{+1}}}.
\end{eqnarray}

For the terms corresponding to intervals in $(\cI^*_{+1})_{\text{ev/odd}}\setminus (\cI^*_{+1})'_{\text{ev/odd}}$ we do nothing. Thus, we have
\begin{eqnarray}
\nonumber
&&(\ref{diff1})+(\ref{diff2})\\
&=& P^{(+)}_{\mathcal{I}^*_{{+1}}}\xi\cdot t\,\Big[\,\sum_{\mathcal{J}^*\in(\cI^*_{+1})_{\text{odd}}'}\Big( \langle\, V^{\cI}_{\overline{\mathcal{J}^*}}\,\rangle_{\Psi^A_{\overline{\cJ^*}}} -\langle\, V^{\cI}_{\overline{\mathcal{J}^*}}\,\rangle_{\Psi^B_{\overline{\cJ^*}}}\Big)P^{(-),A}_{\overline{\mathcal{J}^*}\cup \tau_1(\overline{\mathcal{J}^*})}\Big]P^{(+)}_{\mathcal{I}^*_{{+1}}} \label{diff1+}\\
&&+P^{(+)}_{\mathcal{I}^*_{{+1}}}\xi\cdot t\,\Big[\,\sum_{\mathcal{J}^*\in(\cI^*_{+1})_{\text{odd}}'}\Big( \langle\, V^{\cI}_{\overline{\mathcal{J}^*}}\,\rangle_{\Psi^A_{\overline{\cJ^*}}} -\langle\, V^{\cI}_{\overline{\mathcal{J}^*}}\,\rangle_{\Psi^B_{\overline{\cJ^*}}}\Big)P^{(-),A}_{\overline{\mathcal{J}^*}}P^{(+)}_{\overline{\mathcal{J}^*}\cup \tau_1(\overline{\mathcal{J}^*})}\Big]P^{(+)}_{\mathcal{I}^*_{{+1}}}\label{diff2+}\\
&&+P^{(+)}_{\mathcal{I}^*_{{+1}}}\xi\cdot t\,\Big[\,\sum_{\mathcal{J}^*\in(\cI^*_{+1})_{\text{ev}}'}\Big( \langle\, V^{\cI}_{\overline{\mathcal{J}^*}}\,\rangle_{\Psi^B_{\overline{\cJ^*}}} -\langle\, V^{\cI}_{\overline{\mathcal{J}^*}}\,\rangle_{\Psi^A_{\overline{\cJ^*}}}\Big)P^{(-),A}_{\overline{\mathcal{J}^*}\cup \tau_{-1}(\overline{\mathcal{J}^*})}\Big]P^{(+)}_{\mathcal{I}^*_{{+1}}}\label{diff1-} \\
&&+P^{(+)}_{\mathcal{I}^*_{{+1}}}\xi\cdot t\,\Big[\,\sum_{\mathcal{J}^*\in(\cI^*_{+1})_{\text{ev}}'}( \langle\, V^{\cI}_{\overline{\mathcal{J}^*}}\,\rangle_{\Psi^B_{\overline{\cJ^*}}} -\langle\, V^{\cI}_{\overline{\mathcal{J}^*}}\,\rangle_{\Psi^A_{\overline{\cJ^*}}})P^{(-),B}_{\overline{\mathcal{J}^*}} P^{(+)}_{\overline{\mathcal{J}^*}\cup \tau_{-1}(\overline{\mathcal{J}^*})}\Big]P^{(+)}_{\mathcal{I}^*_{{+1}}}\label{diff2-}\\
&& + P^{(+)}_{\mathcal{I}^*_{{+1}}} {\xi\cdot t }\,\Big[\,\sum_{\mathcal{J}^*\in(\cI^*_{+1})_{\text{ev}}\setminus (\cI^*_{+1})'_{\text{ev}} } \Big(\langle\, V^{\cI}_{\overline{\mathcal{J}^*}}\,\rangle_{\Psi^B_{\overline{\cJ^*}}} -\langle\, V^{\cI}_{\overline{\mathcal{J}^*}}\,\rangle_{\Psi^A_{\overline{\cJ^*}}}\Big) \,P^{(-),B}_{\overline{\mathcal{J}^*}}\Big] P^{(+)}_{\mathcal{I}^*_{{+1}}}\label{rem1}\\
&&+ P^{(+)}_{\mathcal{I}^*_{{+1}}} {\xi\cdot t }\,\Big[\,\sum_{\mathcal{J}^*\in(\cI^*_{+1})_{\text{odd}}\setminus(\cI^*_{+1})'_{\text{odd}} }  \Big(\langle\, V^{\cI}_{\overline{\mathcal{J}^*}}\,\rangle_{\Psi^A_{\overline{\cJ^*}}}-\langle\, V^{\cI}_{\overline{\mathcal{J}^*}}\,\rangle_{\Psi^B_{\overline{\cJ^*}}}\Big) \,P^{(-),A}_{\overline{\mathcal{J}^*}}\Big] P^{(+)}_{\mathcal{I}^*_{{+1}}}\label{rem2} \,.
\end{eqnarray}
We proceed our study of $(\ref{diff1})+(\ref{diff2})$ by combining different terms in (\ref{diff1+})-(\ref{rem2}). We start with the estimate of the following terms
\begin{itemize}
\item $(\ref{diff1+})+(\ref{diff1-})$. 

\noindent
From (\ref{translints}) we have that for each $\cJ^* \in (\cI^*_{+1})'_{\text{odd}}$, the projection $P^{(-),A}_{\overline{\mathcal{J}^*}\cup \tau_1(\overline{\mathcal{J}^*})}$ is equal to the projection $P^{(-),A}_{\overline{\mathcal{K}^*}\cup \tau_{-1}(\overline{\mathcal{K}^*})}$ where $\cK=\tau_1(\cJ)\in (\cI^*_{+1})'_{\text{even}}$. Thus each term in (\ref{diff1+}) is paired with a term in (\ref{diff1-}); their combination gives
\begin{eqnarray}
\Big( \langle\, V^{\cI}_{\overline{\mathcal{J}^*}}\,\rangle_{\Psi^A_{\overline{\cJ^*}}} -\langle\, V^{\cI}_{\overline{\mathcal{J}^*}}\,\rangle_{\Psi^B_{\overline{\cJ^*}}} + \langle\, V^{\cI}_{\tau_1(\overline{\mathcal{J}^*})}\,\rangle_{\Psi^B_{\tau_1(\overline{\cJ^*})}} -\langle\, V^{\cI}_{\tau_1(\overline{\mathcal{J}^*})}\,\rangle_{\Psi^A_{\tau_1(\overline{\cJ^*})}}\Big)P^{(-),A}_{\overline{\mathcal{J}^*}\cup \tau_1(\overline{\mathcal{J}^*})}
\end{eqnarray}

 Now we recall that (see an analogous identity in (\ref{id-cov-1})-(\ref{id-cov-2}))
\begin{equation}
 \langle\, V^{\cI}_{\overline{\mathcal{J}^*}}\,\rangle_{\Psi^B_{\overline{\mathcal{J}^*}}}= \langle\, V^{\cI}_{\tau_1(\overline{\mathcal{J}^*})}\,\rangle_{\Psi^B_{\tau_{1}(\overline{\mathcal{J}^*})}}, \quad\quad  \langle V^{\cI}_{\overline{\mathcal{J}^*}}\,\rangle_{\Psi^A_{\overline{\mathcal{J}^*}}}= \langle\, V^{\cI}_{\tau_1(\overline{\mathcal{J}^*})}\,\rangle_{\Psi^A_{\tau_{1}(\overline{\mathcal{J}^*})}},
\end{equation}  
which follow from the translation covariance stated in Proposition \ref{trans-cov}. Thus we conclude that 
\begin{equation}
(\ref{diff1+})+(\ref{diff1-})=0\,.
\end{equation}

\item $(\ref{rem1})+(\ref{rem2})$. 

\noindent
For each $K\in\mathbb{N}$, there is at most one $\mathcal{J}\in\mathfrak{I}$ with $\ell(\cJ)=K$ and $\cJ^*\in(\cI^*_{+1})_{\text{odd}}\setminus(\cI^*_{+1})'_{\text{odd}}$; analogously, there is at most one $\cK \in \mathfrak{I}$ with $\ell(\cK)=K$ such that $\cK^*\in(\cI^*_{+1})_{\text{ev}}\setminus(\cI^*_{+1})'_{\text{ev}}$.
Thanks to (\ref{ind-ass}), this implies 
\begin{equation}
\pm ((\ref{rem1})+({\ref{rem2}})) \leq 4\,\xi\cdot t\,\cdot  \sum\limits_{K=1}^{\ell(\cI)-1} C_{J,h}\cdot \,\frac{(\xi\cdot t)^{\frac{K-1}{8}}}{K^2} P^{(+)}_{\cI^*_{+1}}\,.
\end{equation}

\item $(\ref{diff2+})+(\ref{diff2-})$. 

\noindent
We use Lemma \ref{gapestantif} and estimate (the factor $4$ in (\ref{P+term-1}) and (\ref{P+term-2}) comes from applying the estimates in (\ref{first-lem-3.5}) (odd case) and in (\ref{third-lem-3.5}) (even case) to each of the two  terms contained in (\ref{diff2+}) and (\ref{diff2-})):
\begin{itemize} 
\item if $\cI^*$ has an odd number of sites,
\begin{eqnarray}
&&\pm ( (\ref{diff2+})+(\ref{diff2-}))\nonumber\\
&\leq & P^{(+)}_{\mathcal{I}^*_{{+1}}}\xi\cdot t\,\Big[\,\sum_{K=2}^{\ell(\cI)-1}\,\sum_{\mathcal{K}\in(\cI^*_{+1})_{K}}( 4\cdot C_{J,h}\cdot \frac{(\xi\cdot t)^{\frac{K-2}{8}}}{(K-1)^2})P^{(+)}_{\overline{\mathcal{K}^*}}\Big]P^{(+)}_{\mathcal{I}^*_{{+1}}}\nonumber \\
&\leq& P^{(+)}_{\mathcal{I}^*_{{+1}}}\xi\cdot t\,\Big[\,\sum_{K=2}^{\ell(\cI)-1}\,(4\cdot C_{J,h}\cdot  \frac{(\xi\cdot t)^{\frac{K-2}{8}}}{(K-1)^2})(K+1)\frac{1}{2|J|-h}(H^0_{\cI^*_{+1}}-\langle H^0_{\cI^*_{+1}}\rangle_{\Psi^A_{\cI^*_{+1}}})\Big]P^{(+)}_{\mathcal{I}^*_{{+1}}}\,;\quad\quad\quad\quad \label{P+term-1}
\end{eqnarray}

\item if $\cI^*$ has an even number of sites,
\begin{eqnarray}
& &\pm ( (\ref{diff2+})+(\ref{diff2-}))\nonumber\\
&\leq &P^{(+)}_{\mathcal{I}^*_{{+1}}}\xi\cdot t\,\Big[\,\sum_{K=2}^{\ell(\cI)-1}\,\sum_{\mathcal{K}\in(\cI^*_{+1})_{K}}( 4\cdot C_{J,h}\cdot \frac{(\xi\cdot t)^{\frac{K-2}{8}}}{(K-1)^2})P^{(+)}_{\overline{\mathcal{K}^*}}\Big]P^{(+)}_{\mathcal{I}^*_{{+1}}}\nonumber \\
&\leq& P^{(+)}_{\mathcal{I}^*_{{+1}}}\xi\cdot t\,\Big[\,\sum_{K=2}^{\ell(\cI)-1}\,(4\cdot C_{J,h}\cdot  \frac{(\xi\cdot t)^{\frac{K-2}{8}}}{(K-1)^2})(K+1)\frac{1}{2|J|-h}(H^0_{\cI^*_{+1}}-\langle H^0_{\cI^*_{+1}}\rangle_{\Psi^A_{\cI^*_{+1}}}+h)\Big]P^{(+)}_{\mathcal{I}^*_{{+1}}}\,.\quad\quad\quad\quad  \label{P+term-2}
\end{eqnarray}
\end{itemize}
where $(\mathcal{I}^*_{+1})_K$ is defined in item $(ii)$ of Subsection $\ref{ferrogap}$.

\end{itemize}

\noindent
\underline{\emph{Study of the terms $(\ref{gapfirstline})+(\ref{energy1})+(\ref{energy2})$}}
\\

By using $P^{(-)}_{\overline{\mathcal{J}^*}}= \charf - P^{(+)}_{\overline{\mathcal{J}^*}}$, we can write 
\begin{eqnarray}
&&(\ref{energy1})+(\ref{energy2})\\
&=& P^{(+)}_{\mathcal{I}^*_{{+1}}}\xi\cdot t\,\Big[\,\sum_{\mathcal{J}^*\in(\cI^*_{+1})_{\text{ev}}} \langle\, V^{\cI}_{\overline{\mathcal{J}^*}}\,\rangle_{\Psi^A_{\overline{\cJ^*}}}  +\sum_{\mathcal{J}^*\in(\cI^*_{+1})_{\text{odd}}} \langle\, V^{\cI}_{\overline{\mathcal{J}^*}}\,\rangle_{\Psi^B_{\overline{\cJ^*}}}\Big]P^{(+)}_{\mathcal{I}^*_{{+1}}}\label{energytot}\\
&&-  P^{(+)}_{\mathcal{I}^*_{{+1}}}\xi\cdot t\,\Big[\,\sum_{\mathcal{J}^*\in(\cI^*_{+1})_{\text{ev}}} \langle\, V^{\cI}_{\overline{\mathcal{J}^*}}\,\rangle_{\Psi^A_{\overline{\cJ^*}}}P^{(+)}_{\overline{\mathcal{J}^*}}  +\sum_{\mathcal{J}^*\in(\cI^*_{+1})_{\text{odd}}} \langle\, V^{\cI}_{\overline{\mathcal{J}^*}}\,\rangle_{\Psi^B_{\overline{\cJ^*}}}P^{(+)}_{\overline{\mathcal{J}^*}}\Big]P^{(+)}_{\mathcal{I}^*_{{+1}}}\,.\label{compl-en}
\end{eqnarray}
We recall that
\begin{itemize}
\item[(i)] for $\mathcal{J}^*\in(\cI^*_{+1})_{\text{ev}}$\,,\quad $\langle\,V^{\cI}_{\overline{\mathcal{J}^*}}\,\rangle_{\Psi^A_{\overline{\mathcal{J}^*}}}=\langle\,V^{\cI}_{\overline{\mathcal{J}^*}} \, \rangle_{\Psi^A_{\cI^*_{+1}}}$ and $\langle\,V^{\cI}_{\overline{\mathcal{J}^*}}\,\rangle_{\Psi^B_{\overline{\mathcal{J}^*}}}=\langle\,V^{\cI}_{\overline{\mathcal{J}^*}} \, \rangle_{\Psi^B_{\cI^*_{+1}}},$
\item[(ii)] for $\mathcal{J}^*\in(\cI^*_{+1})_{\text{odd}}$\,,\quad $\langle\,V^{\cI}_{\overline{\mathcal{J}^*}}\,\rangle_{\Psi^A_{\overline{\mathcal{J}^*}}}=\langle\,V^{\cI}_{\overline{\mathcal{J}^*}} \, \rangle_{\Psi^B_{\cI^*_{+1}}}$ and  $\langle\,V^{\cI}_{\overline{\mathcal{J}^*}}\,\rangle_{\Psi^B_{\overline{\mathcal{J}^*}}}=\langle\,V^{\cI}_{\overline{\mathcal{J}^*}} \, \rangle_{\Psi^A_{\cI^*_{+1}}}$\,.
\end{itemize}
Consequently, we get
\begin{equation}
(\ref{energytot})= (E^A_{\mathcal{I}^*_{{+1}}} - \langle\,H^0_{\mathcal{I}^*_{{+1}}}\,\rangle_{\Psi^A_{\cI^*_{+1}}})P^{(+)}_{\mathcal{I}^*_{{+1}}}\,.
\end{equation}
Hence, we have 
\begin{eqnarray}
& &(\ref{gapfirstline})+(\ref{energy1})+(\ref{energy2})\\
&=&P^{(+)}_{\mathcal{I}^*_{{+1}}}(H^0_{\mathcal{I}^*_{{+1}}} +E^A_{\mathcal{I}^*_{{+1}}} - \langle\,H^0_{\mathcal{I}^*_{{+1}}}\,\rangle_{\Psi^A_{\cI^*_{+1}}} ) P^{(+)}_{\mathcal{I}^*_{{+1}}}\\
& &+P^{(+)}_{\mathcal{I}^*_{{+1}}}\,(\xi\cdot t \,\,\sum_{\overline{\mathcal{J}^*}\subset \mathcal{I}^*_{{+1}}} P^{(+)}_{\overline{\mathcal{J}^*}}V^{{\cI}}_{\overline{\mathcal{J}^*}}\, P^{(+)}_{\overline{\mathcal{J}^*}})P^{(+)}_{\mathcal{I}^*_{{+1}}}\label{compl-en-0}\\
& & -  P^{(+)}_{\mathcal{I}^*_{{+1}}}\xi\cdot t\,\Big[\,\sum_{\mathcal{J}^*\in(\cI^*_{+1})_{\text{ev}}} \langle\, V^{\cI}_{\overline{\mathcal{J}^*}}\,\rangle_{\Psi^A_{\overline{\cJ^*}}}P^{(+)}_{\overline{\mathcal{J}^*}}  +\sum_{\mathcal{J}^*\in(\cI^*_{+1})_{\text{odd}}} \langle\, V^{\cI}_{\overline{\mathcal{J}^*}}\,\rangle_{\Psi^B_{\overline{\cJ^*}}}P^{(+)}_{\overline{\mathcal{J}^*}}\Big]P^{(+)}_{\mathcal{I}^*_{{+1}}}\,.\label{compl-en}
\end{eqnarray}
Concerning the terms in (\ref{compl-en-0})-(\ref{compl-en}), we must distinguish between odd/even number of sites so as to apply Lemma \ref{gapestantif} similarly to what we have done for (\ref{diff2+})+(\ref{diff2-}).\\

\noindent
\underline{\emph{Conclusion}}
\\

We now collect our estimates obtained so far and, depending on the parity of the number of sites of $\mathcal{I}^*_{{+1}}$, we show how to derive the inequalities in the statement. In the following $C'_{J,h}$ stands for a quantity which depends on $J$ and $h$, and may vary line by line.

For an odd number of sites, we use point 1 in Lemma \ref{gapestantif}, to get
\begin{eqnarray}
\nonumber
& &P^{(+)}_{\mathcal{I}^*_{{+1}}}\,(G_{{\mathcal{I}^*_{{+1}}}}- E^A_{\mathcal{I}^*_{{+1}}}) \, P^{(+)}_{\mathcal{I}^*_{{+1}}}\\
&\geq&  \Big(1- \xi\cdot t\sum_{\ell(\cJ)=1}^{\ell(\cI)-1}\, \frac{C_{J,h}'}{2|J|-h}\cdot (\xi\cdot t)^{\frac{\ell(\cJ)-1}{8}}\,(\ell(\cJ)+1)\,\Big)\,(H^0_{\mathcal{I}^*_{{+1}}}-\langle\,H^0_{\mathcal{I}^*_{{+1}}}\,\rangle_{\Psi^A_{\cI^*_{+1}}} )P^{(+)}_{\mathcal{I}^*_{{+1}}}\label{3-in}\\
& &- \,\xi\cdot t\,\cdot  \sum\limits_{\ell(\cJ)=1}^{\ell(\cI)-1} C_{J,h}'\,\frac{(\xi\cdot t)^{\frac{\ell(\cJ)-1}{8}}}{(\ell(\cJ))^2} P^{(+)}_{\cI^*_{+1}}\,\quad\quad\quad\quad \nonumber\\
&\geq&  \Big(1- \xi\cdot t\sum_{\ell(\cJ)=1}^{\ell(\cI)-1}\, \frac{C_{J,h}'}{2|J|-h}\cdot (\xi\cdot t)^{\frac{\ell(\cJ)-1}{8}}\,(\ell(\cJ)+1)\,\Big)\,2|J|P^{(+)}_{\mathcal{I}^*_{{+1}}}\label{3-fin}\\
& &- \,\xi\cdot t\,\cdot  \sum\limits_{\ell(\cJ)=1}^{\ell(\cI)-1} C_{J,h}'\,\frac{(\xi\cdot t)^{\frac{\ell(\cJ)-1}{8}}}{(\ell(\cJ))^2} P^{(+)}_{\cI^*_{+1}}\,\quad\quad\quad\quad \nonumber
\end{eqnarray}
where we have used (\ref{first-lem-3.5b}) to estimate the term proportional to $H^0_{\mathcal{I}^*_{{+1}}}-\langle\,H^0_{\mathcal{I}^*_{{+1}}}\rangle_{\Psi^A_{\cI^*_{+1}}}$ in (\ref{3-in}).

For an even number of sites, we use point \emph{2}. in Lemma \ref{gapestantif}, to get
\begin{eqnarray}
\nonumber
& &P^{(+)}_{\mathcal{I}^*_{{+1}}}\,(G_{{\mathcal{I}^*_{{+1}}}}- E^A_{\mathcal{I}^*_{{+1}}}) \, P^{(+)}_{\mathcal{I}^*_{{+1}}}\\
&\geq&  \Big[1-\xi\cdot t\sum_{\ell(\cJ)=1}^{\ell(\cI)-1}\, \frac{C_{J,h}'}{2|J|-h}\cdot (\xi\cdot t)^{\frac{\ell(\cJ)-1}{8}}\,(\ell(\cJ)+1)\Big](H^0_{\mathcal{I}^*_{{+1}}}-\langle\,H^0_{\mathcal{I}^*_{{+1}}}\,\rangle_{\Psi^A_{\cI^*_{+1}}} )P^{(+)}_{\mathcal{I}^*_{{+1}}} \label{3-bis}\\
\nonumber
&&- \,\xi\cdot t\,\cdot  \sum\limits_{\ell(\cJ)=1}^{\ell(\cI)-1} C_{J,h}' \cdot\frac{(\xi\cdot t)^{\frac{\ell(\cJ)-1}{8}}}{(\ell(\cJ))^2} P^{(+)}_{\cI^*_{+1}}
\nonumber\\
&&+ \, h \, \xi\cdot t\cdot \Big[ - \sum_{\ell(\cJ)=1}^{\ell(\cI)-1}\, \frac{C_{J,h}'}{2|J|-h}\cdot (\xi\cdot t)^{\frac{\ell(\cJ)-1}{8}}\,(\ell(\cJ)+1)\Big]P^{(+)}_{\mathcal{I}^*_{{+1}}}\,\nonumber \\
&\geq&  \Big[1-\xi\cdot t\sum_{\ell(\cJ)=1}^{\ell(\cI)-1}\, \frac{C_{J,h}'}{2|J|-h}\cdot (\xi\cdot t)^{\frac{\ell(\cJ)-1}{8}}\,(\ell(\cJ)+1)\Big](2|J|-2 h)P^{(+)}_{\mathcal{I}^*_{{+1}}} \label{3-bis-bis}\\
\nonumber
&&- \,\xi\cdot t\,\cdot  \sum\limits_{\ell(\cJ)=1}^{\ell(\cI)-1} C_{J,h}' \cdot\frac{(\xi\cdot t)^{\frac{\ell(\cJ)-1}{8}}}{(\ell(\cJ))^2} P^{(+)}_{\cI^*_{+1}}
\nonumber\\
&&+  \, h \, \xi\cdot t\cdot \Big[ - \sum_{\ell(\cJ)=1}^{\ell(\cI)-1}\, \frac{C_{J,h}'}{2|J|-h}\cdot (\xi\cdot t)^{\frac{\ell(\cJ)-1}{8}}\,(\ell(\cJ)+1)\Big]P^{(+)}_{\mathcal{I}^*_{{+1}}}\,\nonumber
\end{eqnarray}
where we have used (\ref{third-lem-3.5b}) to estimate the term proportional to $H^0_{\mathcal{I}^*_{{+1}}}-\langle\,H^0_{\mathcal{I}^*_{{+1}}}\rangle_{\Psi^A_{\cI^*_{+1}}}$ in (\ref{3-bis}).

\noindent
From the inequalities in (\ref{3-fin}) and (\ref{3-bis-bis}) we easily deduce the statement in (\ref{final-eq-1}) regarding $G_{{\mathcal{I}^*_{{+1}}}}-E^A_{\mathcal{I}^*_{{+1}}}$.

We can prove the estimate concerning $P^{(+)}_{\mathcal{I}^*_{{+1}}}\,(G_{{\mathcal{I}^*_{{+1}}}}-E^B_{{\mathcal{I}^*_{{+1}}}})\,P^{(+)}_{{\mathcal{I}^*_{{+1}}}}$ by repeating almost verbatim the proof used for $P^{(+)}_{\mathcal{I}^*_{{+1}}}\,(G_{{\mathcal{I}^*_{{+1}}}}-E^A_{{\mathcal{I}^*_{{+1}}}})\,P^{(+)}_{{\mathcal{I}^*_{{+1}}}}$. Namely, in order to get inequalities analogous to (\ref{3-fin}) and (\ref{3-bis-bis}) and finally  the statement in (\ref{final-eq-1}), we have to use  (\ref{second-lem-3.5}) (in point \emph{1}) or (\ref{third-lem-3.5}) (in point \emph{2}) of Lemma \ref{gapestantif} for a number of sites of $\cI^*_{+1}$  odd or even, respectively:

\noindent
\emph{\underline{odd number of sites}}
\begin{eqnarray}
\nonumber
& &P^{(+)}_{\mathcal{I}^*_{{+1}}}\,(G_{{\mathcal{I}^*_{{+1}}}}- E^B_{\mathcal{I}^*_{{+1}}}) \, P^{(+)}_{\mathcal{I}^*_{{+1}}}\\
&\geq&P^{(+)}_{\mathcal{I}^*_{{+1}}}(H^0_{\mathcal{I}^*_{{+1}}} - \langle\,H^0_{\mathcal{I}^*_{{+1}}}\,\rangle_{\Psi^B_{\cI^*_{+1}}} ) P^{(+)}_{\mathcal{I}^*_{{+1}}} \\
\nonumber
&& + \xi\cdot t P^{(+)}_{\mathcal{I}^*_{{+1}}}\Big[ - \sum_{\ell(\cJ)=1}^{\ell(\cI)-1}\, \frac{C_{J,h}'}{2|J|-h}\cdot (\xi\cdot t)^{\frac{\ell(\cJ)-1}{8}}\,(\ell(\cJ)+1)\Big]\,(H^0_{\mathcal{I}^*_{{+1}}}-\langle\,H^0_{\mathcal{I}^*_{{+1}}}\,\rangle_{\Psi^B_{\cI^*_{+1}}} + 2h)P^{(+)}_{\mathcal{I}^*_{{+1}}}\\
\nonumber
&& - \,\xi\cdot t\,\cdot  \sum\limits_{\ell(\cJ)=1}^{\ell(\cI)-1} C_{J,h}'\cdot\,\frac{(\xi\cdot t)^{\frac{\ell(\cJ)-1}{8}}}{(\ell(\cJ))^2} P^{(+)}_{\cI^*_{+1}}\\
\nonumber
\nonumber
&\geq  &\Big[1-\xi\cdot t\sum_{\ell(\cJ)=1}^{\ell(\cI)-1}\, \frac{C_{J,h}'}{2|J|-h}\cdot (\xi\cdot t)^{\frac{\ell(\cJ)-1}{8}}\,(\ell(\cJ)+1)\,\Big](2|J|-2 h)P^{(+)}_{\mathcal{I}^*_{{+1}}}\\
\nonumber
&& - \,\xi\cdot t\,\cdot  \sum\limits_{\ell(\cJ)=1}^{\ell(\cI)-1} C_{J,h}'\cdot\,\frac{(\xi\cdot t)^{\frac{\ell(\cJ)-1}{8}}}{(\ell(\cJ))^2} P^{(+)}_{\cI^*_{+1}}\\
\nonumber
&&+  2\,h\, \xi\cdot t\cdot \Big[ - \sum_{\ell(\cJ)=1}^{\ell(\cI)-1}\, \frac{C_{J,h}'}{2|J|-h}\cdot (\xi\cdot t)^{\frac{\ell(\cJ)-1}{8}}\,(\ell(\cJ)+1)\Big]P^{(+)}_{\mathcal{I}^*_{{+1}}}\,;
\end{eqnarray}

\noindent
\emph{\underline{even number of sites}}
\begin{eqnarray}
\nonumber 
& &P^{(+)}_{\mathcal{I}^*_{{+1}}}\,(G_{{\mathcal{I}^*_{{+1}}}}- E^B_{\mathcal{I}^*_{{+1}}}) \, P^{(+)}_{\mathcal{I}^*_{{+1}}}\\
&\geq&P^{(+)}_{\mathcal{I}^*_{{+1}}}(H^0_{\mathcal{I}^*_{{+1}}} - \langle\,H^0_{\mathcal{I}^*_{{+1}}}\,\rangle_{\Psi^B_{\cI^*_{+1}}} ) P^{(+)}_{\mathcal{I}^*_{{+1}}} \\
\nonumber
&& + \xi\cdot t P^{(+)}_{\mathcal{I}^*_{{+1}}}\Big[ - \sum_{\ell(\cJ)=1}^{\ell(\cI)-1}\, \frac{C_{J,h}'}{2|J|-h}\cdot (\xi\cdot t)^{\frac{\ell(\cJ)-1}{8}}\,(\ell(\cJ)+1)\Big]\,(H^0_{\mathcal{I}^*_{{+1}}}-\langle\,H^0_{\mathcal{I}^*_{{+1}}}\,\rangle_{\Psi^B_{\cI^*_{+1}}} + h)P^{(+)}_{\mathcal{I}^*_{{+1}}}\\
\nonumber
&& - \,\xi\cdot t\,\cdot  \sum\limits_{\ell(\cJ)=1}^{\ell(\cI)-1} C_{J,h}'\cdot\,\frac{(\xi\cdot t)^{\frac{\ell(\cJ)-1}{8}}}{(\ell(\cJ))^2} P^{(+)}_{\cI^*_{+1}}\\
\nonumber
\nonumber
&\geq&  \Big[1-\xi\cdot t\sum_{\ell(\cJ)=1}^{\ell(\cI)-1}\, \frac{C_{J,h}'}{2|J|-h}\cdot (\xi\cdot t)^{\frac{\ell(\cJ)-1}{8}}\,(\ell(\cJ)+1)\,\Big](2|J|-2 h)P^{(+)}_{\mathcal{I}^*_{{+1}}}\\
\nonumber
&& - \,\xi\cdot t\,\cdot  \sum\limits_{\ell(\cJ)=1}^{\ell(\cI)-1} C_{J,h}'\,\frac{(\xi\cdot t)^{\frac{\ell(\cJ)-1}{8}}}{(\ell(\cJ))^2} P^{(+)}_{\cI^*_{+1}}\\
\nonumber
&&+  \,h\, \xi\cdot t\cdot \Big[ - \sum_{\ell(\cJ)=1}^{\ell(\cI)-1}\, \frac{C_{J,h}'}{2|J|-h}\cdot (\xi\cdot t)^{\frac{\ell(\cJ)-1}{8}}\,(\ell(\cJ)+1)\Big]P^{(+)}_{\mathcal{I}^*_{{+1}}}\,.
\end{eqnarray}
\hfill \qed

\subsection{Estimates of operator norms of potentials and main theorem}\label{normsests}
In this section we prove our main results. Namely, in Theorem \ref{th-norms}  we collect the preliminary ingredients and show by induction the crucial bounds on the operator norms of the potentials assumed in some of the previous arguments. This proves that the block-diagonalization can be implemented up to the last step. In Theorem \ref{final-thm} we draw the conclusions about the low lying spectrum of the XXZ Hamiltonian in (\ref{XXZ-ham}). 

In order to make Theorem \ref{th-norms} as straightforward as possible, we prepare the ground in Sections \ref{hooked-pot} and \ref{hooked-proj} below, where we defer parts of the proof by induction (of Theorem \ref{th-norms});  namely we study some of the expressions entering the algorithm,  in order to estimate their operator norms in terms of the norms of the effective potentials at a given step.  Similarly, the control of the Lie-Schwinger series (which is part of the induction) is deferred to Lemma \ref{control-LS}.
\subsubsection{Hooked potentials}\label{hooked-pot}
Assuming the inductive hypothesis in (\ref{ind-ass})  and the bounds in (\ref{bound-V}), (\ref{bound-S1}), and (\ref{bound-S2}) proven in Lemma \ref{control-LS}, for $t$ sufficiently small we readily derive the following relations.
\begin{itemize}
\item \underline{Ferromagnetic case}
\begin{eqnarray}
& &\Big\|\sum_{n=1}^{\infty}\frac{1}{n!}\,ad^{n}Z_{\mathcal{I}^*}(V^{\mathcal{I}_{-1}}_{\overline{\mathcal{K}^*}})\Big\|\label{comp-in}\\
&\leq &C \cdot  \frac{A_1}{J+h} \cdot \xi\cdot t \cdot \|V^{\mathcal{I}_{-1}}_{\mathcal{I}}\|\cdot \|V^{\mathcal{I}_{-1}}_{\overline{\mathcal{K}^*}}\|\\
&\leq &C_{J,h} \cdot C\cdot  \frac{A_1}{J+h}\cdot \xi\cdot t\cdot \|V^{\mathcal{I}_{-1}}_{\mathcal{I}}\|\cdot \|V^{\mathcal{I}_{-1}}_{\mathcal{K}}\|\,.
\end{eqnarray}
where $C$ and $A_1$ are universal constants, and we recall that $C_{J,h}$ is defined in (\ref{ind-ass}).

By similar steps we can prove that

\begin{equation}
\Big\|\sum_{n=1}^{\infty}\frac{1}{n!}\,ad^{n}Z_{\mathcal{I}^*}(V^{\cI_{-1}}_{\mathcal{K}})\Big\|\leq C\cdot  \frac{A_1}{J+h}\cdot \xi\cdot t\cdot 
\|V^{\mathcal{I}_{-1}}_{\mathcal{I}}\|\cdot \|V^{\mathcal{I}_{-1}}_{\mathcal{K}}\|\,.
\end{equation}

\item \underline{Antiferromagnetic case}

\begin{eqnarray}
& &\Big\|\sum_{n=1}^{\infty}\frac{1}{n!}\,ad^{n}Z_{\mathcal{I}^*}(V^{\mathcal{I}_{-1}}_{\overline{\mathcal{K}^*}})\Big\|\\
&\leq &C\cdot  \frac{A_1}{|J|-h} \cdot \xi\cdot t\cdot 
\|V^{\cI_{-1}}_{\mathcal{I}}\|\cdot \|V^{\mathcal{I}_{-1}}_{\overline{\mathcal{K}^*}}\|\\
&\leq &C_{J,h} \cdot C\cdot  \frac{A_1}{|J|-h}\cdot \xi\cdot t\cdot 
\|V^{\mathcal{I}_{-1}}_{\mathcal{I}}\|\cdot \|V^{\mathcal{I}_{-1}}_{\mathcal{K}}\|\,
\end{eqnarray}

and

\begin{equation}
\Big\|\sum_{n=1}^{\infty}\frac{1}{n!}\,ad^{n}Z_{\mathcal{I}^*}(V^{\cI_{-1}}_{\mathcal{K}})\Big\|\leq C\cdot  \frac{A_1}{|J|-h}\cdot \xi\cdot t\cdot 
\|V^{\mathcal{I}_{-1}}_{\mathcal{I}}\|\cdot \|V^{\mathcal{I}_{-1}}_{\mathcal{K}}\|\,.\label{comp-fin}
\end{equation}
\end{itemize}

\noindent

\subsubsection{Off-diagonal part of the hooked Ising terms}\label{hooked-proj}
In this section we explain how to treat the hooked Ising terms. To this end, we provide a refined estimate of the commutator appearing in expression  (\ref{uno}) below.

\begin{lem}\label{Lemma-LR}
Assuming  \textit{S}1) and \textit{S}2) of Theorem \ref{th-norms}  in step $\cI_{-1}$, the following inequality holds true
\begin{equation} \label{bound-hooking-proj}
\Big\|P^{(+)}_{\overline{\mathcal{I}^*}}\,\Big(ad\,Z_{\mathcal{I}^*}(\,\frac{\sigma^z_{i^*_{-}-1}\sigma^z_{i^*_{-}}}{\xi \cdot t})\Big)\,P^{(-)}_{{\overline{\mathcal{I}^*}}}\Big\|\leq \mathcal{O}((\xi \cdot t)^{\frac{1}{2}}\cdot \|V^{\mathcal{I}_{-1}}_{\mathcal{I}}\|)\,
\end{equation}
where $\xi$ is assumed to be odd and  larger than or equal to $9$ in the antiferromagnetic case, and larger than or equal to $\max\{6\,;\,\frac{J}{h}\}$ in the ferromagnetic one.
\end{lem}

\noindent
\emph{Proof}

\noindent
We treat the antiferromagnetic case first and then explain how to recover the ferromagnetic case from it. 

Recall the formulae introduced in Section \ref{conj-formulae} and write
\begin{eqnarray}
P^{(+)}_{{\overline{\mathcal{I}^*}}}\,\Big(ad\,Z_{\mathcal{I}^*}(\,{\frac{\sigma^z_{i^*_{-}-1}\sigma^z_{i^*_{-}}}{\xi \cdot t}})\Big)\,P^{(-)}_{{\overline{\mathcal{I}^*}}}
&=&P^{(+)}_{{\overline{\mathcal{I}^*}}}\,\Big[\,Z_{\mathcal{I}^*}\,,\,{\frac{\sigma^z_{i^*_{-}-1}\sigma^z_{i^*_{-}}}{\xi \cdot t}}\Big]\,P^{(-)}_{{\overline{\mathcal{I}^*}}}\label{uno}\\
&=&P^{(+)}_{{\overline{\mathcal{I}^*}}}\,\Big[\,Z_{\mathcal{I}^*}\,,\,{\frac{\sigma^z_{i^*_{-}-1}\sigma^z_{i^*_{-}}}{\xi \cdot t}}+\frac{1}{\xi \cdot t}\Big]\,P^{(-)}_{{\overline{\mathcal{I}^*}}}\label{uno-prime}\\
&=&-P^{(+)}_{{\overline{\mathcal{I}^*}}}\,({\frac{\sigma^z_{i^*_{-}-1}\sigma^z_{i^*_{-}}}{\xi \cdot t}}+\frac{1}{\xi \cdot t})\,Z_{\mathcal{I}^*}\,P^{(-)}_{{\overline{\mathcal{I}^*}}}\label{uno-bis}\\
&=&-\sum_{j=1}^{\infty}{t^{\frac{ j}{2}}}P^{(+)}_{{\overline{\mathcal{I}^*}}}\,({\frac{\sigma^z_{i^*_{-}-1}\sigma^z_{i^*_{-}}}{\xi \cdot t}}+\frac{1}{\xi \cdot t})\,(Z_{\mathcal{I}^*})_j\,P^{(-)}_{{\overline{\mathcal{I}^*}}}\,,\label{due-bis}
\end{eqnarray}
where in the step from (\ref{uno}) to (\ref{uno-bis}) we have used $$({\frac{\sigma^z_{i^*_{-}-1}\sigma^z_{i^*_{-}}}{\xi \cdot t}}+\frac{1}{\xi \cdot t})\,P^{(-)}_{{\overline{\mathcal{I}^*}}}=0\,.$$
In expression (\ref{due-bis}) above we can discard all the terms of the series starting from $j=2$, i.e., 
\begin{equation}\label{partial-expression}
-\sum_{j=2}^{\infty}(\xi \cdot t)^j P^{(+)}_{\overline{\mathcal{I}^*}}\,(\frac{\sigma^z_{i^*_{-}-1}\sigma^z_{i^*_{-}}}{\xi \cdot t}+\frac{1}{\xi \cdot t})\,(Z_{\mathcal{I}^*})_j\,P^{(-)}_{{\overline{\mathcal{I}^*}}}\,,
\end{equation}
 since $\|(\ref{partial-expression})\|$  is bounded by $\mathcal{O}(\xi \cdot t\cdot \|V^{\mathcal{I}_{-1}}_{\mathcal{I}}\|^2)$ and, consequently, the bound in  (\ref{bound-hooking-proj}) is fulfilled 
thanks to \textit{S}1) of Theorem \ref{th-norms}  in step $\cI_{-1}$. Now we focus on the remaining quantity
 \begin{eqnarray}\label{leading-expression}
& &-t^{\frac{ 1}{2}}P^{(+)}_{\overline{\mathcal{I}^*}}\,({\frac{\sigma^z_{i^*_{-}-1}\sigma^z_{i^*_{-}}}{\xi \cdot t}}+\frac{1}{\xi \cdot t})\,(Z_{\mathcal{I}^*})_1\,P^{(-)}_{\overline{\mathcal{I}^*}}\,
\end{eqnarray}
which, in the antiferromagnetic case, corresponds to 
\begin{eqnarray}
& &-P^{(+)}_{\overline{\mathcal{I}^*}}\,(\sigma^z_{i^*_{-}-1}\sigma^z_{i^*_{-}}+1)\,\frac{1}{G_{\mathcal{I}^*}-E^A_{\mathcal{I}^*}}P^{(+)}_{\mathcal{I}^*}\,V^{\mathcal{I}_{-1}}_{\mathcal{I}}\,P^{(-),A}_{\mathcal{I}^*}\,P^{(-)}_{\overline{\mathcal{I}^*}}\,\label{due}\\
&&- P^{(+)}_{\overline{\mathcal{I}^*}}\,(\sigma^z_{i^*_{-}-1}\sigma^z_{i^*_{-}}+1)\,\frac{1}{G_{\mathcal{I}^*}-E^B_{\mathcal{I}^*}}P^{(+)}_{\mathcal{I}^*}\,V^{\mathcal{I}_{-1}}_{\mathcal{I}}\,P^{(-),B}_{\mathcal{I}^*}\,P^{(-)}_{\overline{\mathcal{I}^*}}\,.\label{tre}
\end{eqnarray}
We make use of the identity
\begin{equation}
\frac{1}{G_{\mathcal{I}^*}-E^{A/B}_{\mathcal{I}^*}}P^{(+)}_{\mathcal{I}^*}=\frac{1}{G_{\mathcal{I}^*}-E^{A/B}_{\mathcal{I}^*}+i\delta_{\xi \cdot t}}P^{(+)}_{\mathcal{I}^*}+\frac{i\delta_{\xi \cdot t}}{G_{\mathcal{I}^*}-E^{A/B}_{\mathcal{I}^*}} \,\frac{1}{G_{\mathcal{I}^*}-E^{A/B}_{\mathcal{I}^*}+i\delta_{\xi \cdot t} }P^{(+)}_{\mathcal{I}^*}
\end{equation}
where $\delta_{\xi \cdot t}$ is set equal to $(\xi \cdot t)^{\frac{1}{2}}$. Then, thanks to the gap bound (\ref{final-eq-1}) proven in Lemma \ref{gap-bound} (in step $\mathcal{I}_{-1}$),  we can write
\begin{eqnarray}
(\ref{due})+(\ref{tre})&=&-P^{(+)}_{{\overline{\mathcal{I}^*}}}\,(\sigma^z_{i^*_{-}-1}\sigma^z_{i^*_{-}}+\charf)\,\frac{1}{G_{\mathcal{I}^*}-E^A_{\mathcal{I}^*}+i\delta_{\xi \cdot t} }P^{(+)}_{\mathcal{I}^*}\,V^{\mathcal{I}_{-1}}_{\mathcal{I}}\,P^{(-),A}_{\mathcal{I}^*}\,P^{(-)}_{{\overline{\mathcal{I}^*}}} \quad \label{intermediate-0}\\
&& -P^{(+)}_{{\overline{\mathcal{I}^*}}}\,(\sigma^z_{i^*_{-}-1}\sigma^z_{i^*_{-}}+\charf)\,\frac{1}{G_{\mathcal{I}^*}-E^B_{\mathcal{I}^*}+i\delta_{\xi \cdot t} }P^{(+)}_{\mathcal{I}^*}\,V^{\mathcal{I}_{-1}}_{\mathcal{I}}\,P^{(-),B}_{\mathcal{I}^*}\,P^{(-)}_{{\overline{\mathcal{I}^*}}}\quad \\
& &+R_1\label{intermediate}
\end{eqnarray}
with $\|R_1\|\leq \mathcal{O}(\frac{\delta_{\xi \cdot t}}{2|J|-2h} \cdot  \|V^{\mathcal{I}_{-1}}_{\mathcal{I}}\|)$. 
The remainder term, $R_1$, appearing in (\ref{intermediate})  fulfills the bound in (\ref{bound-hooking-proj}), whereas the first and second terms require some further manipulation explained below. Since the latter ones are estimated in the same way, we will proceed by analyzing the right hand side of (\ref{intermediate-0}) only. For this purpose, we use the Neumann expansion displayed below
\begin{eqnarray}
\label{neumannexp}
\nonumber
&&\frac{1}{G_{\mathcal{I}^*}-E^A_{\mathcal{I}^*}+i\delta_{\xi \cdot t} }P^{(+)}_{\mathcal{I}^*} \nonumber \\
&=&\frac{1}{P^{(+)}_{\mathcal{I}^*}(G_{\mathcal{I}^*}-E^A_{\mathcal{I}^*}+i\delta_{\xi \cdot t} )P^{(+)}_{\mathcal{I}^*}}P^{(+)}_{\mathcal{I}^*} \nonumber \\
&=& \frac{1}{H^{0}_{\mathcal{I}^*}-\langle H^{0}_{\mathcal{I}^*}\rangle_{\Psi^A_{\cI^*}}+i\delta_{\xi \cdot t} }P^{(+)}_{\mathcal{I}^*}+\\
\nonumber
&&+\frac{1}{H^{0}_{\mathcal{I}^*}-\langle H^{0}_{\mathcal{I}^*}\rangle_{\Psi^A_{\cI^*}}+i{ \delta_{\xi \cdot t}} }\sum_{j=1}^{\infty}\Big[\,\Big(\, -{\xi \cdot t }\,\,\sum_{\overline{\mathcal{J}^*}\subset \mathcal{I}^*} P^{(+)}_{\mathcal{I}^*}(V^{\cI_{-1}}_{\overline{\mathcal{J}^*}}-\langle V^{\cI_{-1}}_{\overline{\mathcal{J}^*}}\rangle_{\Psi^A_{\mathcal{I}^*}})P^{(+)}_{\mathcal{I}^*}\Big)\frac{1}{H^{0}_{\mathcal{I}^*}-\langle H^{0}_{\mathcal{I}^*}\rangle_{\Psi^A_{\cI^*}}+i\delta_{\xi \cdot t} }\Big]^j\,P^{(+)}_{\mathcal{I}^*}\,
\end{eqnarray}
which is well defined due to arguments as in the proof of Lemma \ref{gap-bound} (see in particular (\ref{AFHC})), and we can estimate
\begin{eqnarray}
& &\Big\|\frac{1}{H^{0}_{\mathcal{I}^*}-\langle H^{0}_{\mathcal{I}^*}\rangle_{\Psi^A_{\cI^*}}+i\delta_{\xi \cdot t} }\sum_{j=1}^{\infty}\Big[\,\Big(\, -{\xi \cdot t }\,\,\sum_{\overline{\mathcal{J}^*}\subset \mathcal{I}^*} P^{(+)}_{\mathcal{I}^*}(V^{\cI_{-1}}_{\overline{\mathcal{J}^*}}-\langle V^{\cI_{-1}}_{\overline{\mathcal{J}^*}} \rangle_{\Psi^A_{\cI^*}})P^{(+)}_{\mathcal{I}^*}\Big)\frac{1}{H^{0}_{\mathcal{I}^*}-\langle H^{0}_{\mathcal{I}^*}\rangle_{\Psi^A_{\cI^*}}+i\delta_{\xi \cdot t} }\Big]^j\, P^{(+)}_{\mathcal{I}^*}\Big\| \nonumber\\
&\leq& \mathcal{O}(\xi \cdot t)\,.
\end{eqnarray}
Hence we write
\begin{eqnarray}
& &-P^{(+)}_{{\overline{\mathcal{I}^*}}}\,(\sigma^z_{i^*_{-}-1}\sigma^z_{i^*_{-}}+1)\,\frac{1}{G_{\mathcal{I}^*}-E^A_{\mathcal{I}^*}+i\delta_{\xi \cdot t} }P^{(+)}_{\mathcal{I}^*}\,V^{\mathcal{I}_{-1}}_{\mathcal{I}}\,P^{(-),A}_{\mathcal{I}^*}\,P^{(-)}_{{\overline{\mathcal{I}^*}}} \\
&=&-P^{(+)}_{{\overline{\mathcal{I}^*}}}\,(\sigma^z_{i^*_{-}-1}\sigma^z_{i^*_{-}}+1)\,\,\frac{1}{H^{0}_{\mathcal{I}^*}-\langle H^{0}_{\mathcal{I}^*}\rangle_{\Psi^A_{\cI^*}}+i\delta_{\xi \cdot t} }\,P^{(+)}_{\mathcal{I}^*}\,V^{\mathcal{I}_{-1}}_{\mathcal{I}}\,P^{(-),A}_{\mathcal{I}^*}\,P^{(-)}_{{\overline{\mathcal{I}^*}}}\label{H0-left}\\
& &+R_2\,,
\end{eqnarray}
where $\|R_2\|\leq \mathcal{O}(\xi \cdot t\cdot \|V^{\mathcal{I}_{-1}}_{\mathcal{I}}\|)$.\\

\noindent
In expression (\ref{H0-left}),  first we substitute $P^{(+)}_{\mathcal{I}^*}=\charf -P^{(-)}_{\mathcal{I}^*}$, then we exploit  that $V^{\cI_{-1}}_{\cI}$ is block diagonal w.r.t. $P^{(-),A/B}_{\cI^*}$ and write
\begin{eqnarray}
& &-P^{(+)}_{{\overline{\mathcal{I}^*}}}\,(\sigma^z_{i^*_{-}-1}\sigma^z_{i^*_{-}}+1)\,\frac{1}{H^{0}_{\mathcal{I}^*}-\langle H^{0}_{\mathcal{I}^*}\rangle_{\Psi^A_{\cI^*}}+i\delta_{\xi \cdot t} }P^{(+)}_{\mathcal{I}^*}\,V^{\mathcal{I}_{-1}}_{\mathcal{I}}\,P^{(-),A}_{\mathcal{I}^*}\,P^{(-)}_{{\overline{\mathcal{I}^*}}} \\
&=&-P^{(+)}_{{\overline{\mathcal{I}^*}}}\,(\sigma^z_{i^*_{-}-1}\sigma^z_{i^*_{-}}+1)\,\frac{1}{H^{0}_{\mathcal{I}^*}-\langle H^{0}_{\mathcal{I}^*}\rangle_{\Psi^A_{\cI^*}}+i\delta_{\xi \cdot t} }\,V^{\mathcal{I}_{-1}}_{\mathcal{I}}\,P^{(-),A}_{\mathcal{I}^*}\,P^{(-)}_{{\overline{\mathcal{I}^*}}}\\
& &+P^{(+)}_{{\overline{\mathcal{I}^*}}}\,(\sigma^z_{i^*_{-}-1}\sigma^z_{i^*_{-}}+1)\,\frac{1}{H^{0}_{\mathcal{I}^*}-\langle H^{0}_{\mathcal{I}^*}\rangle_{\Psi^A_{\cI^*}}+i\delta_{\xi \cdot t} }P^{(-),A}_{\mathcal{I}^*}\,V^{\mathcal{I}_{-1}}_{\mathcal{I}}\,P^{(-),A}_{\mathcal{I}^*}\,P^{(-)}_{{\overline{\mathcal{I}^*}}}\,.\label{remainder-2}
\end{eqnarray}
Next, we recall the following three identities: 

\noindent
1)
\begin{equation}
\frac{1}{H^{0}_{\mathcal{I}^*}-\langle H^{0}_{\mathcal{I}^*}\rangle_{\Psi^A_{\cI^*}}+i\delta_{\xi \cdot t} }\,{ P^{(-),A}_{\mathcal{I}^*}} V^{\mathcal{I}_{-1}}_{\mathcal{I}}\,P^{(-),A}_{\mathcal{I}^*}\,P^{(-)}_{{\overline{\mathcal{I}^*}}}=\frac{1}{i\delta_{\xi \cdot t}}P^{(-),A}_{\mathcal{I}^*}\,V^{\mathcal{I}_{-1}}_{\mathcal{I}}\,P^{(-),A}_{\mathcal{I}^*}\,P^{(-)}_{{\overline{\mathcal{I}^*}}}
\end{equation}
which holds since $(H^{0}_{\mathcal{I}^*}-\langle H^{0}_{\mathcal{I}^*}\rangle_{\Psi^A_{\cI^*}})P^{(-),A}_{\mathcal{I}^*}=0$; 

\noindent
2)
\begin{equation}
P^{(-),A}_{\mathcal{I}^*}\,V^{\mathcal{I}_{-1}}_{\mathcal{I}}\,P^{(-),A}_{\mathcal{I}^*}\,P^{(-)}_{{\overline{\mathcal{I}^*}}}\,=\, \langle\, V^{\mathcal{I}_{-1}}_{\mathcal{I}}\,\rangle_{\Psi^A_{\cI^*}} \, P^{(-),A}_{\mathcal{I}^*}\,P^{(-)}_{{\overline{\mathcal{I}^*}}}=\langle\, V^{\mathcal{I}_{-1}}_{\mathcal{I}}\,\rangle_{\Psi^A_{\cI^*}} \,  \,P^{(-),A}_{{\overline{\mathcal{I}^*}}}\,
\end{equation}
which holds since $\overline{\mathcal{I}^*}$ is an extension  of $\mathcal{I}^*$ by two sites both on the left and on the right for bulk-intervals, and by two sites either on the left or on the right for boundary-intervals;

\noindent
3)
\begin{equation}\label{eigenvalue}
(\sigma^z_{i^*_{-}-1}\sigma^z_{i^*_{-}}+1){ P^{(-)}_{\overline{\mathcal{I}^*}}}=0\,,
\end{equation}
see Remark \ref{rem-LB}.
From 1), 2), and 3) above we deduce
\begin{equation}
(\ref{remainder-2})=0\, .
\end{equation}
Finally,  we show how to control
\begin{equation}
-P^{(+)}_{{\overline{\mathcal{I}^*}}}\,{(\sigma^z_{i^*_{-}-1}\sigma^z_{i^*_{-}}+1)}\,\frac{1}{H^{0}_{\mathcal{I}^*}-\langle H^{0}_{\mathcal{I}^*}\rangle_{\Psi^A_{\cI^*}}+i\delta_{\xi \cdot t} }\,V^{\mathcal{I}_{-1}}_{\mathcal{I}}\,P^{(-),A}_{\mathcal{I}^*}\,P^{(-)}_{{\overline{\mathcal{I}^*}}}\label{hooking-2}\,.
\end{equation}
We re-write
\begin{equation}\label{res-int}
\frac{1}{H^{0}_{\mathcal{I}^*}-\langle H^{0}_{\mathcal{I}^*}\rangle_{\Psi^A_{\cI^*}}+i\delta_{\xi \cdot t}}=-i\,\int_{0}^{t^{-\frac{1}{3}}}\,e^{i\Big(H^{0}_{\mathcal{I}^*}-\langle H^{0}_{\mathcal{I}^*}\rangle_{\Psi^A_{\cI^*}}+i\delta_{\xi \cdot t}\Big) s}\,ds-i\,\int_{t^{-\frac{1}{3}}}^{+\infty}\,e^{i\Big(H^{0}_{\mathcal{I}^*}-\langle H^{0}_{\mathcal{I}^*}\rangle_{\Psi^A_{\cI^*}}+i\delta_{\xi \cdot t}\Big) s}\,ds
\end{equation}
and define
\begin{equation}
R_4:=i\,P^{(+)}_{\overline{\mathcal{I}^*}}\,(\sigma^z_{i^*_{-}-1}\sigma^z_{i^*_{-}}+1)\,\int_{t^{-1/3}}^{+\infty}\,e^{i\Big(H^{0}_{\mathcal{I}^*}-\langle H^{0}_{\mathcal{I}^*}\rangle_{\Psi^A_{\cI^*}}+i\delta_{\xi \cdot t}\Big) s}\,ds\,V^{\mathcal{I}_{-1}}_{\mathcal{I}}\,P^{(-),A}_{\mathcal{I}^*}\,P^{(-)}_{\overline{\mathcal{I}^*}}
\end{equation}
with $\|R_4\|\leq \mathcal{O}(\frac{e^{-\delta_{\xi \cdot t}\cdot t^{-\frac{1}{3}}}}{\delta_{\xi \cdot t}}\|V^{\mathcal{I}_{-1}}_{\mathcal{I}}\|) $.
By using the identities in (\ref{res-int}) and in (\ref{eigenvalue}), we write
\begin{eqnarray}
& &-P^{(+)}_{\overline{\mathcal{I}^*}}\,(\sigma^z_{i^*_{-}-1}\sigma^z_{i^*_{-}}+1)\,\frac{1}{H^{0}_{\mathcal{I}^*}-\langle H^{0}_{\mathcal{I}^*}\rangle_{\Psi^A_{\cI^*}}+i\delta_{\xi \cdot t} }\,V^{\mathcal{I}_{-1}}_{\mathcal{I}}\,P^{(-),A}_{\mathcal{I}^*}\,P^{(-)}_{\overline{\mathcal{I}^*}}-R_4\label{leading-2}\\
&=&i\cdot \int_{0}^{t^{-1/3}}\,P^{(+)}_{\overline{\mathcal{I}^*}}\,(\sigma^z_{i^*_{-}-1}\sigma^z_{i^*_{-}}+1)\,e^{i\Big(H^{0}_{\mathcal{I}^*}-\langle H^{0}_{\mathcal{I}^*}\rangle_{\Psi^A_{\cI^*}}+i\delta_{\xi \cdot t}\Big) s}\,V^{\mathcal{I}_{-1}}_{\mathcal{I}}\,P^{(-),A}_{\overline{\mathcal{I}^*}}\,ds\\
&=&i\cdot  \int_{0}^{t^{-1/3}}\,P^{(+)}_{\overline{\mathcal{I}^*}}\,(\sigma^z_{i^*_{-}-1}\sigma^z_{i^*_{-}}+1)\,e^{-\delta_{\xi \cdot t} \cdot s}\, e^{i \Big(H^{0}_{\mathcal{I}^*}-\langle H^{0}_{\mathcal{I}^*}\rangle_{\Psi^A_{\cI^*}}\Big)s}\,V^{\mathcal{I}_{-1}}_{\mathcal{I}}\,e^{-i\Big(H^{0}_{\mathcal{I}^*}-\langle H^{0}_{\mathcal{I}^*}\rangle_{\Psi^A_{\cI^*}}\Big)s}P^{(-),A}_{\overline{\mathcal{I}^*}}\,ds\nonumber\\
&=&i\cdot \int_{0}^{t^{-1/3}}\,P^{(+)}_{\overline{\mathcal{I}^*}}\,\,e^{-\delta_{\xi \cdot t} \cdot s}\, \Big[(\sigma^z_{i^*_{-}-1}\sigma^z_{i^*_{-}}+1)\,,\,e^{i\,H^{0}_{\mathcal{I}^*}s}\,V^{\mathcal{I}_{-1}}_{\mathcal{I}}\,e^{-i\, H^{0}_{\mathcal{I}^*} s}\Big]\,P^{(-),A}_{\overline{\mathcal{I}^*}}\,ds\,=\,0\,
\end{eqnarray}
since for $\xi\geq 6$  the commutator is identically zero by exploiting that the Hamiltonian $H^{0}_{\mathcal{I}^*}$  is the sum of commuting terms which are on-site or supported in intervals of size $1$  at most, in the microscopic unit. 
\begin{rem}
We observe that in this paper the parameter $\xi$ can be chosen to be  $t$-independent since the propagation speed of the unperturbed dynamics is zero. In the antiferromagnetic case we can then assume $\xi$ equal to the smallest odd number such that $\mathbb{N} \ni \xi/3\geq 2$.
\end{rem}

\noindent
By collecting all the estimates the bound in (\ref{bound-hooking-proj}) is proven. 
\begin{rem}\label{rem-LB}
Concerning the relation between a single Ising term and  the ground-state vectors of the unperturbed Hamiltonian of the entire chain,  we observe that in the estimate of the commutator  in (\ref{bound-hooking-proj}) 
we only use that they are eigenvectors of each Ising term.
\end{rem}
Regarding the ferromagnetic case, in the analogous proof starting from (\ref{uno-prime}) we must replace $$\frac{\sigma^z_{i^*_{-}-1}\sigma^z_{i^*_{-}}}{\xi \cdot t}+\frac{1}{\xi \cdot t}\quad \text{with}\quad \frac{\sigma^z_{i^*_{-}-1}\sigma^z_{i^*_{-}}}{\xi \cdot t}-\frac{1}{\xi \cdot t}\,;$$
the rest of the proof is indeed simpler due to the one-dimensionality of the ground-state subspace of $H^{0}_{\mathcal{I}^*}$.
\hfill \qed

\subsubsection{Main theorems}

We are now ready to collect all the ingredients and prove our main results.
\begin{thm}\label{th-norms}
{ There exists $\bar{t}>0$ independent of $N$, such that for all $|t|<\bar{t}$,}  { for any $\mathfrak{I}\ni\mathcal{L}\preceq \Lambda_{-1}$,  the Hamiltonians $G_{\mathcal{L}^*}$ are well defined}, and the properties below hold true:
\begin{enumerate}
\item[S1)] for any interval  $\mathcal{J}\in\mathfrak{I}$, the following operator norms estimates hold
$$\|V^{\mathcal{L}}_{\mathcal{J}}\| \leq\frac{(\xi \cdot t)^{\frac{\ell(\mathcal{J})-1}{8}}}{\ell(\mathcal{J})^2}\,$$
for $\mathcal{J} \, \succ \, \mathcal{L}$; 

\item[S2)] in the antiferromagnetic case, the spectrum of the operators $(G_{{\mathcal{L}^*_{{+1}}}}-E^{A/B}_{{\mathcal{L}^*_{{+1}}}})$ restricted to $P^{(+)}_{\mathcal{L}^*_{{+1}}}\mathcal{H}^{\Lambda}$ is bounded below by $\Delta_{\mathcal{L}^*_{+1}}\geq \frac{2|J|-2h}{2}$,
where  $G_{\mathcal{L}_{+1}^*}$ is defined in formula (\ref{expression-G}). Analogously, in the ferromagnetic case,  the spectral gap of  $G_{\mathcal{L}_{+1}^*}$ above the ground-state energy is bounded below by $\Delta_{\mathcal{L}^*_{+1}}\geq \frac{2J+2h}{2}$ \,.
\end{enumerate}
\end{thm}

\noindent
\emph{Proof}

\noindent
The proof is by induction  in the interval $\cI$ that labels the block-diagonalization step. Hence for each operator support $\cJ\in\mathfrak{I}$ we shall prove S1) and S2) from step $\cI=\cI_0$ up to step $\cI= \Lambda_{-1}$. That is we assume that S1) holds for all $V^{\mathcal{K}}_{\mathcal{J}}$ with $\mathcal{K} \prec \cI$ and  S2) for all $\mathcal{K} \prec \cI$. Then we show that they hold for all $V^{\cI}_{\mathcal{J}}$ and for  $G_{(\mathcal{I}^*)_{+1}}$, respectively. This implies that $K_{\Lambda}^{\cI}$ and (by Lemma \ref{control-LS}) $Z_{\mathcal{I}^*}$ are well defined operators and, consequently,  (\ref{equality-hams}) holds true.

\noindent
For $\cI= \cI_0$,  $\mathcal{S}1)$ can be verified by direct computation,  indeed $\|V_{\mathcal{J}}^{\cI_0}\|\leq 1$ for $\mathcal{J}$ with $\ell(\mathcal{J})=1$ due to (\ref{0-pot-bis}) and (\ref{V-0-bound}), 
and $\|V_{\mathcal{J}}^{\cI_0}\|=0$ (see (\ref{0-pot})) for $\mathcal{J}$ with $\ell(\mathcal{J})>1$; S2) holds trivially since, by definition,  $G_{(\mathcal{I}_0)_{+1}}=H^{0}_{(\mathcal{I}_0)_{+1}}$ where $(\mathcal{I}_0)_{+1}$ is the interval with $Q((\mathcal{I}_0)_{+1})=\ell((\mathcal{I}_0)_{+1})=1$.

The induction step consists of several parts and for each of them we choose $t(\geq 0)$ in an interval such that the previous parts  and Lemma \ref{control-LS} are verified. By this procedure we may progressively restrict such interval until we determine a $\bar{t}>0$ for which all the parts hold true for $0\leq t <  \bar{t}$.
\\

\noindent
\emph{Induction step in the proof of S1)}

\noindent
Starting from $\mathcal{L}$ down to $(\mathcal{I}_0)_{+1}$, step by step, we relate the norm of $V^{\cI}_{\mathcal{J}}$ to the ones of the operators, in step $\mathcal{I}_{-1}$,  in terms of which  $V^{\cI}_{\mathcal{J}}$ is expressed according to the algorithm. It is then clear that, at fixed $\mathcal{J}$,  for most of the steps the norm is preserved, i.e., $\|V^{\cI}_{\mathcal{J}}\|=\|V^{\mathcal{I}{-1}}_{\mathcal{J}}\|$,  and only for special steps we have nontrivial relations.

\noindent
By the rules of the algorithm displayed in Definition  \ref{def-interactions-multi},   $V^{\mathcal{L}}_{{ \mathcal{J}}}$   is defined only for $ \mathcal{J} \succ\mathcal{L}$,  and this is possible if $\ell(\mathcal{J})> \ell(\mathcal{L})$ or if $\ell(\mathcal{J})= \ell(\mathcal{L})$ and $Q(\mathcal{J})> Q(\mathcal{L})$.
\\

\noindent
\emph{Case $\ell(\mathcal{J})=1$}

\noindent
In this case statement S1) holds since, due to a-1) of Definition  \ref{def-interactions-multi}, by repeated steps back we  readily obtain
\begin{equation}
\|V^{\mathcal{L}}_{\mathcal{J}}\|=\|V^{\cI_0}_{\mathcal{J}}\|=\|V_{\mathcal{J}}\|\leq 1\,.
\end{equation}

\noindent
\emph{Case $\ell(\mathcal{J})\geq 2$}

\noindent
We recall that by construction $V^{\mathcal{I}}_{\mathcal{J}}$ is defined only for $\ell(\cJ)\geq \ell(\mathcal{I})$. {For the re-expansion from step $\cI$ to $\mathcal{I}_{-1}$}, we distinguish the following situations related to cases a), b), and c) of Definition  \ref{def-interactions-multi}:
\begin{itemize}
\item[i)] In case a-1),  and in case c) with the additional requirement that $\widetilde{\mathcal{I}^*}$ does not contain endpoints of $\mathcal{J}$,  we have that
\begin{equation}\label{cons}
\|V^{\cI}_{\mathcal{J}}\|=\|V^{\mathcal{I}_{-1}}_{\mathcal{J}}\|\,;
\end{equation} 
indeed, in case a-1) the statement is trivial, while in case  c) only the contribution in (\ref{identity-c}) is nonzero due to requirement that $\widetilde{\mathcal{I}^*}$ does not contain endpoints of $\mathcal{J}$.
\item[ii)]
In case c), if $\widetilde{\mathcal{I}^*}$  contains only one of the endpoints of $\mathcal{J}$, the contributions are those in (\ref{identity-c}), (\ref{A-map-1-bis}),  and (\ref{Valgo3}), hence we can estimate
\begin{eqnarray}\label{arg}
\|V^{\cI}_{\mathcal{J}}\|&\leq& \|V^{\cI_{-1}}_{\mathcal{J}}\|\\
& &+\sum_{\mathcal{K}\in [\mathcal{G}^{\cI}_{\mathcal{J}}]_1}\,\Big\|\sum_{n=1}^{\infty}\frac{1}{n!}\,ad^{n}Z_{\mathcal{I}^*}(V^{\cI_{-1}}_{\mathcal{K}})\Big\|\label{hook-V}\\
& &+\sum_{\mathcal{K}^*\in [\mathcal{G}^{\cI}_{\mathcal{J}}]_2}\,\Big\|\sum_{n=1}^{\infty}\frac{1}{n!}\,ad^{n}Z_{\mathcal{I}^*}(V^{\mathcal{I}_{-1}}_{\overline{\mathcal{K}^*}})\Big\|\label{hook-V^*}
\end{eqnarray}
where for the convenience of the reader we recall that
\begin{eqnarray}
[\mathcal{G}^{\mathcal{I}}_{\mathcal{J}}]_1 &:=&\Big\{ \, \mathcal{K} \in\mathfrak{I} \,\,\vert \,\,\mathcal{K} \succ \mathcal{I} \,,\,\mathcal{K} \cap \mathcal{I}^*\neq \emptyset , \,\\
&&\,\mathcal{K} \neq \mathcal{J} \,,\,\text{and} \,\, \widetilde{\mathcal{I}^* }\cup \mathcal{K} =\mathcal{J} \,\,\Big\}\,\nonumber 
\end{eqnarray}
\begin{eqnarray}
[ \mathcal{G}^{\mathcal{I}}_{\mathcal{J}}]_2  &:=& 
\Big\{ \, \mathcal{K}^* \in\mathfrak{I}^* \,\,\vert \,\,\mathcal{I} \succ \mathcal{K}\,,\,\mathcal{K}^* \cap \mathcal{I}^*\neq \emptyset \,, \mathcal{K}^* \nsubset\mathcal{I}^*\nonumber\\ 
&& \text{and} \,\,
\widetilde{\mathcal{I}^*}\cup \widetilde{\mathcal{K}^*}=\mathcal{J}\,\,\Big\}\,. \,\nonumber
\end{eqnarray}
For the estimate of $(\ref{hook-V})+(\ref{hook-V^*})$, we invoke the computations in (\ref{comp-in})-(\ref{comp-fin}) of Section \ref{hooked-pot}, the inductive hypothesis S1), and we split our study into three parts corresponding to the following subcases (see Remark \ref{bound.-bulk}): 
ii-1) $\mathcal{J}$ is a bulk-interval (hence also $\mathcal{I}$ is a bulk-interval since $\mathcal{I}^* \subset \mathcal{J}$); 
ii-2)  $\mathcal{J}$  and $\cI$ are both boundary intervals; 
ii-3) $\mathcal{J}$ is a boundary interval but $\cI$ is a bulk-interval.

\noindent
We also use the symbol $C'_{J,h}$ for any positive constants that depend on $J$ and $h$. They may change from line to line.
\begin{enumerate}
\item[ii-1)]
If $\mathcal{J}$ is a bulk-interval then $\ell(\mathcal{J})-\ell(\mathcal{I})\geq 3$ and $\ell(\mathcal{K})\geq \ell(\mathcal{J})-\ell(\mathcal{I})-2$ in the summation concerning (\ref{hook-V^*}). Hence,  we can estimate (see (\ref{comp-in})-(\ref{comp-fin}))
\begin{eqnarray}
(\ref{hook-V})+(\ref{hook-V^*})&\leq & C'_{J,h} \cdot C\cdot  A_1\cdot \xi \cdot t\cdot \sum_{\ell(\mathcal{K})=\ell(\mathcal{J})-\ell(\mathcal{I})-2}^{\ell(\mathcal{J})-1}\, \frac{(\xi \cdot t)^{\frac{\ell(\cI)-1}{8}}}{\ell(\cI)^2}\cdot \frac{(\xi \cdot t)^{\frac{\ell(\mathcal{K})-1}{8}}}{\ell(\mathcal{K})^2} \\
&\leq & C'_{J,h}\cdot C\cdot  A_1\cdot \xi \cdot t\cdot \sum_{m=0}^{\ell(\cI)+1}\, \frac{(\xi \cdot t)^{\frac{\ell(\cI)-1}{8}}}{\ell(\cI)^2}\cdot \frac{(\xi \cdot t)^{\frac{\ell(\cJ)-\ell(\cI)+m-3}{8}}}{(\ell(\cJ)-\ell(\cI)-2+m)^2}\\
&= & C'_{J,h}\cdot C\cdot  A_1\cdot \xi \cdot t\cdot (\xi \cdot t)^{\frac{\ell(\cJ)-4}{8}} \sum_{m=0}^{\ell(\cI)+1}\,  \frac{(\xi \cdot t)^{\frac{m}{8}}}{\ell(\cI)^2\cdot (\ell(\cJ)-\ell(\cI)-2+m)^2}\quad\quad\quad  \\
&\leq &C'_{J,h}\cdot (\xi \cdot t)^{\frac{5}{8}}\cdot \frac{(\xi \cdot t)^{\frac{\ell(\cJ)-1}{8}}}{\ell(\cI)^2\cdot (\ell(\cJ)-\ell(\cI)-2)^2}\,.
\end{eqnarray}
\item[ii-2)]
If $\mathcal{J}$ is a boundary-interval but  $\cI$ is a bulk-interval then $\ell(\mathcal{J})-\ell(\mathcal{I})\geq 3$ and $\ell(\mathcal{K})\geq \ell(\mathcal{J})-\ell(\mathcal{I})-2$ in the summation concerning (\ref{hook-V^*}). In this case we get an analogous estimate
\begin{eqnarray}
(\ref{hook-V})+(\ref{hook-V^*})&\leq &C'_{J,h}\cdot (\xi \cdot t)^{\frac{5}{8}}\cdot \frac{(\xi \cdot t)^{\frac{\ell(\cJ)-1}{8}}}{\ell(\cI)^2\cdot (\ell(\cJ)-\ell(\cI)-2)^2}\,.
\end{eqnarray}

\item[ii-3)]
If both $\mathcal{J}$ and $\cI$ are boundary-intervals then $\ell(\mathcal{J})-\ell(\mathcal{I})\geq 2$ and $\ell(\mathcal{K})\geq \ell(\mathcal{J})-\ell(\mathcal{I})-1$ in the summation concerning (\ref{hook-V^*}).  In this case
we estimate
\begin{eqnarray}
(\ref{hook-V})+(\ref{hook-V^*})&\leq &C'_{J,h} \cdot C\cdot  A_1\cdot \xi \cdot t\cdot \sum_{\ell(\mathcal{K})=\ell(\mathcal{J})-\ell(\mathcal{I})-1}^{\ell(\mathcal{J})-1}\, \frac{(\xi \cdot t)^{\frac{\ell(\cI)-1}{8}}}{\ell(\cI)^2}\cdot \frac{(\xi \cdot t)^{\frac{\ell(\mathcal{K})-1}{8}}}{\ell(\mathcal{K})^2} \\
&\leq & C'_{J,h}\cdot C\cdot  A_1\cdot \xi \cdot t\cdot \sum_{m=0}^{\ell(\cI)}\, \frac{(\xi \cdot t)^{\frac{\ell(\cI)-1}{8}}}{\ell(\cI)^2}\cdot \frac{(\xi \cdot t)^{\frac{\ell(\cJ)-\ell(\cI)+m-2}{8}}}{(\ell(\cJ)-\ell(\cI)-1+m)^2}\quad\quad\quad \\
&= & C'_{J,h}\cdot C\cdot  A_1\cdot \xi \cdot t\cdot (\xi \cdot t)^{\frac{\ell(\cJ)-3}{8}} \sum_{m=0}^{\ell(\cI)}\,  \frac{(\xi \cdot t)^{\frac{m}{8}}}{\ell(\cI)^2\cdot (\ell(\cJ)-\ell(\cI)-1+m)^2}\quad\quad \\
&\leq &C'_{J,h}\cdot (\xi \cdot t)^{\frac{3}{4}}\cdot \frac{(\xi \cdot t)^{\frac{\ell(\cJ)-1}{8}}}{\ell(\cI)^2\cdot (\ell(\cJ)-\ell(\cI)-1)^2}\,.
\end{eqnarray}
\end{enumerate}

\item[iii)]
In case c), if $\widetilde{\cI^*}$  contains both the two endpoints of $\mathcal{J}$, i.e., $\widetilde{\cI^*}=\mathcal{J}$, the possible contributions are all those in (\ref{identity-c})-(\ref{b-212}) of Definition \ref{def-interactions-multi}, and we can estimate
\begin{eqnarray}\label{arg}
\|V^{\cI}_{\mathcal{J}}\|&\leq& \|V^{\cI_{-1}}_{\mathcal{J}}\|\\
& &+\|(\ref{A-map-1-bis})\|+\|(\ref{Valgo3})\| \label{0-tilde-case}\\
& &+\|(\ref{Valgo67})\|+\|(\ref{lshighorder})\| + \| (\ref{b-21})\|+\| (\ref{b-212})\|\,.  \label{tilde-case}
\end{eqnarray}

\noindent
Notice that the terms in (\ref{0-tilde-case}) are those in (\ref{hook-V}) and (\ref{hook-V^*}). Concerning $(\ref{tilde-case})$, together with the inductive hypothesis S1) we invoke: 

\noindent
a) either (\ref{bound-S1}) or (\ref{bound-S2}) (depending on the sign of $J$) and (\ref{bound-j}) in Lemma \ref{control-LS} for the estimate of  $ \|(\ref{Valgo67})\|$, $\|(\ref{lshighorder}\|$, and  $ \| (\ref{b-21})\|$; 

\noindent
b) Lemma \ref{Lemma-LR}  for the estimate of  $\| (\ref{b-212})\|$. 

\noindent
In order to study the terms above we split the study into three parts.
\begin{enumerate}
\item[iii-1)]  If $\mathcal{\cJ}$ is a bulk-interval then $\ell(\cJ)- \ell(\cI)=2$, the contribution of (\ref{hook-V^*}) is absent (i.e., the corresponding set $[\mathcal{G}^{\cI}_{\mathcal{J}}]_2$ is empty),  and $\ell(\mathcal{K})\geq \ell(\mathcal{J})-\ell(\mathcal{I})-1$ in the sum in (\ref{hook-V}). Hence, we estimate 
\begin{eqnarray}
& &(\ref{0-tilde-case})+(\ref{tilde-case})\\
&\leq &C'_{J,h} \cdot C\cdot  A_1\cdot \xi \cdot t\cdot \sum_{\ell(\mathcal{K})=\ell(\mathcal{J})-\ell(\mathcal{I})-1}^{\ell(\mathcal{J})-1}\, \frac{(\xi \cdot t)^{\frac{\ell(\cI)-1}{8}}}{\ell(\cI)^2}\cdot \frac{(\xi \cdot t)^{\frac{\ell(\mathcal{K})-1}{8}}}{\ell(\mathcal{K})^2} \\
& &+C'_{J,h}\cdot (\xi \cdot t)^{\frac{1}{2}}\cdot \frac{(\xi \cdot t)^{\frac{\ell(\cJ)-3}{8}}}{(\ell(\cJ)-2)^2}\\
&\leq & C'_{J,h}\cdot C\cdot  A_1\cdot \xi \cdot t\cdot \sum_{m=0}^{\ell(\cI)}\, \frac{(\xi \cdot t)^{\frac{\ell(\cI)-1}{8}}}{\ell(\cI)^2}\cdot \frac{(\xi \cdot t)^{\frac{\ell(\cJ)-\ell(\cI)+m- 2}{8}}}{(\ell(\cJ)-\ell(\cI)-1+m)^2}\\
& &+C'_{J,h}\cdot (\xi \cdot t)^{\frac{1}{2}}\cdot \frac{(\xi \cdot t)^{\frac{\ell(\cJ)-3}{8}}}{(\ell(\cJ)-2)^2}\\
&= & C'_{J,h}\cdot C\cdot  A_1\cdot \xi \cdot t\cdot (\xi \cdot t)^{\frac{\ell(\cJ)-3}{8}} \sum_{m=0}^{\ell(\cI)}\,  \frac{(\xi \cdot t)^{\frac{m}{8}}}{\ell(\cI)^2\cdot (\ell(\cJ)-\ell(\cI)-1+m)^2}\quad\quad \\
& &+C'_{J,h}\cdot (\xi \cdot t)^{\frac{1}{2}}\cdot \frac{(\xi \cdot t)^{\frac{\ell(\cJ)-3}{8}}}{(\ell(\cJ)-2)^2}\\
&\leq &C'_{J,h}\cdot (\xi \cdot t)^{\frac{3}{4}}\cdot \frac{(\xi \cdot t)^{\frac{\ell(\cJ)-1}{8}}}{\ell(\cI)^2\cdot (\ell(\cJ)-\ell(\cI)-1)^2}+C'_{J,h}\cdot (\xi \cdot t)^{\frac{1}{2}}\cdot \frac{(\xi \cdot t)^{\frac{\ell(\cJ)-3}{8}}}{(\ell(\cJ)-2)^2}\quad\quad\quad 
\end{eqnarray}
where for the origin of the summand $C'_{J,h}\cdot (\xi \cdot t)^{\frac{1}{2}}\cdot \frac{(\xi \cdot t)^{\frac{\ell(\cJ)-3}{8}}}{(\ell(\cJ)-2)^2}$ see the explanations in a) and b) above.
\item[iii-2)]  If both $\mathcal{\cJ}$ and $\cI$ are boundary-intervals then $\ell(\cJ)- \ell(\cI)=1$,
the contribution of (\ref{hook-V^*}) is absent and $\ell(\mathcal{K})\geq \ell(\mathcal{J})-\ell(\mathcal{I})$ in the sum in (\ref{hook-V}). Hence, similarly to subcase iii-1),   we can estimate 
\begin{eqnarray}
\|(\ref{tilde-case})\|&\leq &C'_{J,h} \cdot C\cdot  A_1\cdot \xi \cdot t\cdot \sum_{\ell(\mathcal{K})=\ell(\mathcal{J})-\ell(\mathcal{I})}^{\ell(\mathcal{J})-1}\, \frac{(\xi \cdot t)^{\frac{\ell(\cI)-1}{8}}}{\ell(\cI)^2}\cdot \frac{(\xi \cdot t)^{\frac{\ell(\mathcal{K})-1}{8}}}{\ell(\mathcal{K})^2} \\
& &+C'_{J,h}\cdot (\xi \cdot t)^{\frac{1}{2}}\cdot \frac{(\xi \cdot t)^{\frac{\ell(\cJ)-2}{8}}}{(\ell(\cJ)-1)^2}\\
&\leq & C'_{J,h} \cdot C\cdot  A_1\cdot \xi \cdot t\cdot \sum_{m=0}^{\ell(\cI)-1}\, \frac{(\xi \cdot t)^{\frac{\ell(\cI)-1}{8}}}{\ell(\cI)^2}\cdot \frac{(\xi \cdot t)^{\frac{\ell(\cJ)-\ell(\cI)+m-1}{8}}}{(\ell(\cJ)-\ell(\cI)+m)^2}\\
& &+C'_{J,h}\cdot (\xi \cdot t)^{\frac{1}{2}}\cdot \frac{(\xi \cdot t)^{\frac{\ell(\cJ)-2}{8}}}{(\ell(\cJ)-1)^2}\\
&= & C'_{J,h} \cdot C\cdot  A_1\cdot \xi \cdot t\cdot (\xi \cdot t)^{\frac{\ell(\cJ)-2}{8}} \sum_{m=0}^{\ell(\cI)-1}\,  \frac{(\xi \cdot t)^{\frac{m}{8}}}{\ell(\cI)^2\cdot (\ell(\cJ)-\ell(\cI)+m)^2}\quad\quad \\
& &+C'_{J,h}\cdot (\xi \cdot t)^{\frac{1}{2}}\cdot \frac{(\xi \cdot t)^{\frac{\ell(\cJ)-2}{8}}}{(\ell(\cJ)-1)^2}\\
&\leq &C'_{J,h}\cdot (\xi \cdot t)^{\frac{7}{8}}\cdot \frac{(\xi \cdot t)^{\frac{\ell(\cJ)-1}{8}}}{\ell(\cI)^2\cdot (\ell(\cJ)-\ell(\cI))^2}+C'_{J,h}\cdot (\xi \cdot t)^{\frac{1}{2}}\cdot \frac{(\xi \cdot t)^{\frac{\ell(\cJ)-2}{8}}}{(\ell(\cJ)-1)^2}\,.\quad\quad\quad 
\end{eqnarray}
\item[iii-3)]
If $\cJ$ is a boundary-interval but $\cI$ is a bulk-interval then $\ell(\mathcal{J})-\ell(\mathcal{I})= 2$ and the contribution of (\ref{hook-V^*}) is not absent and in the corresponding sum we have $\ell(\mathcal{K})\geq \ell(\mathcal{J})-\ell(\mathcal{I})-1$.  Thus we estimate 
\begin{eqnarray}
\|(\ref{tilde-case})\|&\leq &C'_{J,h}\cdot (\xi \cdot t)^{\frac{3}{4}}\cdot \frac{(\xi \cdot t)^{\frac{\ell(\cJ)-1}{8}}}{\ell(\cI)^2\cdot (\ell(\cJ)-\ell(\cI)-1)^2}\\
& &+C'_{J,h}\cdot (\xi \cdot t)^{\frac{1}{2}}\cdot \frac{(\xi \cdot t)^{\frac{\ell(\cJ)-3}{8}}}{(\ell(\cJ)-2)^2}\,.
\end{eqnarray}

\end{enumerate}
\end{itemize}
\noindent
At fixed $\cJ$ and fixed $\ell(\cI)$,  the situation described in ii) happens only three times at most, namely it happens three times when $\cJ$ is a boundary-interval. At fixed $\cJ$, the situation described in iii) happens only for two intervals  $\cI$ at most, namely it happens twice when $\cJ$ is a boundary-interval. Hence, for fixed $\mathcal{L}$, by re-expanding back down to level $\cI_0$ we can estimate for $t$ sufficiently small
\begin{eqnarray}
\|V^{\mathcal{L}}_{\mathcal{J}}\|&\leq& \|V^{\cI_0}_{\mathcal{J}}\|\\
& &+\charf (\ell(\cJ)-2)\sum_{K=1}^{\ell(\cJ)-1}3\cdot C'_{J,h}\cdot (\xi \cdot t)^{\frac{7}{8}}\cdot \frac{(\xi \cdot t)^{\frac{\ell(\cJ)-1}{8}}}{K^2\cdot (\ell(\cJ)-K)^2}\\
& &+\charf (\ell(\cJ)-3)\sum_{K=1}^{\ell(\cJ)-1}3\cdot C'_{J,h}\cdot (\xi \cdot t)^{\frac{3}{4}}\cdot \frac{(\xi \cdot t)^{\frac{\ell(\cJ)-1}{8}}}{K^2\cdot (\ell(\cJ)-K-1)^2}\\
& &+\charf (\ell(\cJ)-4)\sum_{K=1}^{\ell(\cJ)-1}3\cdot C'_{J,h}\cdot (\xi \cdot t)^{\frac{5}{8}}\cdot \frac{(\xi \cdot t)^{\frac{\ell(\cJ)-1}{8}}}{K^2\cdot (\ell(\cJ)-K-2)^2}\\
&&+ C'_{J,h}\cdot (\xi \cdot t)^{\frac{1}{4}}\cdot \frac{(\xi \cdot t)^{\frac{\ell(\cJ)-1}{8}}}{(\ell(\cJ)-1)^2}\\
&\leq &\frac{(\xi \cdot t)^{\frac{\ell(\cJ)-1}{8}}}{\ell(\cJ)^2}
\end{eqnarray}
where $\charf (z)$ is the characteristic function of $[0,+\infty]$ and in the last step  we have used that $ \|V^{\cI_0}_{\mathcal{J}}\|=0$ for $\ell(\cJ)>1$.

\noindent
\emph{Induction step in the proof of S2)}

\noindent
The statement follows from Lemmata \ref{gap-bound-ferr} and \ref{gap-bound} where we assume S2) in step $\mathcal{L}_{-1}$ and exploit the result just proven for S1) in step $\mathcal{L}$.
\qed

In the next lemma,  which is part of the induction, we (also) estimate the operator norm of a potential after its block-diagonalization. This estimate applies to the subsequent steps as well, since by construction the block-diagonalized potential does not change along the flow; see the algorithm in Definition \ref{def-interactions-multi}.

\begin{lem}\label{control-LS}
Assume that $t>0$ is sufficiently small independently of $N$, and  such that S1) and S2) of Theorem \ref{th-norms} hold true in step $\cI_{-1}$. 
Then the inequalities
\begin{itemize}
\item[(a)]\begin{equation}\label{bound-S1}
\|Z_{\mathcal{I}^*}\|\leq A_1\cdot \frac{\xi \cdot t}{J+h}\cdot
\|V^{\cI_{-1}}_{\mathcal{I}}\|\,,
\end{equation}
 in the ferromagnetic case,
\item[(b)] \begin{equation}\label{bound-S2}
\|Z_{\mathcal{I}^*}\|\leq A_1\cdot \frac{\xi \cdot t}{|J|-h}\cdot 
\|V^{\cI_{-1}}_{\mathcal{I}}\|\,,
\end{equation}
in the antiferromagnetic case,
\end{itemize}
and
\begin{equation}\label{bound-j}
\sum_{j=2}^{\infty}(\xi \cdot t)^{j-1}\,\|(V^{\cI_{-1}}_{\mathcal{I}^*})^{\text{diag}}_j\| \leq \frac{C}{|J|\pm h}\cdot  \xi \cdot t\cdot \|V^{\cI_{-1}}_{\mathcal{I}}\|\,,
\end{equation}
\begin{equation}\label{bound-V}
\|V^{\cI}_{\overline{\mathcal{I}^*}}\|\leq C_{J,h}\cdot \|V^{\cI_{-1}}_{\mathcal{I}}\|\,
\end{equation}
hold true where $\pm$ are referred to the ferromagnetic and the antiferromagnetic case, respectively, and $$C_{J,h}:= 1+2\cdot \frac{A_1}{J+h}$$ in the ferromagnetic case, $$C_{J,h}:=1+2\cdot \frac{A_1}{|J|-h}$$ in the antiferromagnetic case; $A_1$ and $C$ are universal constants.

\end{lem}

\noindent
\emph{Proof} 

\noindent
Concerning  (\ref{bound-S1}), (\ref{bound-S2}), and (\ref{bound-j}) the argument is the same as in  \cite[Lemma A.3]{FP}. The inequality in (\ref{bound-V}) can be obtained directly from (\ref{L-S-series})-(\ref{diaghop2}).
\qed
\begin{figure}
\centering
\includegraphics[width=12cm]{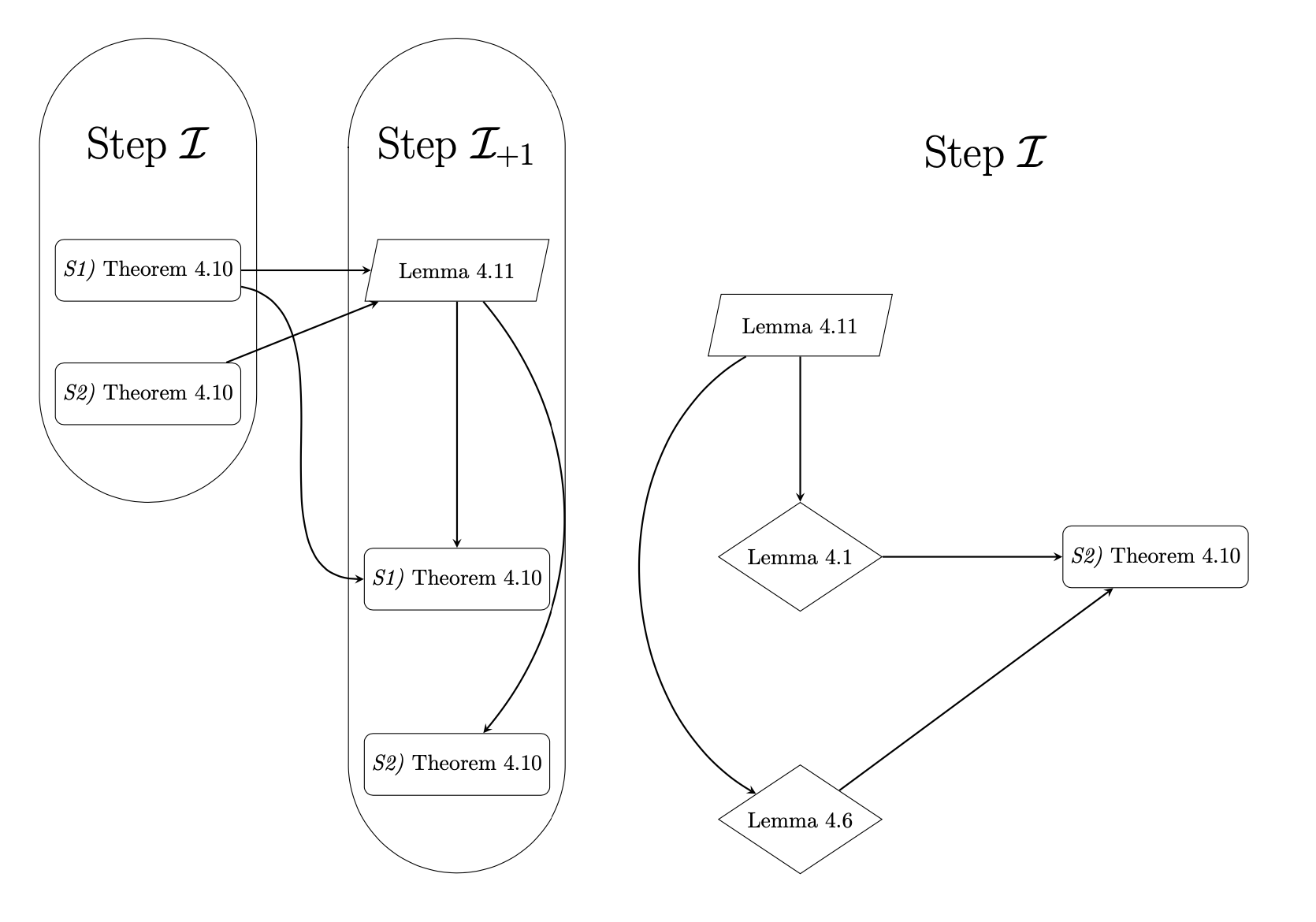}
\caption{On the left, we display the relation between Theorem \ref{th-norms} and Lemma \ref{control-LS} in the inductive step of the block-diagonalization procedure. On the right, we show how Lemma \ref{control-LS} implies \emph{S2)} in Theorem \ref{th-norms} by means of Lemmata \ref{gap-bound-ferr} and \ref{gap-bound}.}
\end{figure}

We can now prove the main result of the paper.
\begin{thm}\label{final-thm}
\begin{itemize}
\noindent
\item[(a)] \emph{If $J>0$, there exists a $\bar{t}>0$ dependent on $J$ and $h$, but independent of $N>\frac{J}{h}+\frac{3}{2}$ such that for all $|t|<\bar{t}$ the ground-state energy $E_{\Lambda}$ of the Hamiltonian $K_{\Lambda}$ in (\ref{XXZ-ham}) is non-degenerate and the spectral gap is bounded below by $2J+2h-\mathcal{O}(\frac{J}{h} \cdot |t|)$}.

\noindent
\item[(b)] \emph{If $J<0$, for $|J|>2h$ there exists a $\bar{t}>0$ dependent on $J$ and $h$, but independent of $N$ such that for all $|t|<\bar{t}$: }

\noindent
\emph{- if $\Lambda$ has an odd number of sites, the set $\mathfrak{S}:=\sigma(K_{\Lambda})\cap [E_{\Lambda}\,,\,E_{\Lambda}+2|J|-\mathcal{O}(|t|)]$,  where $\sigma(K_{\Lambda})$ is the spectrum of $K_{\Lambda}$ and $E_{\Lambda}$ its ground-state energy, consists of two points, $E_{\Lambda}$ and $E'_{\Lambda}$,  with $E'_{\Lambda}-E_{\Lambda}=2h-\mathcal{O}(|t|)$, and the spectral projection associated  with $\mathfrak{S}$ is of rank $2$; }\\
\noindent
\emph{- if $\Lambda$ has an even number of sites, the set $\mathfrak{S}:=\sigma(K_{\Lambda})\cap [E_{\Lambda}\,,\,E_{\Lambda}+2|J|-2h-\mathcal{O}( |t|)]$ consists of at most two points, $E_{\Lambda}$ and $E'_{\Lambda}$,  with $|E'_{\Lambda}-E_{\Lambda}|\leq \mathcal{O}(|t|)$, and the spectral projection associated  with $\mathfrak{S}$ is of rank $2$. }
\end{itemize}
\end{thm}

\noindent
\emph{Proof}

\noindent
By means of the conjugation $e^{Z_{\Lambda_{-1}}}$, we get the transformed Hamiltonian
\begin{eqnarray}
e^{Z_{\Lambda_{-1}}}K_{\Lambda}(t)e^{-Z_{\Lambda_{-1}}}}=G_{\Lambda}+\xi \cdot t\,V_{\Lambda}^{\Lambda_{-1}.
\end{eqnarray}
Next, we implement  a standard Lie-Schwinger block-diagonalization, w.r.t. to the pair of projections $P^{(-)}_\Lambda, P^{(+)}_{\Lambda}$; hence we have that $K_\Lambda(t)$ is unitarily equivalent to
$$\check{K}_{\Lambda}(t):=G_{\Lambda}+\xi \cdot t\,\check{V}_{\Lambda}^{\Lambda}\,,
$$
where $\check{V}_{\Lambda}^{\Lambda}$ is block diagonal w.r.t. $P^{(-)}_\Lambda, P^{(+)}_{\Lambda}$. 

\noindent
By using the results of Theorem \ref{th-norms} (combined with  Lemma \ref{control-LS}), item \emph{(a)} follows from the argument leading to Lemma \ref{gap-bound-ferr} by also including the term $\xi \cdot t \,\check{V}_{\Lambda}^{\Lambda}$. Similarly, item \emph{(b)}  follows from the arguments leading to Lemmata \ref{diffenergy} and \ref{gap-bound} by also including the term $\xi \cdot t\,\check{V}_{\Lambda}^{\Lambda}$, which, despite not being block-diagonal w.r.t. $P^{(-),A}_{\Lambda}, P^{(-),B}_{\Lambda}$, is easily controlled since its norm is estimated much smaller than $\xi \cdot t$ by the procedure of Lemma \ref{control-LS}. In order to get the result in item \emph{(b)} when $\Lambda$ has an odd number of sites,  we also exploit that with regard to $E^{A}_{{\mathcal{I}^*_{{+1}}}}$ the statement of Lemma \ref{gap-bound} can be replaced by $P^{(+)}_{\mathcal{I}^*_{{+1}}}\,(G_{{\mathcal{I}^*_{{+1}}}}-E^{A}_{{\mathcal{I}^*_{{+1}}}})\,P^{(+)}_{{\mathcal{I}^*_{{+1}}}}
\geq\,2|J|\cdot \Big[1- \mathcal{O}(\xi \cdot t) \Big]\,P^{(+)}_{\mathcal{I}^*_{{+1}}}$, as it is evident from the details of the proof. Finally, we note that since for the antiferromagnetic chain $\xi$ is independent of all the parameters defining the model  -- indeed it can be chosen to be equal to $9$ in our procedure -- all the error terms in the statement are $\mathcal{O}(|t|)$. For the ferromagnetic chain, in our procedure we have assumed $\xi > \frac{J}{h}$, hence the error terms are $\mathcal{O}(\frac{J}{h}\cdot |t|)$.
\qed

\begin{rem}\label{ferro-0}
We can treat the ferromagnetic $XXZ$ model without magnetic field similarly to the antiferromagnetic one,  in the sense that we use the algorithm in Definition \ref{def-interactions-multi} and define for any $\mathcal{J}$  the two spectral projections $P^{(-)\,A'}_{\overline{\mathcal{J}^*}}$ and $P^{(-)\,B'}_{\overline{\mathcal{J}^*}}$ which, in this case, are associated with the two vectors
$$\Psi^{A'}_{\overline{\mathcal{J}^*}}:=|\uparrow \uparrow\cdots \uparrow\,\rangle \quad\text{and}\quad \Psi^{B'}_{\overline{\mathcal{J}^*}}:=|\downarrow \downarrow \cdots \downarrow\,\rangle\,,$$
respectively. As for the antiferromagnetic case, we observe that the block-diagonalized potentials (defined in b) of Definition \ref{def-interactions-multi}) are also block-diagonal w.r.t. to $P^{(-)\,A'}_{\overline{\mathcal{J}^*}}$ and $P^{(-)\,B'}_{\overline{\mathcal{J}^*}}$. Furthermore, the two energy levels $E^{A'}_{\mathcal{I}^*_{+1}}$ and $E^{B'}_{\mathcal{I}^*_{+1}}$ of the Hamiltonian $G_{{\mathcal{I}^*_{+1}}}$ (see Section \ref{ferrogap}),  corresponding to the two eigenvectors $\Psi^{A'}_{\mathcal{I}^*_{+1}}$ and  $\Psi^{B'}_{\mathcal{I}^*_{+1}}$,  are defined as in (\ref{def-E-A}) and (\ref{def-E-B}). Next, we fix a perpendicular direction and consider a rotation by $\pi$ around it  of the spin variables, for each site of the chain.  Since all local terms of $K_{\Lambda}$ are invariant under such rotation, and since $\Psi^{A'}_{\mathcal{I}^*_{+1}}$ and  $\Psi^{B'}_{\mathcal{I}^*_{+1}}$ are mapped one to each other, we deduce that $E^{A'}_{\mathcal{I}^*_{+1}}$ and $E^{B'}_{\mathcal{I}^*_{+1}}$ coincide. Hence the statement on the spectral gap follows by essentially the same procedure used in Lemma \ref{gap-bound},  and the control of the block-diagonalization flow can be carried out as shown in this section.
\end{rem}

\newpage

\end{document}